\documentclass[12pt]{article}
\usepackage{amssymb}           
\usepackage{newtxtext,newtxmath} 
\usepackage[margin=1.25in]{geometry}
\usepackage{graphicx}
\usepackage{comment}
\usepackage{amsmath, amssymb}
\usepackage{authblk}
\usepackage{caption}
\usepackage{cite}
\usepackage{setspace}
\usepackage{hyperref}
\usepackage{lineno}
\usepackage{xcolor}
\usepackage{float}

\DeclareCaptionFont{figfont}{\footnotesize}
\usepackage{times}

\title{\normalsize\textbf{Quantum Interference and Rashba Spin-Orbit Coupling in a Chain of Planar Quantum Rings: Effects on Magnetic and Transport Properties}}

\author[1,2]{\normalsize Armen Harutyunyan\thanks{armen.harutyunyan@ysu.am}}

\affil[1]{\small \itshape Center for Modeling and Simulations of Nanostructures,
Laboratory of Solid State Physics,
Yerevan State University,
1 Al. Manoogian, 0025, Yerevan, Armenia,}
\affil[2]{\small \itshape Institute of Geological Sciences, National Academy of Sciences, 24A M. Baghramyan Ave., 0019, Yerevan, Armenia}

\date{} 

\begin{document}
\maketitle
\vspace{-2em} 

\noindent
{\small
Magneto-transport properties of a two-dimensional electron gas in a chain of planar quantum rings are investigated under the Rashba spin-orbit interaction and a transverse homogeneous magnetic field. A modulation potential function models the ring-chain periodicity along one direction and the confinement in the perpendicular one. The electron energy minibands collapse into discrete levels with high degeneracy at specific magnetic field values. The Rashba effect significantly influences the system’s properties. Calculations reveal a transition from diamagnetic to paramagnetic behavior in the spin-difference orbital magnetization at high Rashba coupling strengths. This is consistent with the reversal of the spin-difference persistent current observed at the same Rashba values. Total and spin-difference magnetizations exhibit oscillations linked to miniband nodes.
The longitudinal magnetoconductance component shows oscillations resembling Shubnikov–de Haas behavior, while the transverse component displays a ladder-like profile reminiscent of the quantum Hall effect. However, both phenomena are more closely associated with the periodic collapse of minibands, leading to strong density-of-states oscillations, rather than with the mechanisms behind the quantum Hall effect. This highlights the rich physics of quantum topological phases in nanostructures with non-trivial geometry. At high Rashba coupling, this behavior degrades. Spin magnetization shows pronounced oscillations, indicating complex interplay between the Zeeman and Rashba effects on spin polarization. These results offer insights into experimentally relevant electronic and spin characteristics attainable in modulated semiconductor structures, contributing to the development of advanced 2D-based materials for magneto-transport and spintronics applications.} \vspace{1em}


\section*{\normalsize Introduction}

Two-dimensional (2D) nanomaterials \cite{lin2023recent} play a pivotal role in both fundamental and applied research due to their exceptional physico-chemical properties. A notable example is graphene \cite{novoselov2004electric}, which demonstrates remarkable physical, optical, electronic as well as thermal \cite{kalantari2022thermal} characteristics. Since graphene’s discovery, numerous additional 2D nanomaterials comprising both organic and inorganic structures have been identified, accompanied by the development of novel fabrication methodologies \cite{zhang2015ultrathin}. Among these, the discovery of two-dimensional nanomagnets is particularly noteworthy, opening promising avenues in quantum computing, neuromorphic computing and spintronics \cite{zhang20242d, tokmachev2021two, xu2020two, khan2020recent}.
Beyond physics and electronics, 2D nanomaterials are also finding growing use in water purification and monitoring \cite{fatima2022tunable}, as well as in a range of biomedical applications \cite{murali2021emerging}. Their high surface area, chemical versatility, and biocompatibility make them promising candidates for drug delivery, particularly for challenging therapeutic targets such as brain disorders or cancer. In such contexts, 2D nanomaterials can help overcome biological barriers (e.g., the blood–brain barrier), facilitate tumor penetration, support neural repair \cite{davis20212d, agarwal2018recent, li2023biomedical}. Their unique structure also makes them excellent platforms for studying two-dimensional electron gases (2DEGs). Graphene, transition metal dichalcogenides (TMDCs), quasi-2D semiconductor systems and layered oxides can exhibit highly tunable 2DEG behavior \cite{berger2004ultrathin, mughnetsyan2019rashba,  mughnetsyan2024effect,  mckeown2018arpes, gudmundsson2025spin}, allowing detailed exploration of electron dynamics under external electric and magnetic fields. These systems are suitable for studying quantum Hall effects, spin–orbit coupling, and electronic transport phenomena in reduced dimensions \cite{zhang2018electronic, zarea2006spin, huang2015electronic}.
In parallel with these developments, quantum rings present another intriguing class of quantum nanostructures \cite{viefers2004quantum, FominBook}. Their unique geometry gives rise to quantum interference effects, including the Aharonov–Bohm effect \cite{PhysRev.115.485}, making them powerful tool for probing quantum coherence and electron transport. Advances in nanofabrication have enabled precise engineering of these systems, showing a high degree of tunability through external magnetic fields, spin–orbit coupling \cite{mohamed2018squeezing, ngo2010spin}, and laser fields.
For example, recent studies on GaAs/AlGaAs double quantum rings have demonstrated that the combined application of magnetic fields, intense nonresonant laser fields, and both Rashba and Dresselhaus spin–orbit interactions allows for fine control over energy levels \cite{mora2023double}. This tunability accentuates the potential of quantum rings in optoelectronic and spintronic applications \cite{nagasawa2013control, planelles2007electronic}. In a related study, a two-dimensional quantum ring composed of $Si_{1-x}Ge_x$ was found to sustain significant electromagnetic effects despite reductions in induced current density due to spin–orbit interaction (SOI), further emphasizing its suitability for spintronic applications under the influence of external fields \cite{prasad2023enhancement}.
The exploration of spin–orbit effects and quantum transport in nanostructured systems, where geometry and field-induced tunability play critical roles, is a highly motivating area of research. \cite{bercioux2015quantum, hurand2015field}. For instance, Ibrahim et al. \cite{ibrahim2021effects} investigated a parabolic quantum dot Hamiltonian incorporating Rashba SOI, magnetic fields, and topological defects. Their analysis revealed how variations in confinement strength, Rashba coupling, and defect topology influence energy spectra and drive transitions between diamagnetic and paramagnetic phases. Applying quantum waveguide theory, Du et al. \cite{du2011spin} analyzed spin-dependent transmission in metal/semiconductor/metal double quantum rings connected in series. They demonstrated that transmission coefficients display periodic amplitude oscillations with semiconductor ring size. Importantly, the transmission was shown to be tunable via both Rashba coupling and Aharonov–Bohm flux, features that hold significant promise for the design of future spintronic devices.
Apparently, the investigation of the Rashba effect and magnetically controlled properties in nanosystems is considered an interesting problem. Insights into spin-polarized currents, transport properties such as conductivity, conductance, and superconductivity \cite{liang2009detection, nikipar2012influence}, as well as the system’s magnetization, are of significant importance for both fundamental understanding and potential applications in spintronics and quantum technologies. In this work, we investigate the equilibrium properties of an effectively single-electron system in a one-dimensional chain of planar quantum rings, subject to a transverse homogeneous magnetic field and the Rashba spin–orbit interaction. We analyze the formation of minibands and the origin of their collapse for certain values of magnetic field, along with the system’s spin-polarized current, magnetization and conductunce. These properties provide valuable insight into spin-resolved transport phenomena in low-dimensional systems and are relevant for experimental investigations.

\section*{\normalsize Theory}
\begin{figure}[h!]
    \centering
    \includegraphics[width=0.5\textwidth]{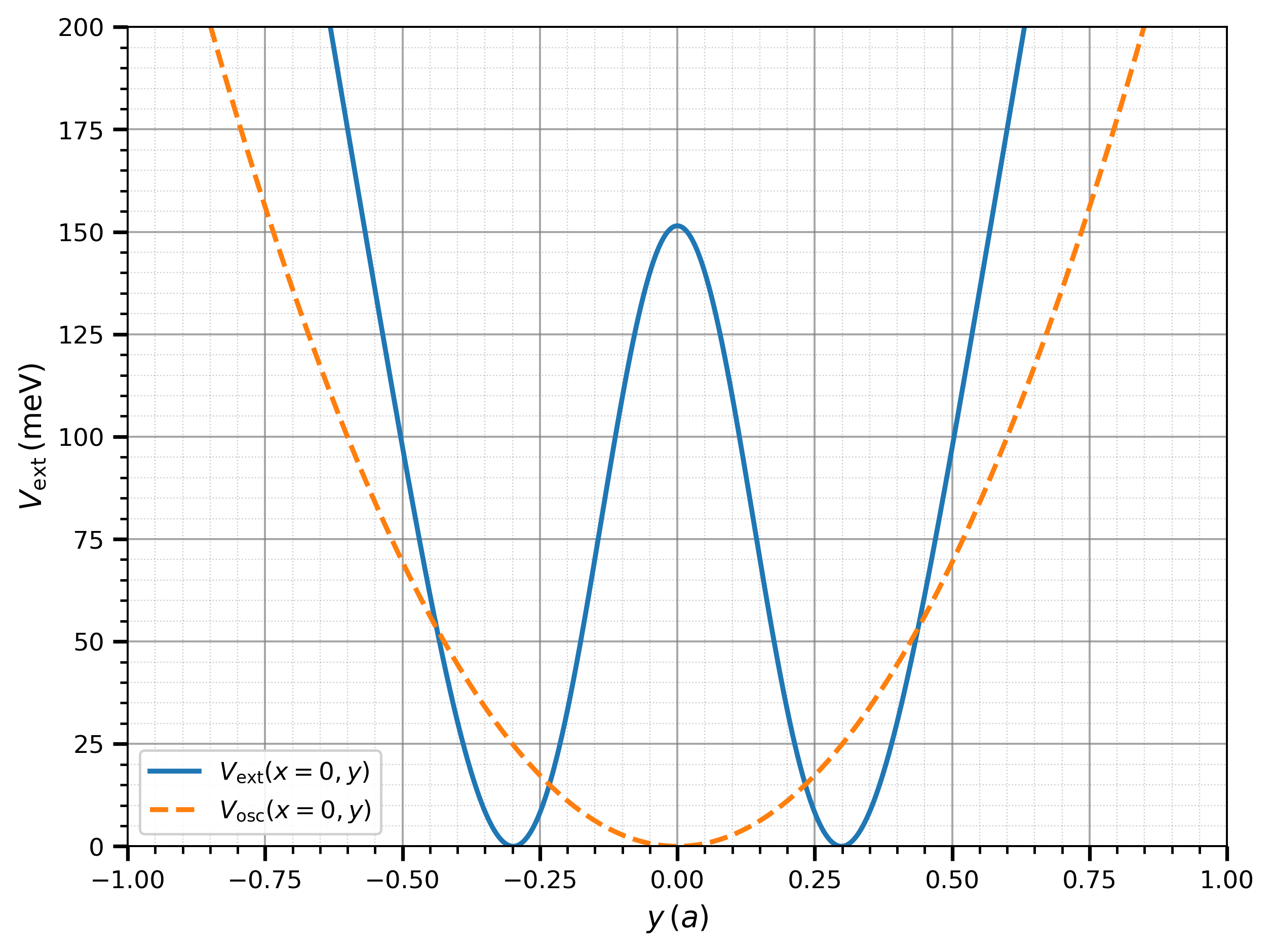}
    \caption{The transverse cross section of potential~\eqref{eq:Vpot} at \( x = 0 \) is illustrated along with the harmonic oscillator potential of corresponding strength.}
    \label{harmonic}
\end{figure}

We use the modulation potential implemented in our previous work \cite{harutyunyan2025magneto}. We refer the reader to that work for a full description and here provide its analytical form for completeness: 
\begin{equation}
V_{\mathrm{ext}}(x, y) = V_0 \left[ -v_1 \cos^2\left(\frac{g x}{2}\right) e^{-\gamma_1 (g y)^2} + v_2 \cos^4\left(\frac{g x}{2}\right) e^{-\gamma_2 (g y)^2} + v_3 (g y)^2 \right]
\label{eq:Vpot}
\end{equation}
where: \( g = \frac{2\pi}{a} \) is the primitive vector of the reciprocal lattice for a chain with period \( a \), and the dimensionless parameters have the following values \( v_1 = 1.0 \), \( v_2 = 1.0 \), \( v_3 = 0.05 \), \( \gamma_1 = 0.15 \), and \( \gamma_2 = 0.45 \). These values provide a periodic structure along the \( x \)-direction with a ring-like confinement in each unit cell and the transverse confinement in the \( y \)-direction. The harmonic oscillator basis is used to solve the Schrödinger equation, taking advantage of the similarity between the confinement of potential~\eqref{eq:Vpot} in the \( y \)-direction and the harmonic oscillator potential evaluated at \( x = 0 \) (see Fig.~\ref{harmonic}).
The full single-particle Hamiltonian including the kinetic energy, external modulation potential, Rashba spin-orbit interaction, and Zeeman coupling in the presence of a uniform external magnetic field \( \mathbf{B} \) applied perpendicular to the system is given by Eq.~\eqref{eq:full-hamiltonian}, where \( \hat{\mathbf{p}} = -i\hbar\nabla \) is the canonical momentum operator, \( \mathbf{A} \) is the magnetic vector potential (with \( \nabla \times \mathbf{A} = \mathbf{B} \)), \( \boldsymbol{\sigma} = (\sigma_x, \sigma_y, \sigma_z) \) is the vector of Pauli matrices, \( \alpha \) is the Rashba coupling strength, \( g \) is the effective Landé g-factor, and \( \mu_{\mathrm{B}}\) is the Bohr magneton. The wavefunction \( \psi(\mathbf{r}) \) is a two-component spinor that accounts for the spin states of the electron:
\begin{equation}
\left[
\frac{1}{2m} \left( \hat{\mathbf{p}} + e\mathbf{A} \right)^2 
+ V_{ext}(\mathbf{r}) 
+ \frac{\alpha}{\hbar} \, \hat{\mathbf{z}} \cdot \left( \boldsymbol{\sigma} \times \left( \hat{\mathbf{p}} + e\mathbf{A} \right) \right) 
+ \frac{g \mu_B}{2} \, \boldsymbol{\sigma} \cdot \mathbf{B}
\right] \psi(\mathbf{r}) = E \psi(\mathbf{r})
\label{eq:full-hamiltonian}
\end{equation}
We adopt the Landau gauge, \( \mathbf{A} = (-By, 0, 0) \), which generally reflects the symmetry of the considered system. The hamiltonian matrix  \emph{H} acts on the spinor in the following way:
\begin{equation}
\mathcal{H} 
\begin{pmatrix} \psi_\uparrow \\ \psi_\downarrow \end{pmatrix}
= E 
\begin{pmatrix} \psi_\uparrow \\ \psi_\downarrow \end{pmatrix},
\end{equation}
where
\begin{equation}
\mathcal{H} = 
\begin{pmatrix} 
H_{11} & H_{12} \\ 
H_{21} & H_{22} 
\end{pmatrix}.
\end{equation}
After some algebraic manipulations, we obtain the following expressions for the matrix elements of the Hamiltonian:
\begin{equation}
\scalebox{0.85}
H_{11} = \frac{1}{2m} \left( p_x^2 - 2eBy\, p_x + e^2 B^2 y^2 \right) + V_{\text{ext}}(x,y) + \frac{g \mu_B B}{2}
\end{equation}
\begin{equation}
\scalebox{0.85}
H_{22} = \frac{1}{2m} \left( p_x^2 - 2eBy\, p_x + e^2 B^2 y^2 \right) + V_{\text{ext}}(x,y) - \frac{g \mu_B B}{2}
\end{equation}
\begin{equation}
\scalebox{0.85}
H_{12} =\frac{\alpha}{\hbar} \left( \hat{p}_y + i\hat{p}_x - i eBy \right)
\end{equation}
\begin{equation}
\scalebox{0.85}
H_{21} =\frac{\alpha}{\hbar} \left( \hat{p}_y - i\hat{p}_x + i eBy \right)
\end{equation}
We seek the solution of the Hamiltonian in the following form:
\begin{equation}
    \psi_{\uparrow,\downarrow}(x, y) = \sum_{n, G} C_{\uparrow,\downarrow}^{(n, G)} \, \frac{e^{i (k + G) x}}{\sqrt{a}} \, \phi_n(y),
\end{equation}
where the arrows \(\uparrow\) and \(\downarrow\) denote spin-up and spin-down states, respectively. The reciprocal lattice vector \(G\) is given by
\[
G = g n_x, \quad n_x \in \mathbb{Z}.
\]
The function \(\phi_n(y)\) is the \(n^\text{th}\) eigenfunction of the harmonic oscillator:
\begin{equation}
    \phi_n(y) = \frac{1}{\sqrt{2^n n!}} \left(\frac{m \omega}{\pi \hbar}\right)^{\frac{1}{4}} e^{-\frac{m \omega y^2}{2 \hbar}} H_n\left(\sqrt{\frac{m \omega}{\hbar}} y\right),
\end{equation}
where \(\omega\) is the oscillator frequency and \(H_n\) is the Hermite polynomial of order \(n\). After integration, we obtain the following Hamiltonian matrix components:
\begin{equation}
\begin{aligned}
H_{ss} = \sum_{n, G} C_{s}^{(n, G)} \Bigg(&
    \left[ \frac{\hbar^2 (k+G)^2}{2m} + \hbar \omega \left(\frac{1}{2} + n\right)
    + \sigma_s \frac{g \mu_B B}{2} \right] \delta_{G', G} \delta_{n, n'} \\
    & - \delta_{G', G} \frac{\hbar \omega}{2} \int \xi^2 f_{n'}^*(\xi(y)) f_n(\xi(y)) \, \mathrm{d}y  
      + \int V_{G' - G}(y) f_{n'}^*(\xi(y)) f_n(\xi(y)) \, \mathrm{d}y
\Bigg)
\end{aligned}
\end{equation}
\begin{flushleft}
\noindent where \( \sigma_s = +1 \) for spin \( s = \uparrow \) and \( \sigma_s = -1 \) for \( s = \downarrow \).
\end{flushleft}
\begin{equation}
H_{s s'} = i \alpha\sum_{n, G} C_{s, s'}^{(n, G)} \, \delta_{G', G} \,   \Bigg[
      \sqrt{\frac{m \omega}{\hbar}} \int \xi(y) \, f_{n'}^*(\xi(y)) f_{n}(\xi(y)) \, \mathrm{d}y 
    - \sqrt{2n}  \sqrt{\frac{m \omega}{\hbar}} \, \delta_{n', n-1} 
    + \sigma_{s s'}   (k+G) \, \delta_{n', n}
\Bigg]
\end{equation}
\noindent where \( \sigma_{s s'} = +1 \) for \( (s, s') = (\downarrow, \uparrow) \), and \( \sigma_{s s'} = -1 \) for \( (s, s') = (\uparrow, \downarrow) \),
\begin{equation}
    \sqrt{\frac{m \omega}{\hbar}} y \equiv \xi(y),
\end{equation}
\begin{equation}
   \frac{1}{\sqrt{2^n n!}} \left(\frac{m \omega}{\pi \hbar}\right)^{\frac{1}{4}}  e^{-\frac{\xi(y)^2}{2}} H_n(\xi(y)) \equiv  f_n(\xi(y)) 
\end{equation}
$V_{G}$ is the external potential Fourier transform:
\begin{equation}
\begin{aligned}
V_{G} = V_0 \Bigg\{ 
& -v_1 e^{-\gamma_1 (g y)^2} \cdot \frac{1}{4} (2 \delta_{n_x,0} + \delta_{n_x,1} + \delta_{n_x,-1}) \\
& + v_2 e^{-\gamma_2 (g y)^2} \cdot \frac{1}{4} \left( \frac{3}{2} \delta_{n_x,0} + \delta_{n_x,1} + \delta_{n_x,-1} + \frac{\delta_{n_x,2} + \delta_{n_x,-2}}{4} \right) 
 + v_3 (g y)^2 \cdot \delta_{n_x,0} 
\Bigg\}
\end{aligned}
\label{vGy}
\end{equation}
$\delta_{i,j}$ is the Kronecker delta.\\\\
The probability current density \( \mathbf{j}(\mathbf{r}, t) \) associated with a wave function \( \psi(\mathbf{r}, t) \) is given by:
\begin{equation}
\mathbf{j}(\mathbf{r}, t) = \frac{\hbar}{2mi} \left[ \psi^*(\mathbf{r}, t) \nabla \psi(\mathbf{r}, t) - \psi(\mathbf{r}, t) \nabla \psi^*(\mathbf{r}, t) \right],
\end{equation}
The magnetization of the spin-oriented system can be calculated as follows:
\begin{equation}
\langle M_{s} \rangle = \frac{1}{2A_{uc}} \int_{\mathrm{UC}} \left( \mathbf{r} \times \langle \mathbf{j}^{n,\mathbf{k}}_s(\mathbf{r}) \rangle \right) \, d^{2}\mathbf{r}.
\end{equation}
Here, \( s \) corresponds to the spin-up (\( \uparrow \)) and spin-down (\( \downarrow \)) states, \( n \) is the band index, and \( \mathbf{k} \) is the wave vector.

The Kubo-Greenwood formula for a non-interacting system is used to obtain conductivity tensor \cite{huhtinen2023conductivity}:
\begin{equation}
\sigma_{\mu\nu}(\omega) = \frac{e^2\hbar}{iA_{uc}} \sum_{\mathbf{k}} \sum_{n,m} 
\frac{n_F(\epsilon_n(\mathbf{k})) - n_F(\epsilon_m(\mathbf{k}))}{\epsilon_n(\mathbf{k}) - \epsilon_m(\mathbf{k})} 
\cdot 
\frac{[v_\mu(\mathbf{k})]_{nm} [v_\nu(\mathbf{k})]_{mn}}{\epsilon_n(\mathbf{k}) - \epsilon_m(\mathbf{k}) + \hbar\omega + i\eta}.
\end{equation}
Here, \( e \) is the elementary charge, \( \hbar \) is the reduced Planck constant, and \( A_{uc} \) is the effective area of the unit cell with sufficient width in the \( y \)-direction to fully accommodate the particle within the system, \( \mathbf{k} \) is the quasi-momentum, \( n \) and \( m \) are band indices, and \( \epsilon_n(\mathbf{k}) \) is the energy of the \(n\)-th band at momentum \( \mathbf{k} \). The function \( n_F(\epsilon) \) is the Fermi-Dirac distribution, \( [v_\mu(\mathbf{k})]_{nm} \) is the matrix element of the velocity operator in the \( \mu \)-direction (where \( \mu = x \) or \( y \)), and \( \eta \) is a small positive parameter that accounts for broadening effects (e.g., due to scattering). 

To calculate the spin magnetization \( \mathbf{S} \), we use the following formula:
\begin{equation}
\langle S \rangle = -\mu_B \cdot g \cdot \frac{1}{A_{\text{uc}}} \int_{\text{uc}} P(\mathbf{r}) \, d^2\mathbf{r}
\label{eq:spin}
\end{equation}
where the spin polarization is:
\begin{equation}
    P(\mathbf{r}) = \frac{\rho_\uparrow(\mathbf{r}) - \rho_\downarrow(\mathbf{r})}{\rho_\uparrow(\mathbf{r}) + \rho_\downarrow(\mathbf{r})}
\end{equation}
and \( \rho_\uparrow \) and \( \rho_\downarrow \) represent the spin-up and spin-down electron densities, respectively.
This yields the net spin magnetization per unit cell due to the spin texture induced by Rashba spin-orbit coupling and the external magnetic field. 

\section*{\normalsize Disscusion}
\begin{figure}[H]
    \centering
    \setlength{\tabcolsep}{1pt} 
    \renewcommand{\arraystretch}{0.5} 
    \begin{tabular}{cccc}
        \includegraphics[width=0.195\textwidth]{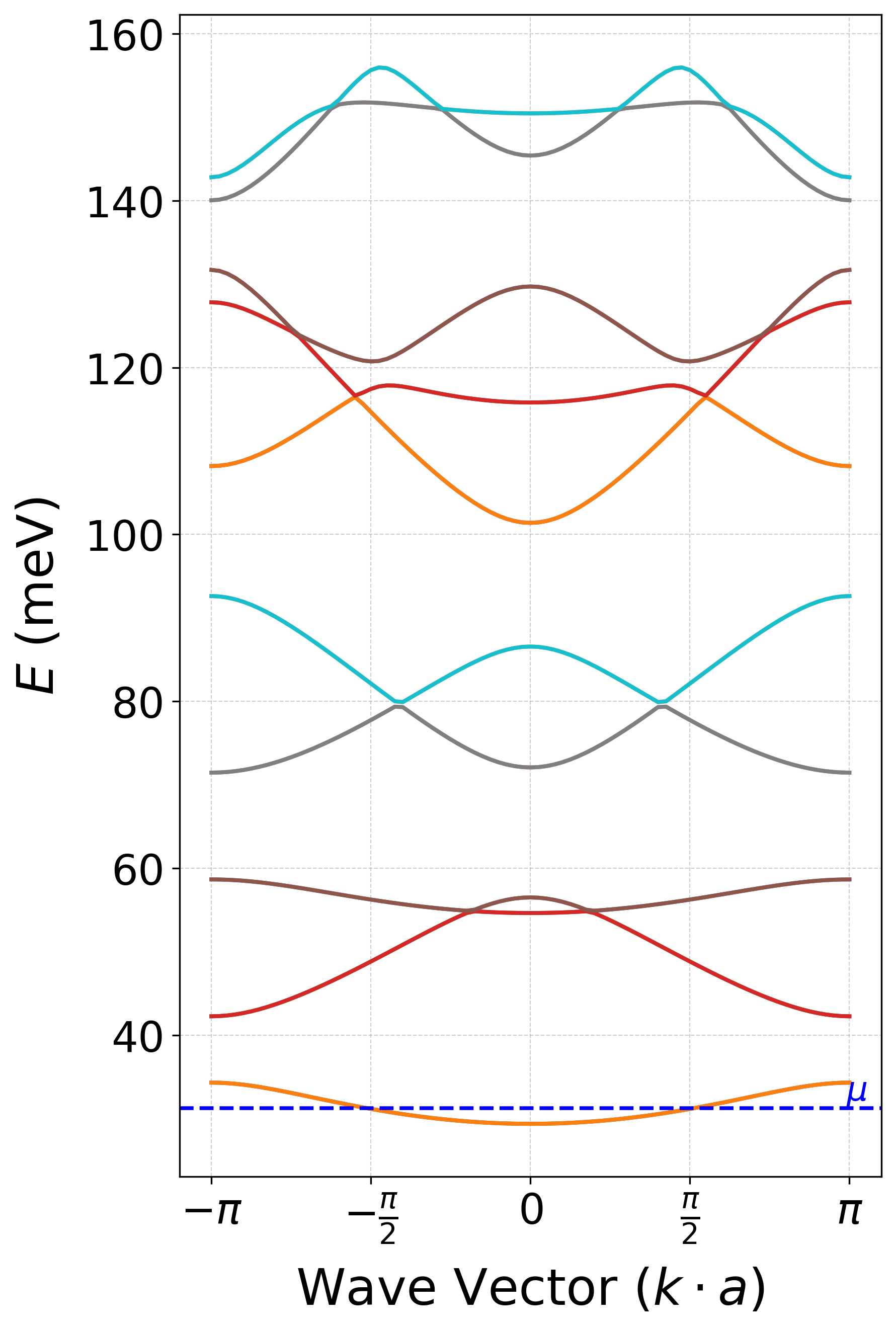} &
        \includegraphics[width=0.195\textwidth]{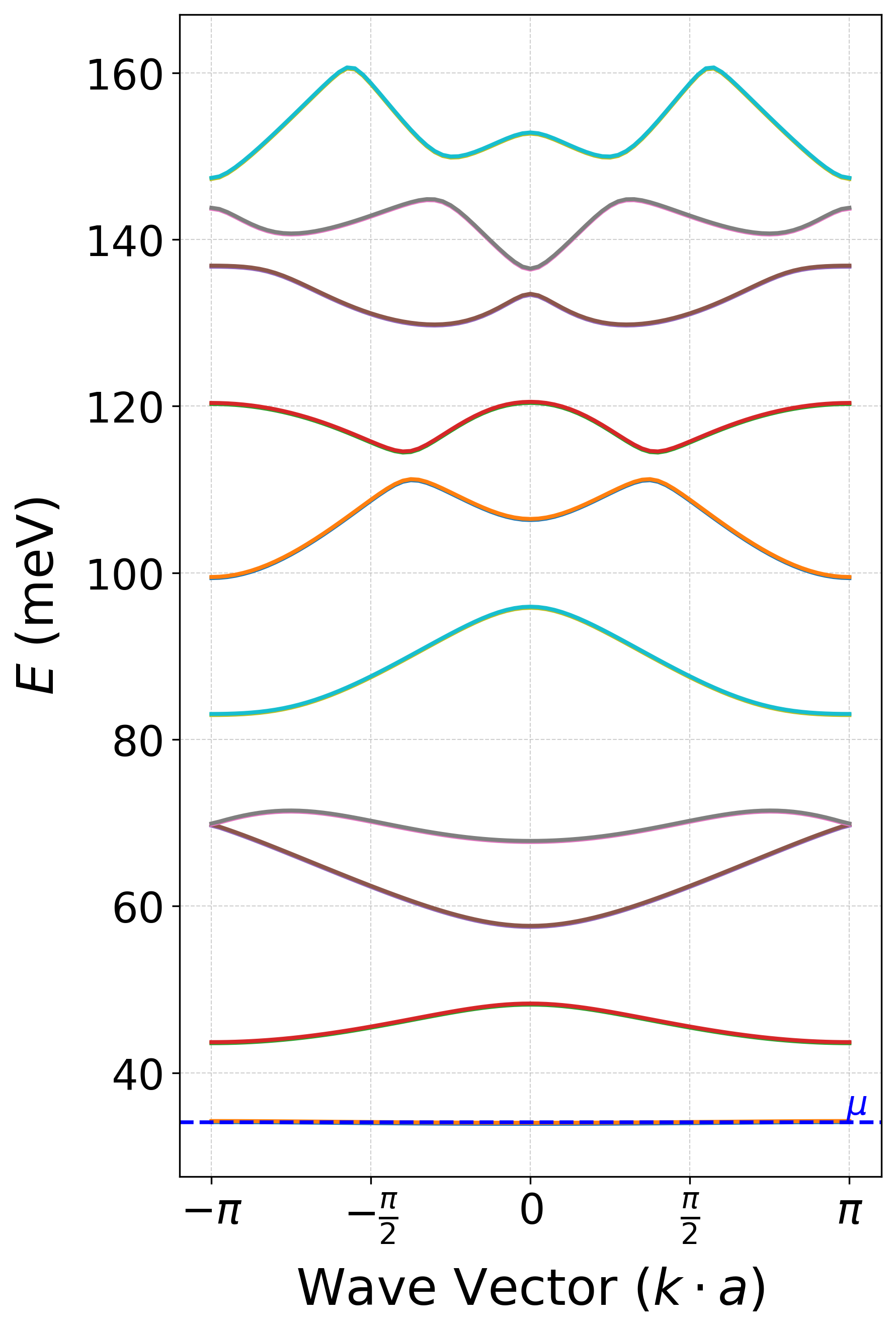} &
        \includegraphics[width=0.195\textwidth]{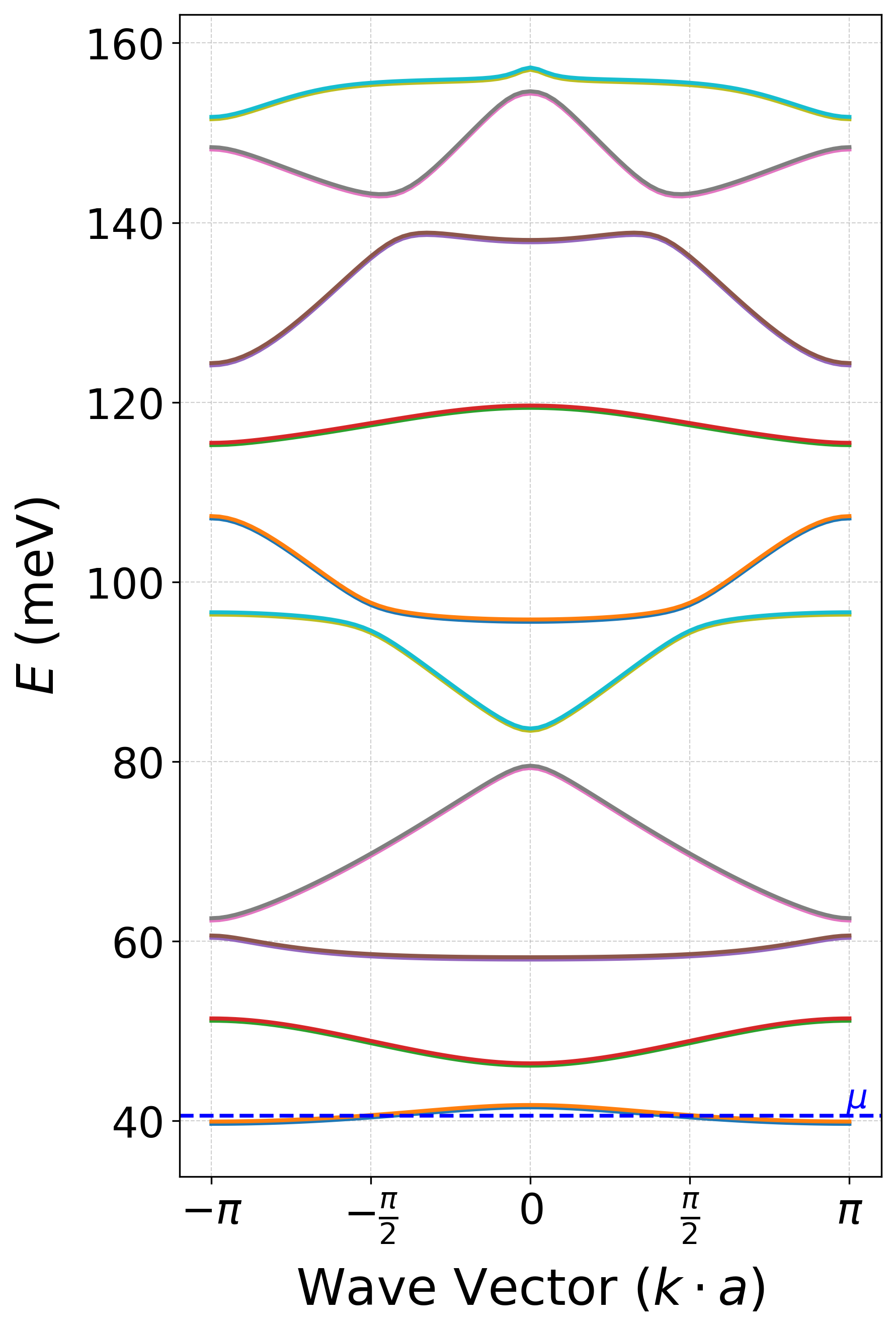} &
        \includegraphics[width=0.195\textwidth]{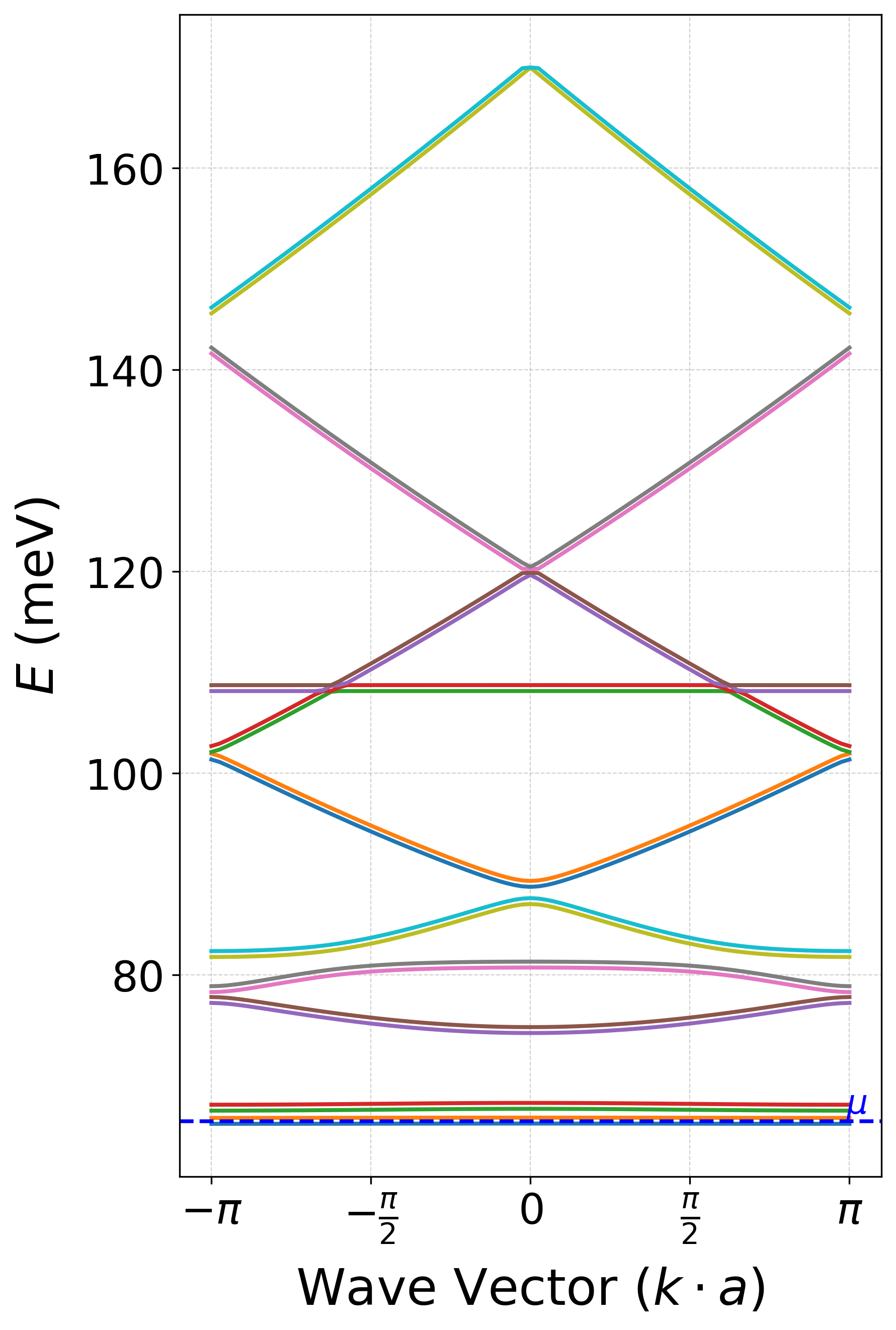} \\
        \includegraphics[width=0.195\textwidth]{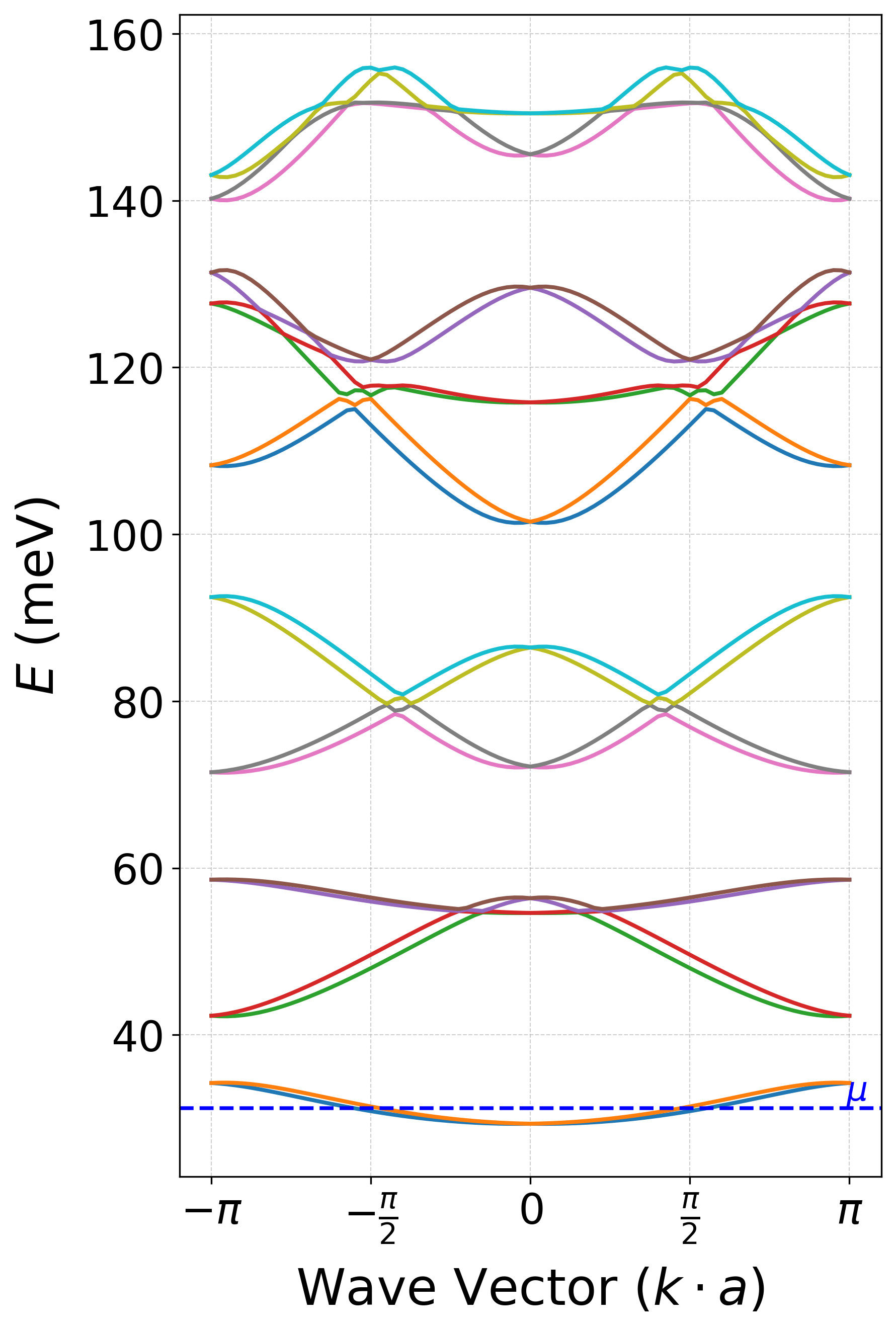} &
        \includegraphics[width=0.195\textwidth]{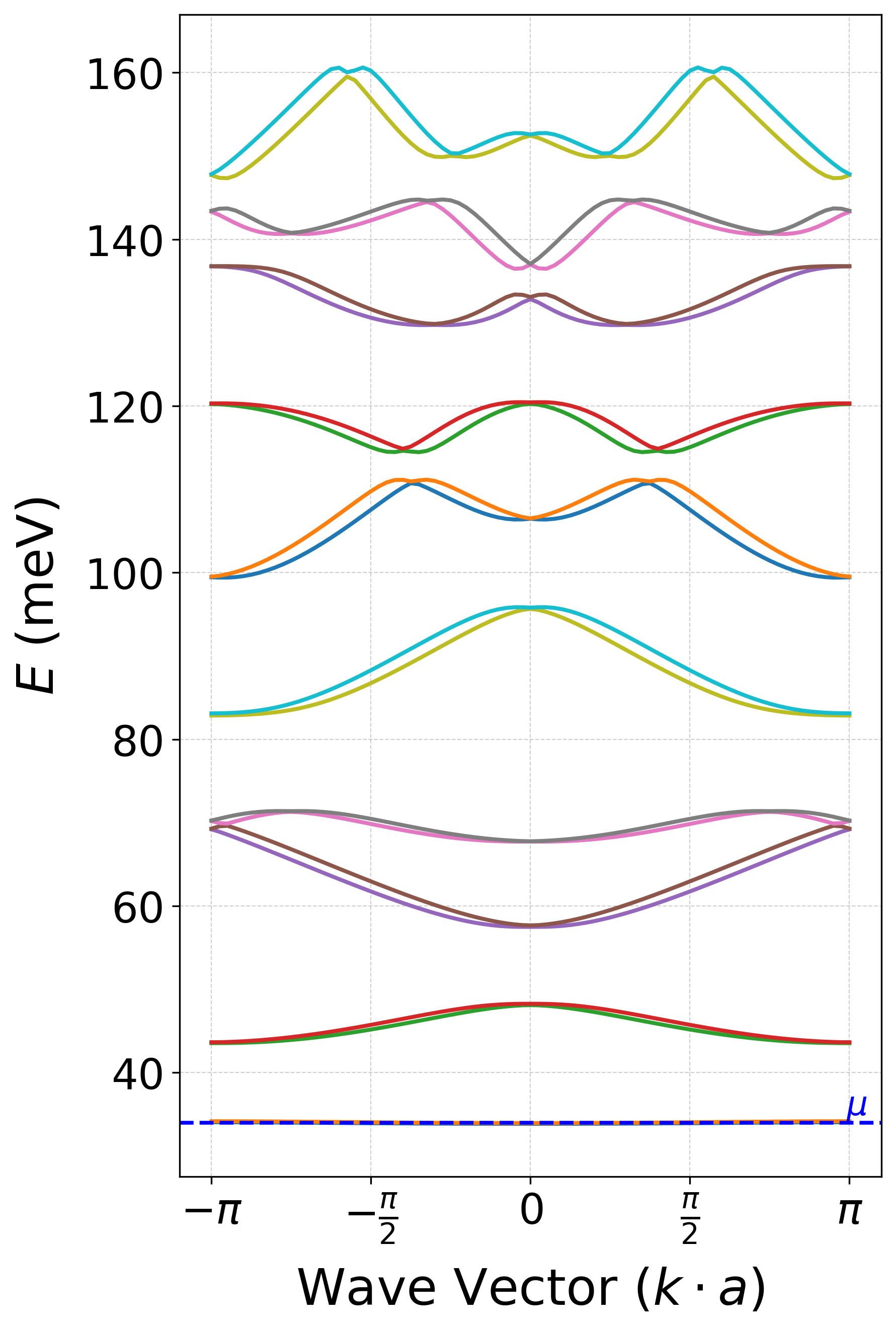} &
        \includegraphics[width=0.195\textwidth]{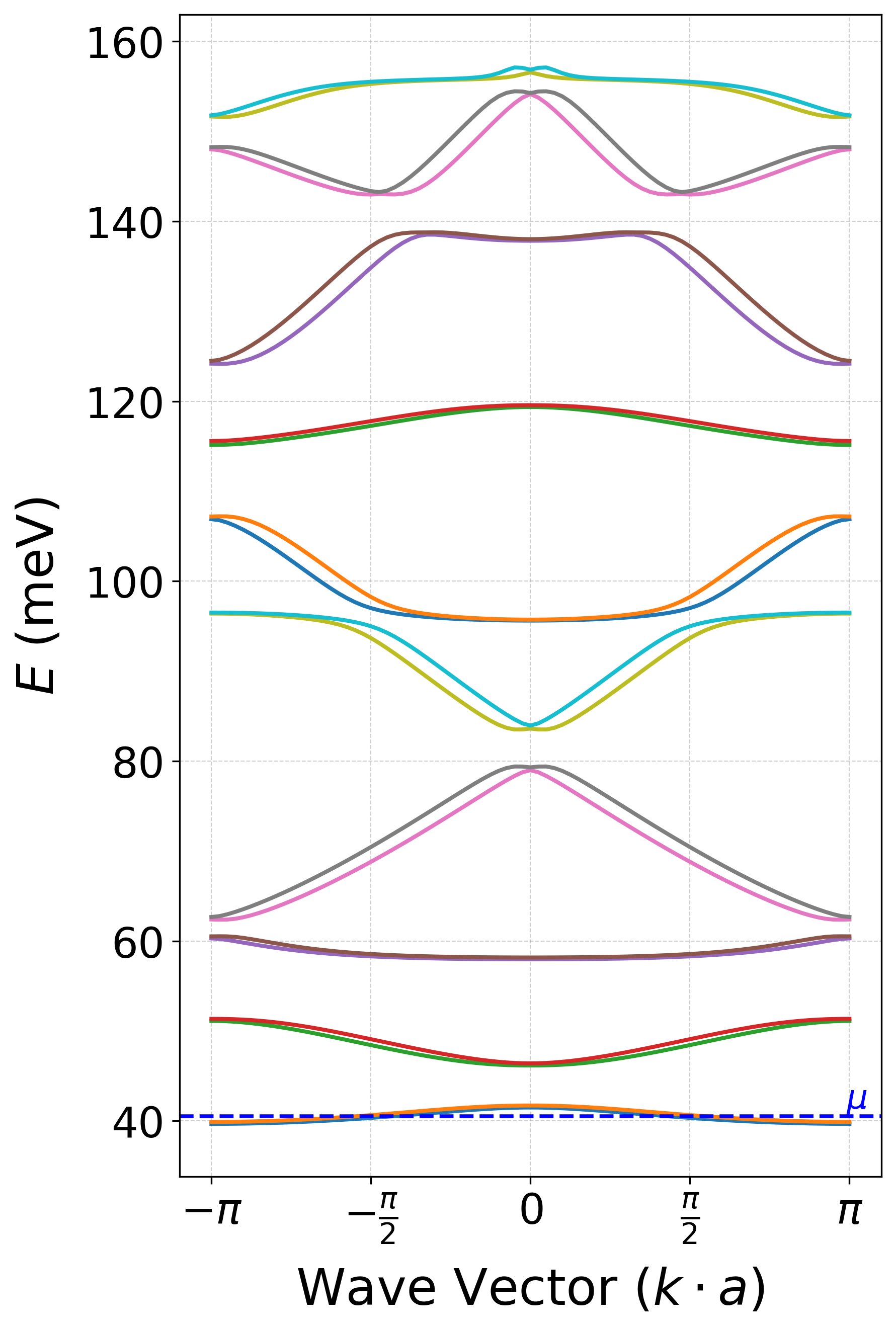} &
        \includegraphics[width=0.195\textwidth]{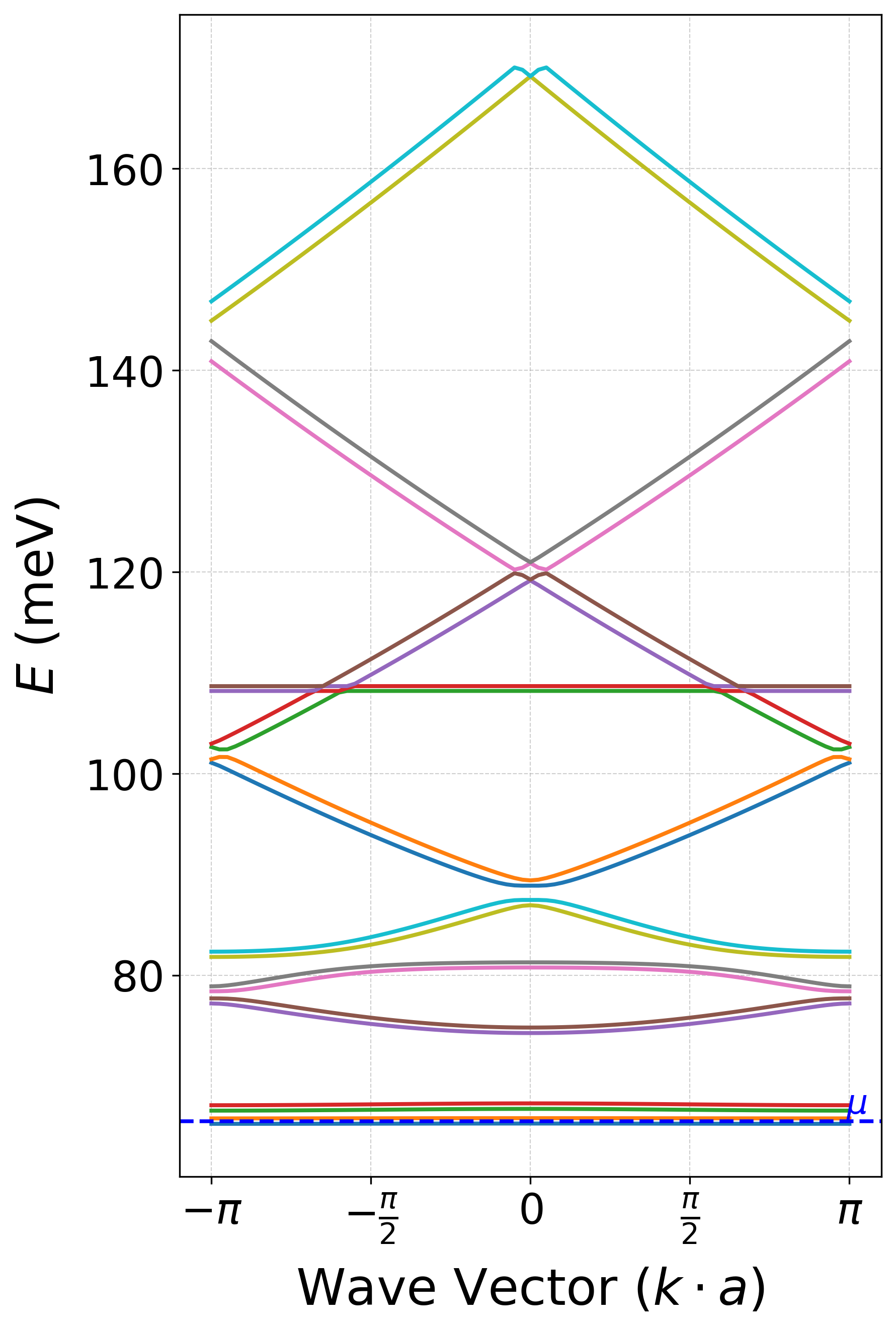} \\
        \includegraphics[width=0.195\textwidth]{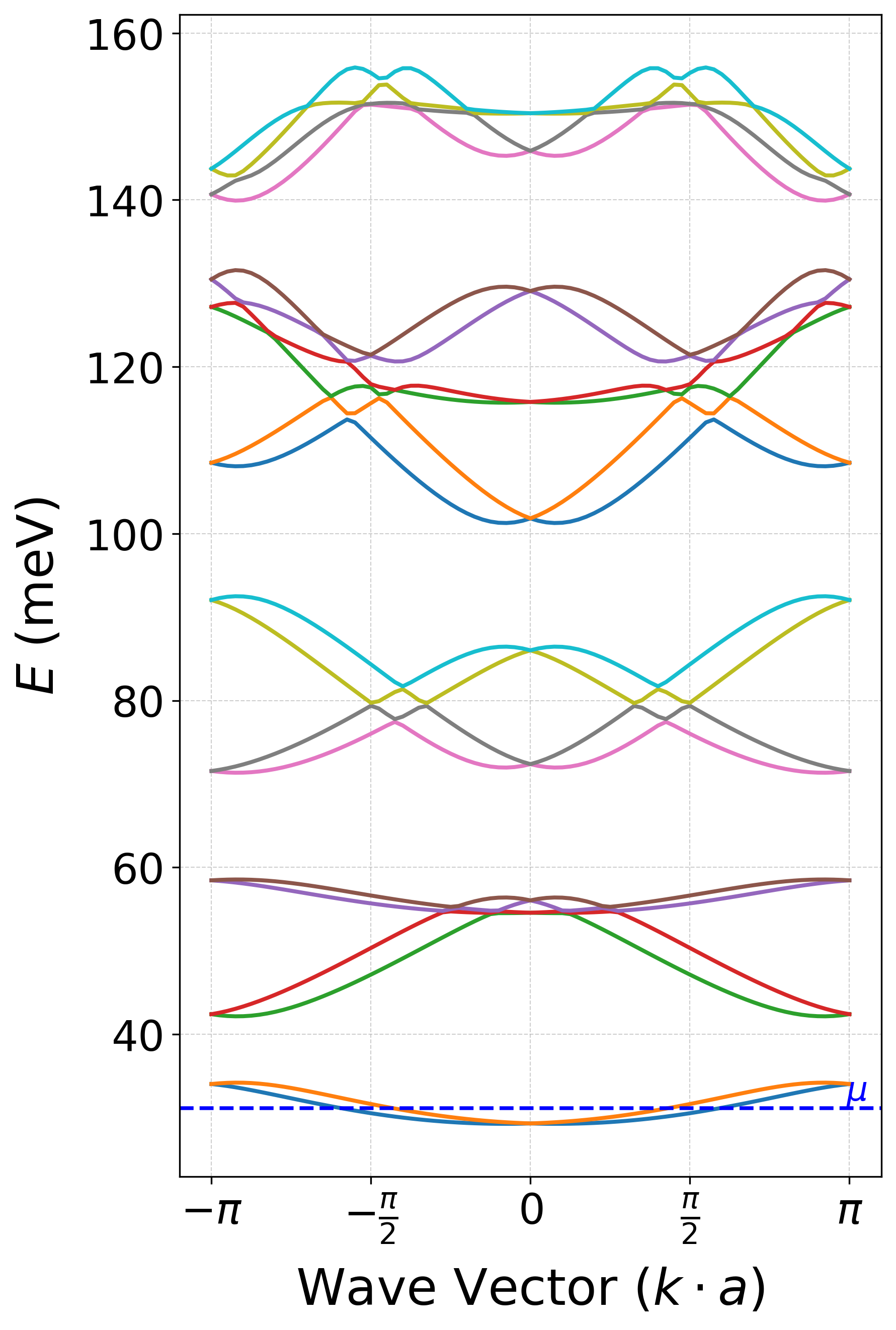} &
        \includegraphics[width=0.195\textwidth]{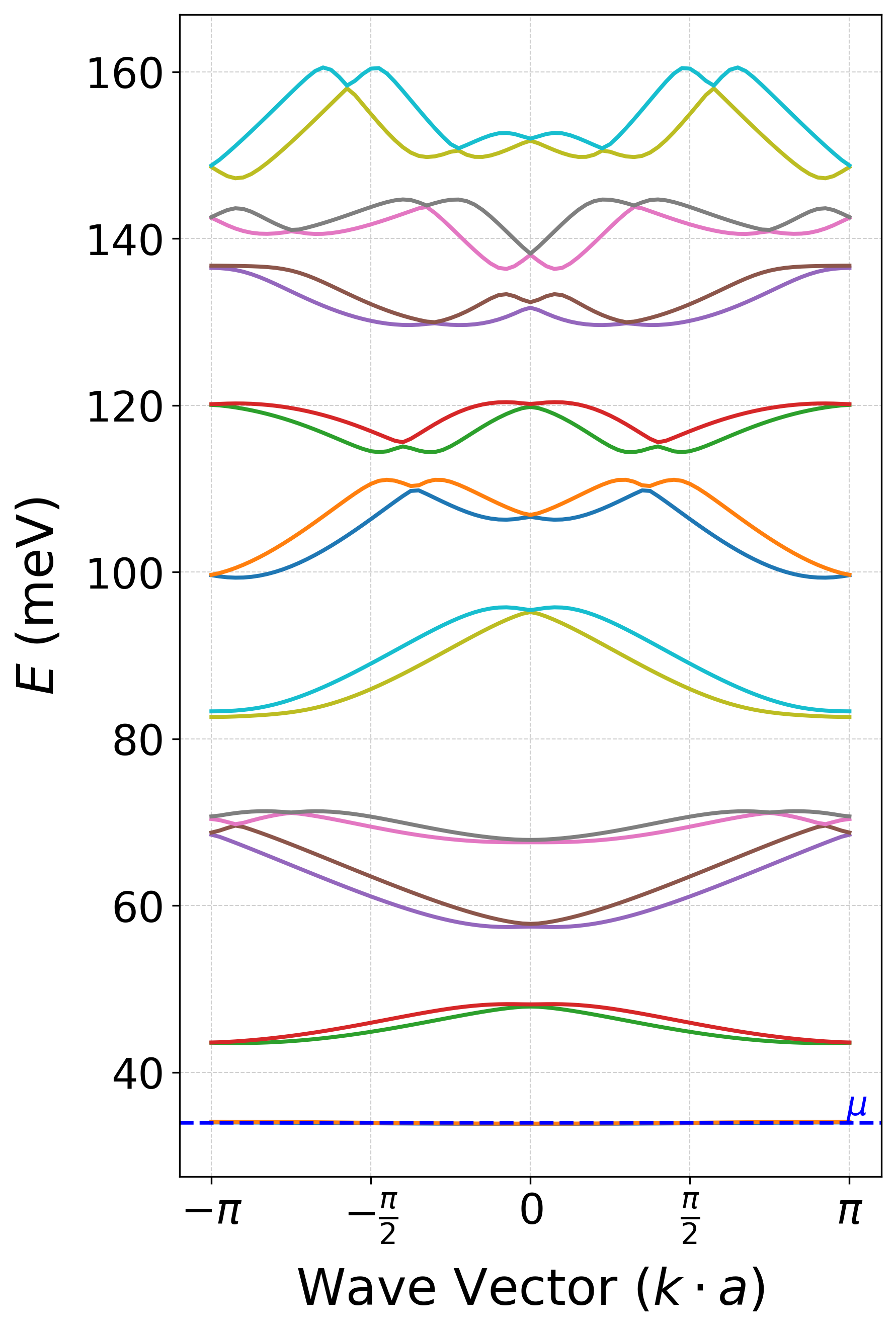} &
        \includegraphics[width=0.195\textwidth]{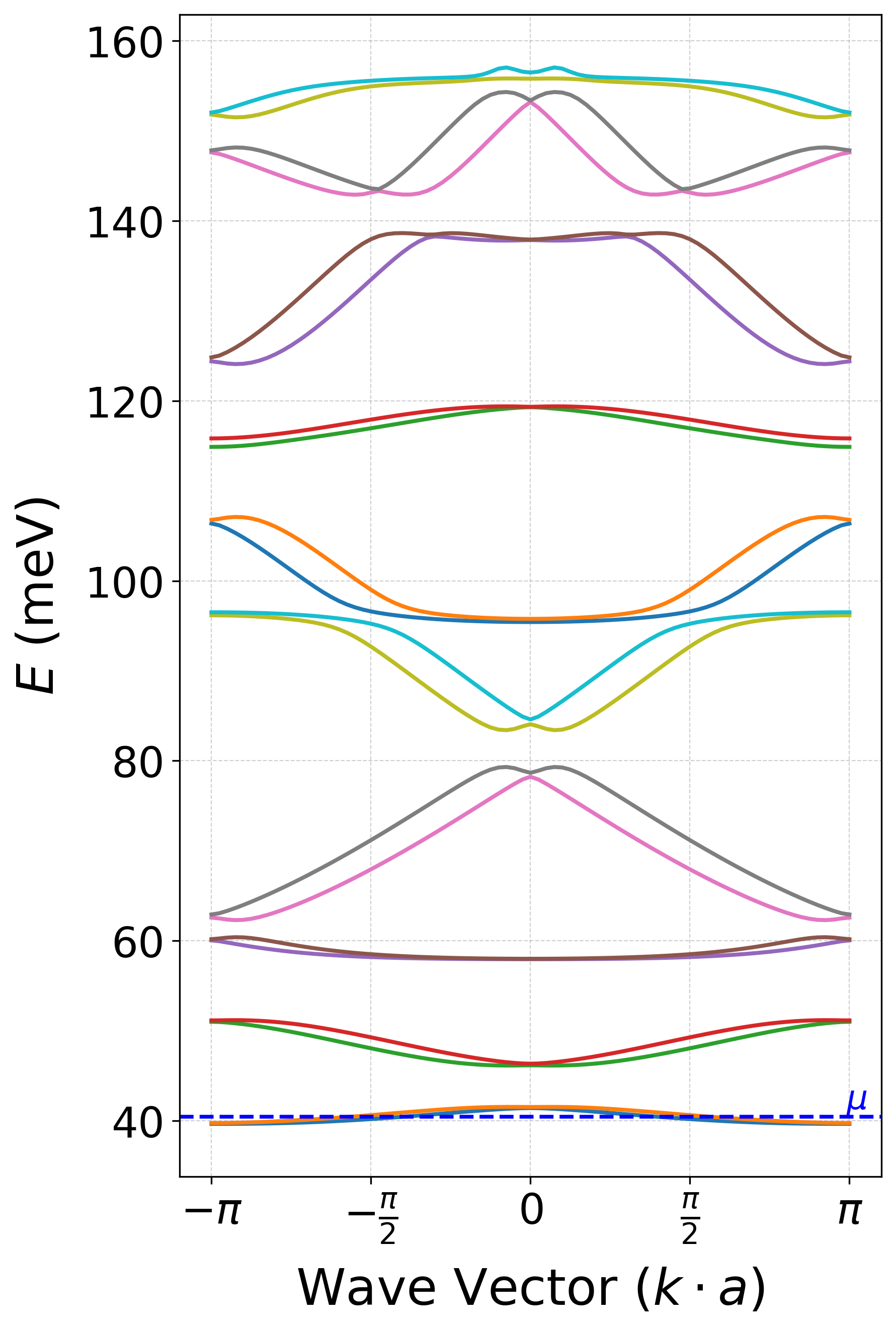} &
        \includegraphics[width=0.195\textwidth]{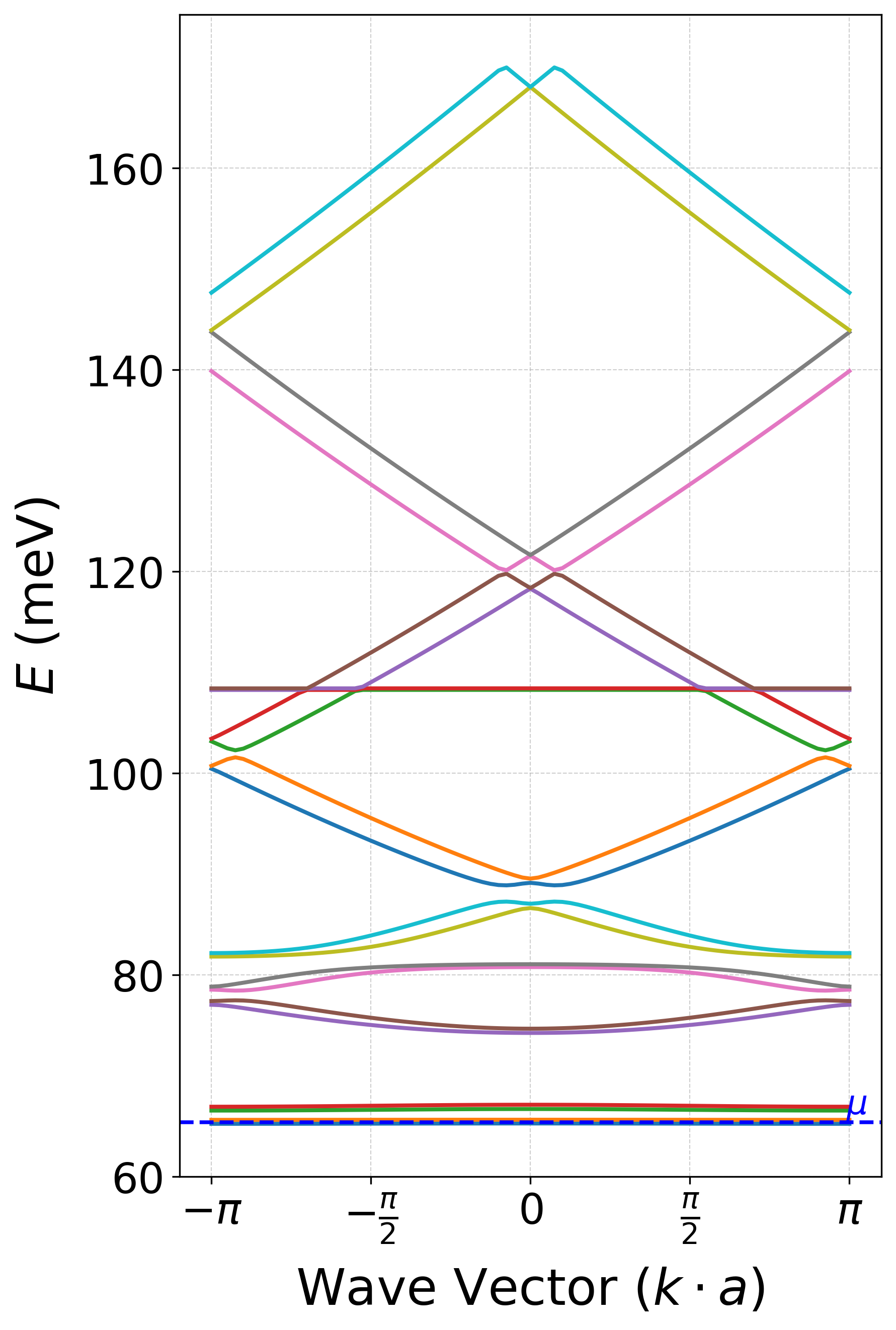} \\
        \includegraphics[width=0.195\textwidth]{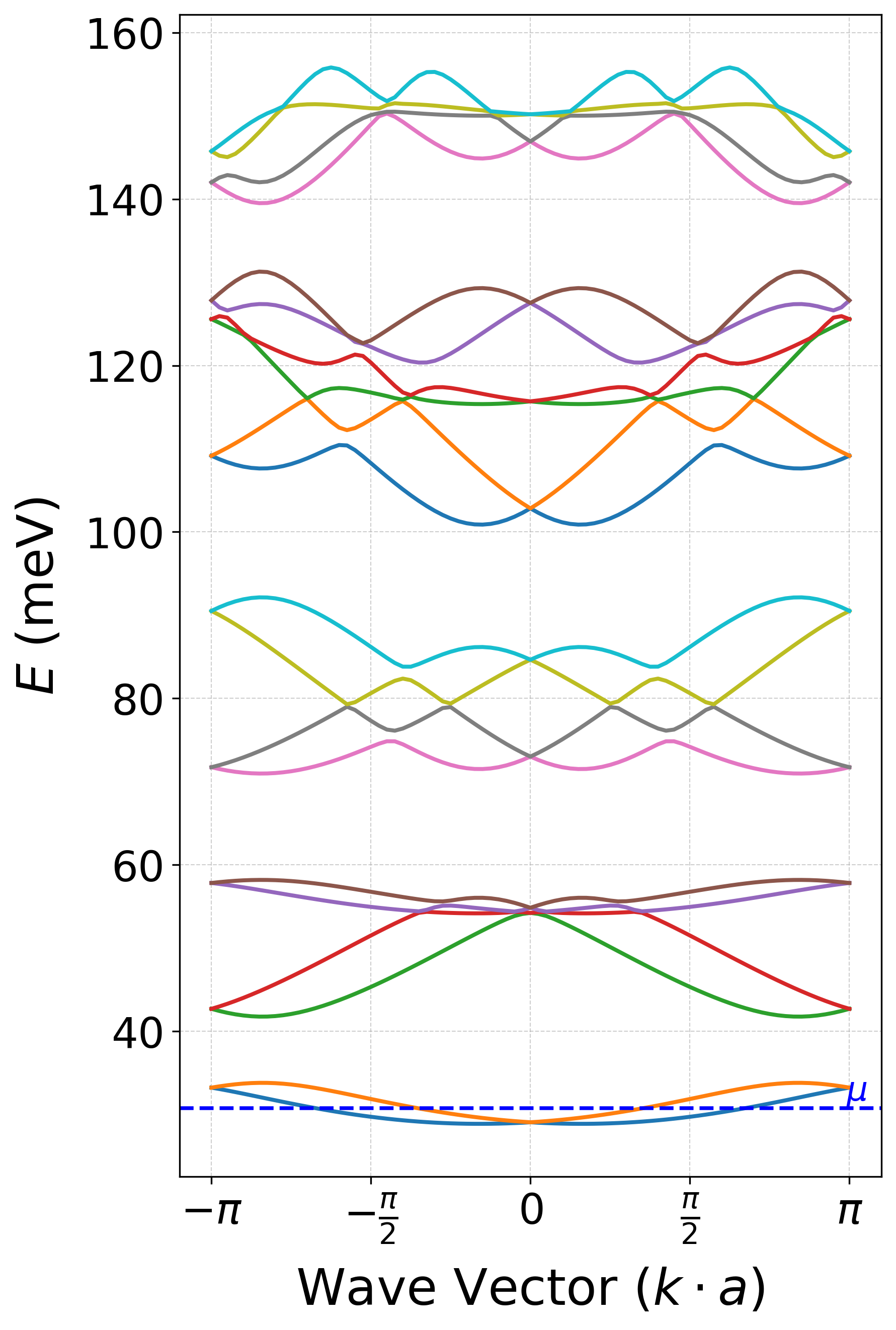} &
        \includegraphics[width=0.195\textwidth]{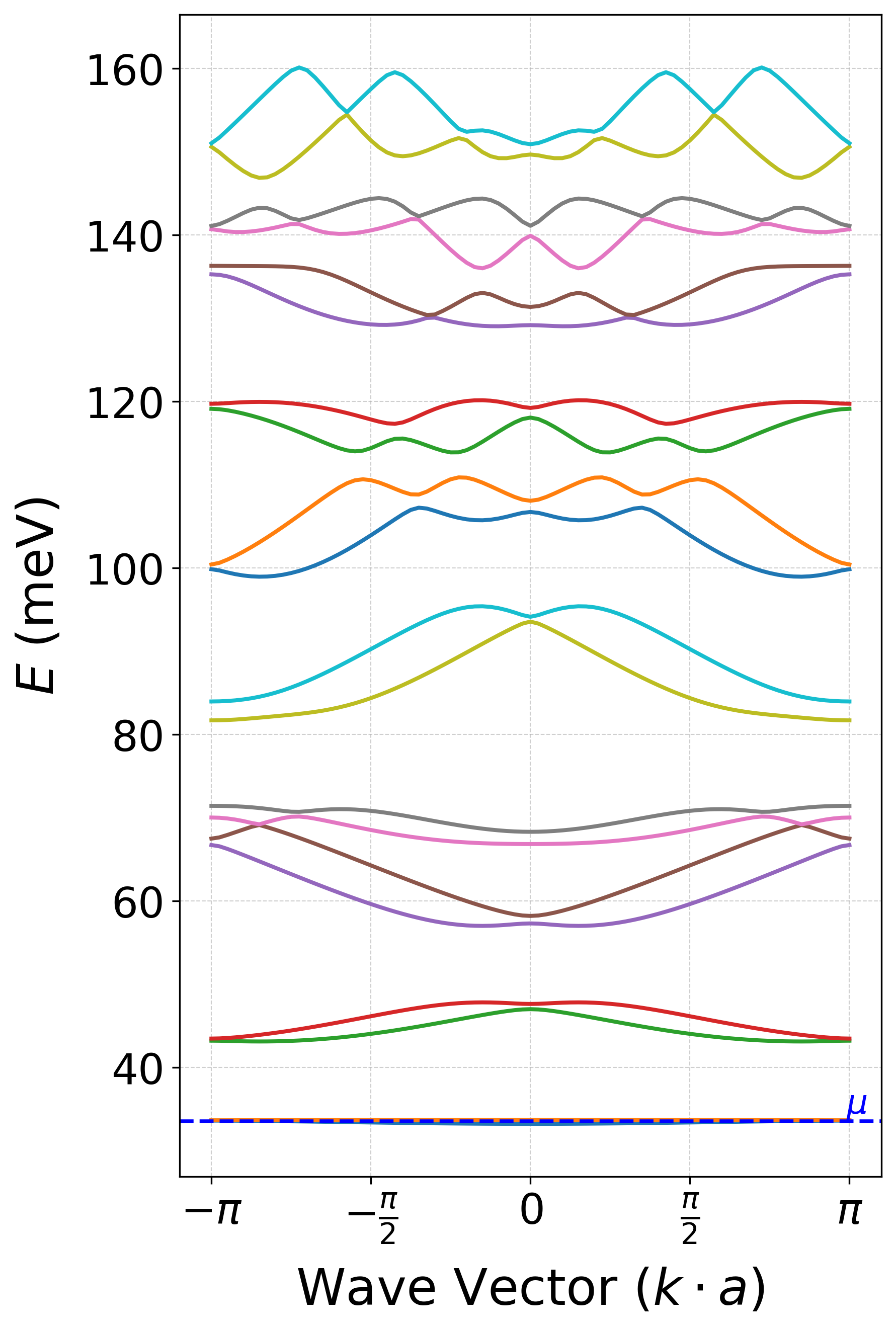} &
        \includegraphics[width=0.195\textwidth]{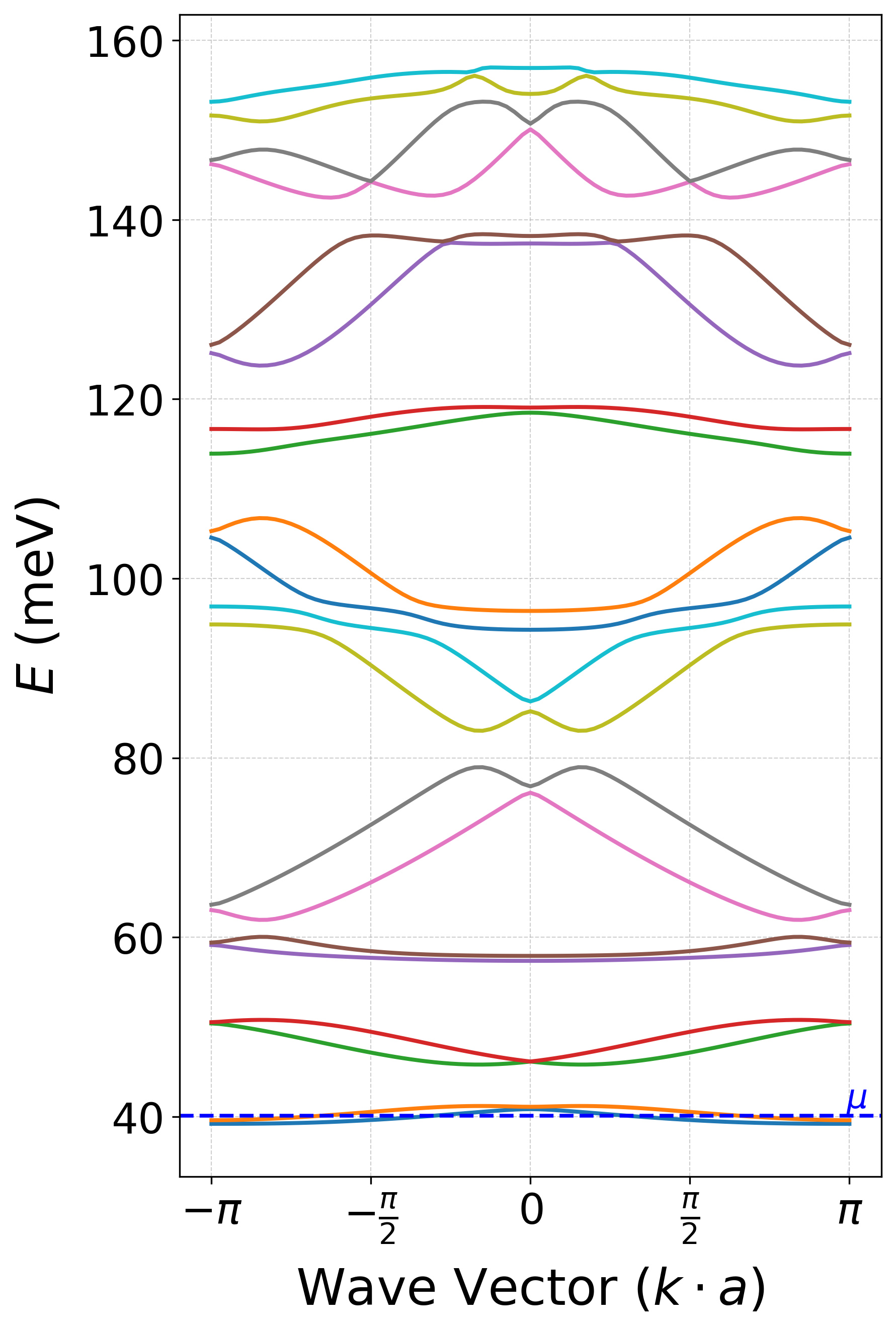} &
        \includegraphics[width=0.195\textwidth]{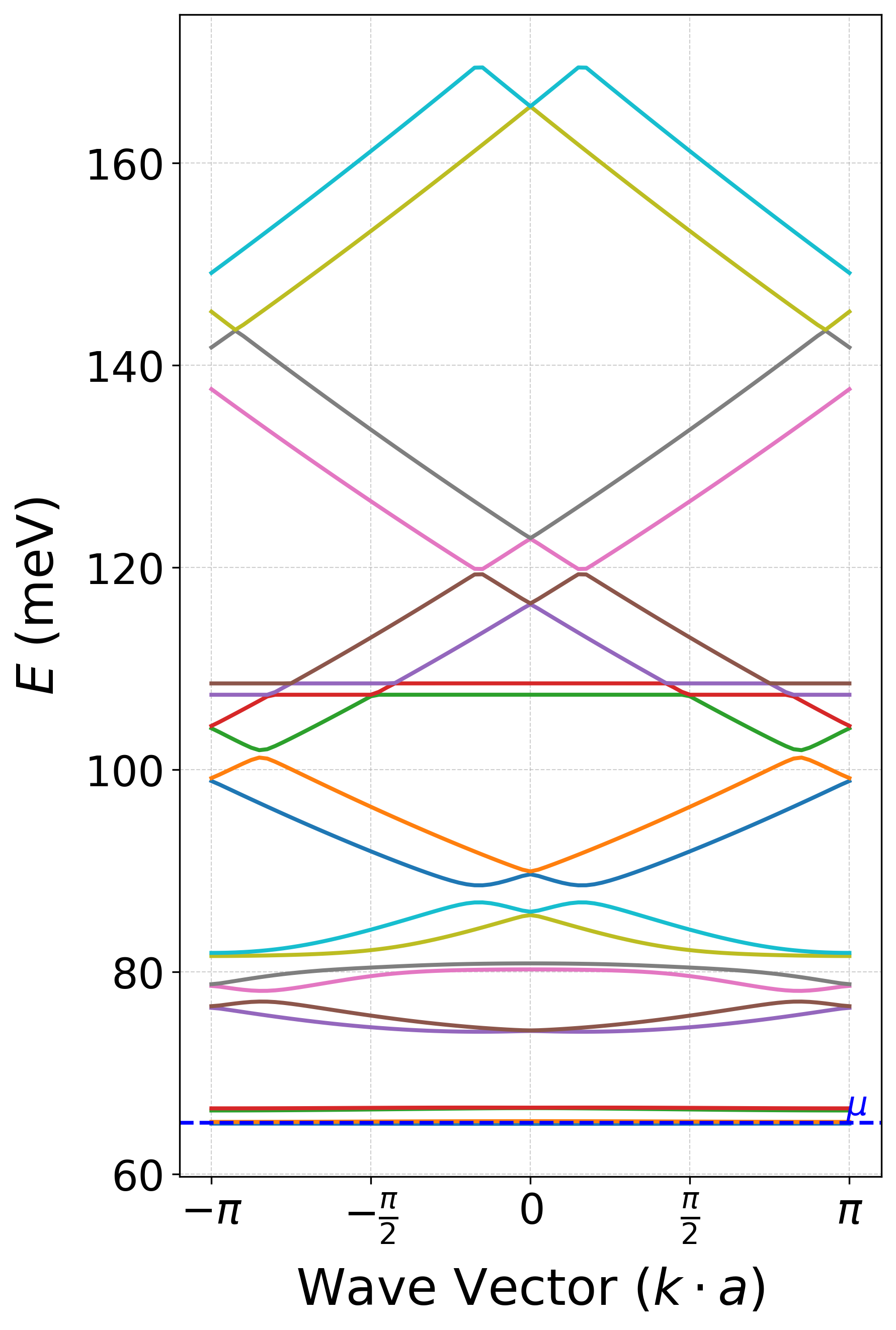} \\
    \end{tabular}
    \caption{Each row corresponds to a different Rashba coupling constant (0, 10, 20, and 40~meV$\cdot$nm, from top to bottom), and each column to a different magnetic field value (0, 5, 10, and 23~T, from left to right).}
    \label{fig:bandstructure}
\end{figure}

\begin{figure}[H]
    \centering
    \begin{minipage}[t]{0.5\textwidth}
        \centering
        \includegraphics[width=\linewidth]{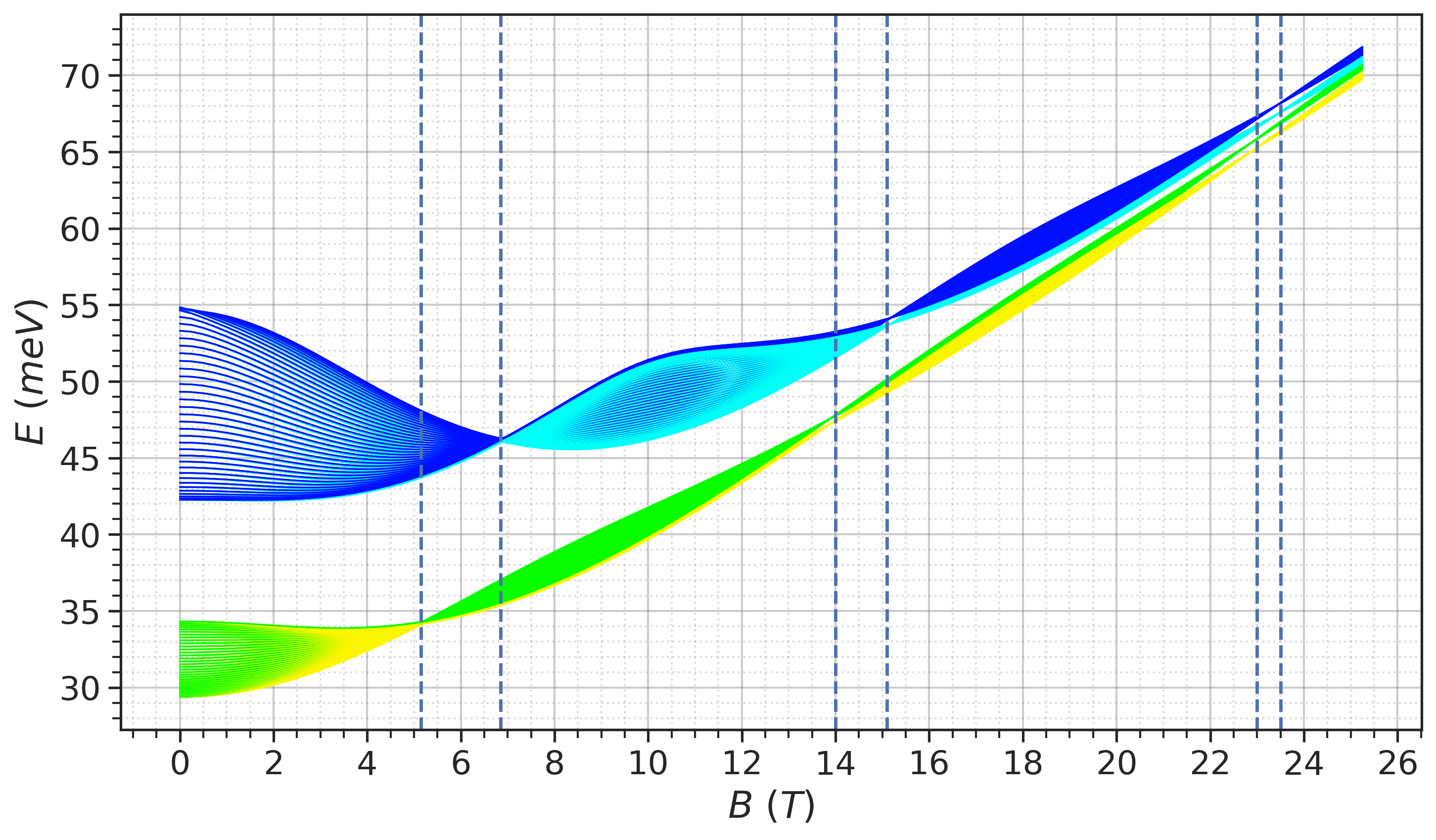}
    \end{minipage}%
    \hfill
    \begin{minipage}[t]{0.5\textwidth}
        \centering
        \includegraphics[width=\linewidth]{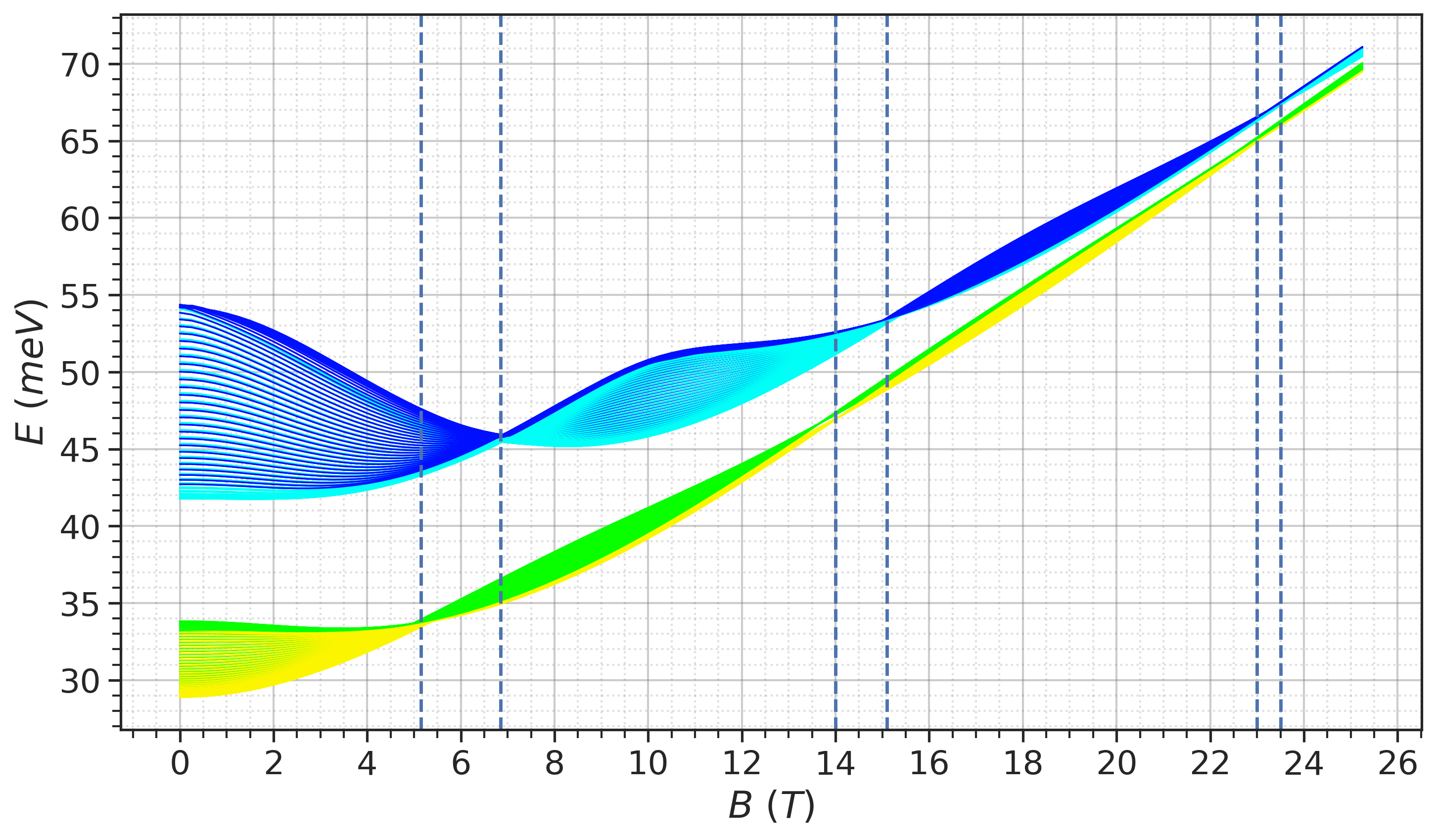}
    \end{minipage}
    \\[0.5cm]
    \caption{Lowest two miniband energies as functions of magnetic field, shown for Rashba coupling constants of $0~\text{meV}\cdot\text{nm}$ (left panel) and $40~\text{meV}\cdot\text{nm}$ (right panel).}
    \label{fig:minibands}
\end{figure}

The numerical calculations are performed for the following values of parameters: \( a = 42\,\mathrm{nm} \), \( V_0 = 250\,\mathrm{meV} \), and \( m = 0.022\,m_e \), where \( m_e \) is the free electron mass. The value for the effective mass $m$ corresponds to InAs material where a significant Rashba interaction can be realized in experiment. The temperature was set to \( T = 4\,\mathrm{K} \), corresponding to an energy of \( 0.342\,\mathrm{meV} \). We considered four different Rashba coupling constants: \( 0 \), \( 10 \), \( 20 \), and \( 40\,\mathrm{meV\cdot nm} \), to obtain the band structures under different conditions, as shown in Fig.~\ref{fig:bandstructure}. The dotted horizontal lines indicate the chemical potential for each case. Along each column, it can be observed that the Rashba constant does not cause significant differences in the chemical potential values; instead, the chemical potential varies noticeably across each row due to the magnetic field.
In Fig.~\ref{fig:minibands}, we observe the reproducibility of the nodes previously reported in our work. This indicates that the miniband nodes are robust with respect to the Rashba interaction. It is worth noting that the band structure in our previous work was obtained using Landau magnetic eigenfunctions as a basis. The dotted vertical lines mark the magnetic field values at which the dispersion disappears; the magnetic fields at \( 5\,\mathrm{T} \) and \( 23\,\mathrm{T} \) were chosen to illustrate this behavior in the band structures.
For the quasi-momentum \( k \), we used 81 uniformly spaced points within the first Brillouine zone (FBZ). Along the first row in the Fig.~\ref{fig:bandstructure}, we observe that in the absence of the Rashba effect, the Zeeman effect is negligible and becomes noticeable only at large magnetic field values. For the second and fourth columns of Fig.~\ref{fig:bandstructure}, the values of the magnetic field are chosen such that the nodes of the first miniband appear almost as a straight line, indicating negligible energy dispersion in those cases. In all subfigures, the chemical potential line passes through the middle of the first band, demonstrating that the first band is half-filled. This situation corresponds to one electron per unit cell of the structure. Under this condition the interaction between electrons does not have any significant effect on the results and can lead to small corrections only, which has been shown in \cite{harutyunyan2025magneto}.
As the Rashba coupling strength increases, the energy splitting becomes more pronounced for all considered magnetic field values. However, the Rashba effect vanishes at \( k = 0, \pm \pi \) due to vanishing group velocity of electron at the center and the edges of the FBZ. The look through the forth column shows that at low and intermediate energies, the bands exhibit smooth and periodic dispersion, due to the interplay of the periodic potential and magnetic field. However, at higher energies (above~100 meV), the dispersion loses the smoothness and for some cases visually resembles intersecting straight lines due to the dominance of magnetic quantization effects over the superlattice potential modulation. We have already indicated that modeling a chain of quantum rings with an oscillatory basis under a magnetic field produces the same energy degeneracy at certain magnetic field values, as described in our previous papers. In the absence of magnetic field a Bloch electron acquires a phase when passing from one unit cell to another in a periodic potential:
\begin{equation}
    \psi_{n, k}(x+a) = e^{ika}\psi_{n, k}(x)
\end{equation}
\noindent
In the presence of a magnetic field, the kinetic momentum is expressed in terms of the canonical one by the following substitution:
\begin{equation}
    \mathbf{p} \to \mathbf{p} + e\mathbf{A}
\end{equation}
As a result, the electron wavefunction acquires the following phase in a lattice under a magnetic field:
\begin{equation}
    \psi_{n,k}(x+a) = \psi_{n,k}(x) \exp\left[ i\left( k a + \frac{e}{\hbar} \int \mathbf{A} \cdot d\mathbf{l} \right) \right]
    \label{eq:wavefunction_magnetic}
\end{equation}
where the line integral is evaluated along the \( x \)-direction, from \( -a/2 \) to \( a/2 \) and the path of integration depends on the value of quasi-momentum. Eq.~\eqref{eq:wavefunction_magnetic} implies that, for certain values of the magnetic field, the average value of \( y(k) \) can adjust in such a way that the phase shift becomes independent of the quasi-momentum \( k \). In this case, the phase in Eq.~\eqref{eq:wavefunction_magnetic} takes the quantized values \( 2\pi n \), where \( n \in \mathbb{Z} \). Consequently, as the magnetic field increases, quasi-periodically repeating nodes are observed.

\begin{figure}[H]
    \centering
    \setlength{\tabcolsep}{1pt} 
    \renewcommand{\arraystretch}{0.5} 
    \begin{tabular}{cccc}
        \includegraphics[width=0.27\textwidth]{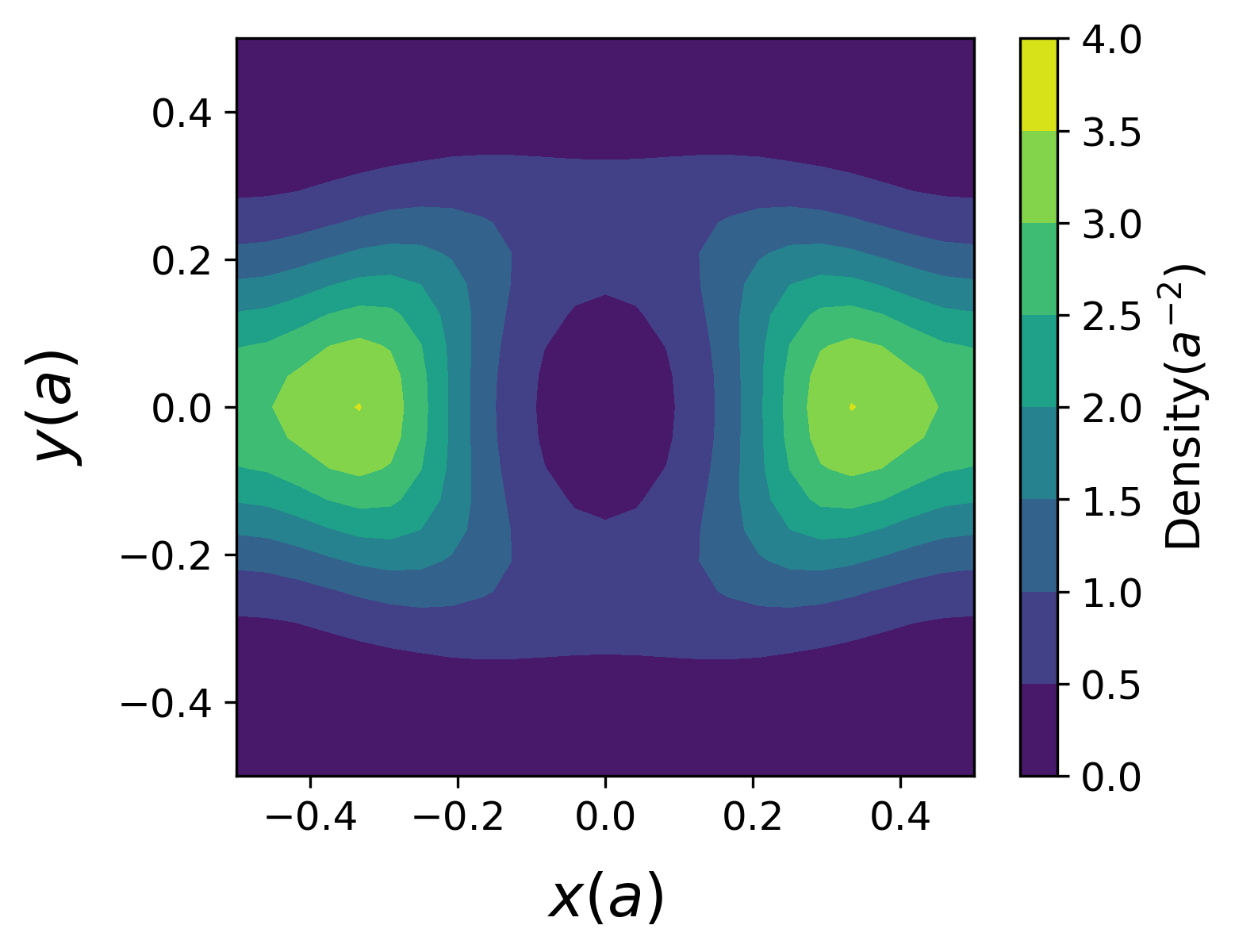} &
        \includegraphics[width=0.27\textwidth]{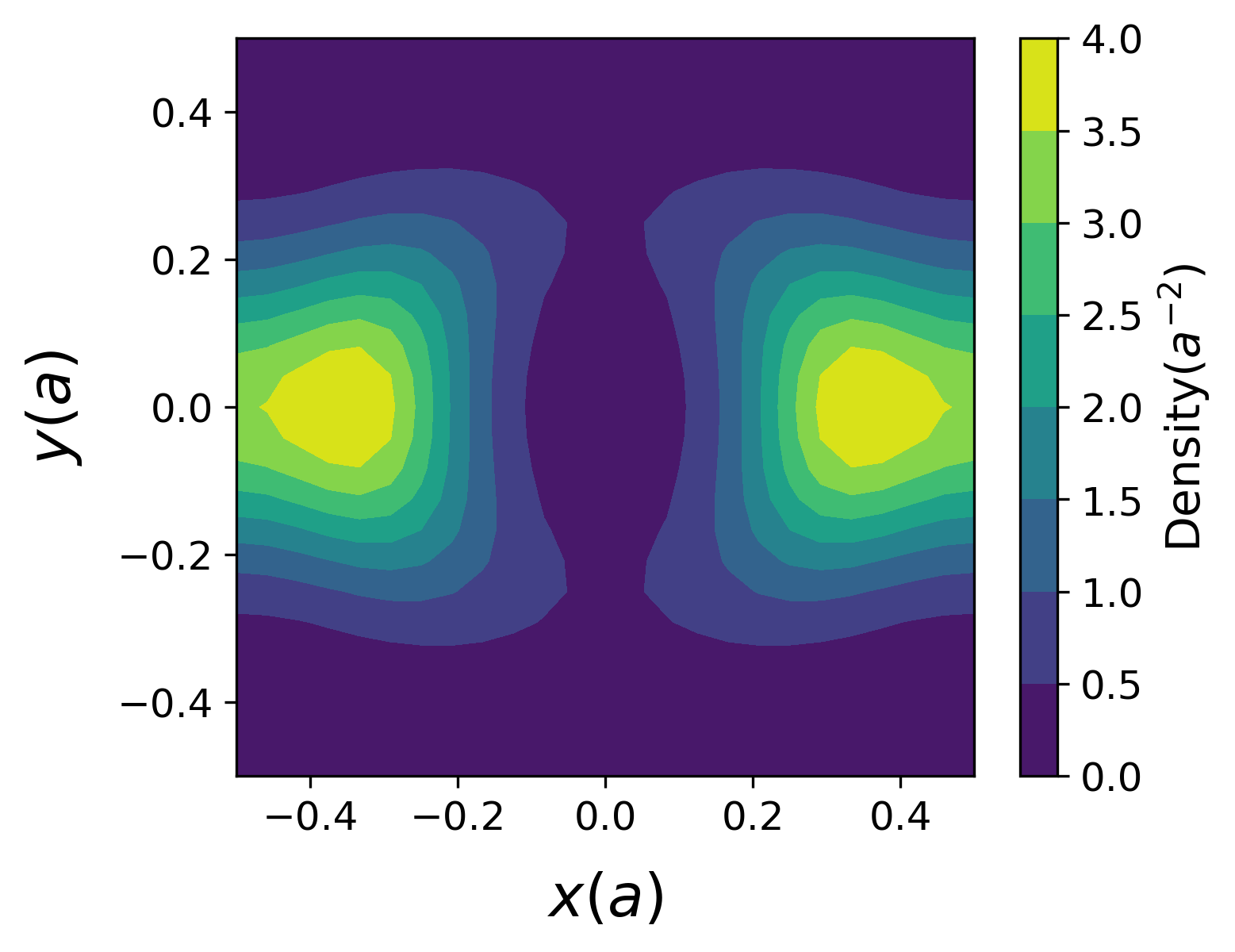} &
        \includegraphics[width=0.27\textwidth]{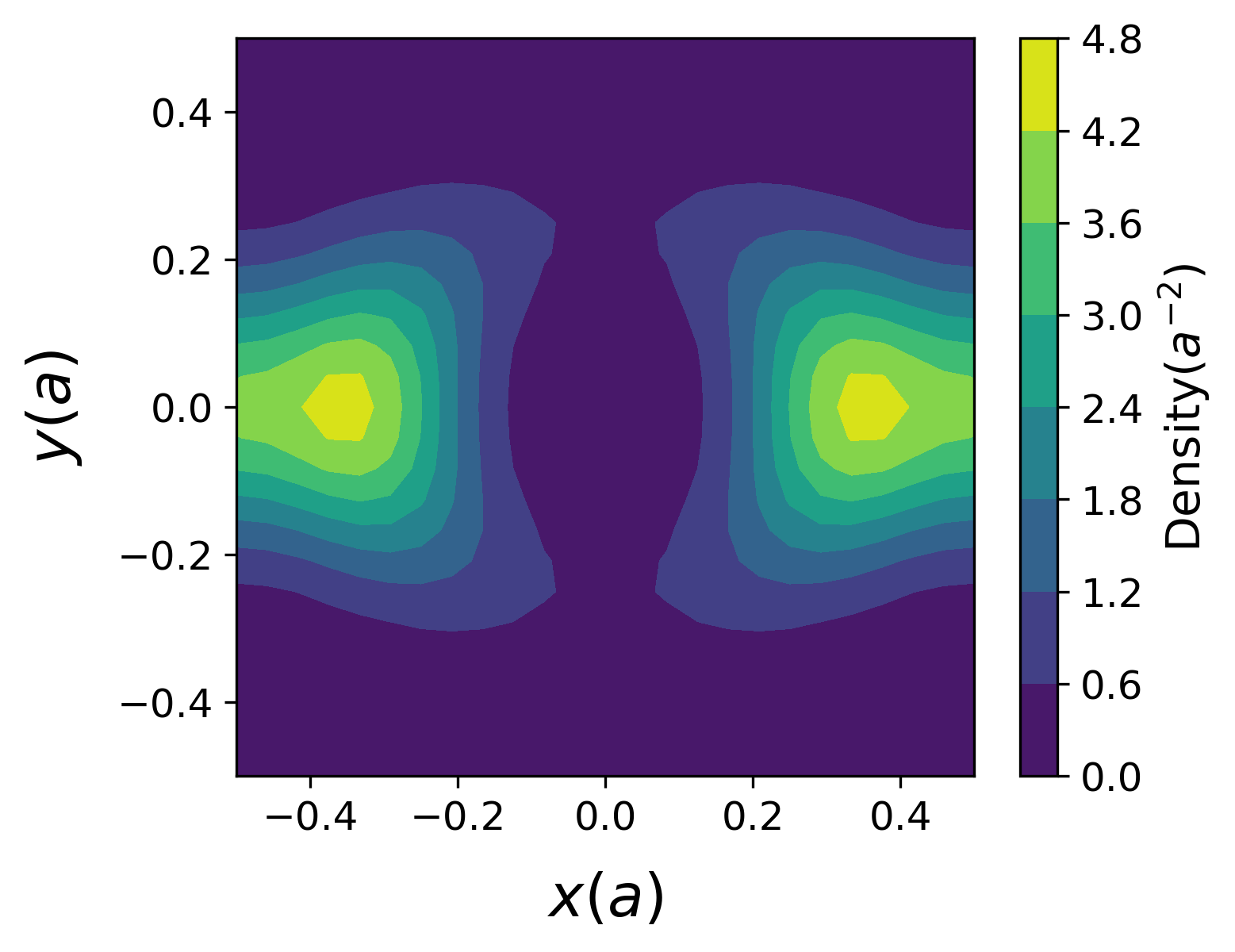} &
        \includegraphics[width=0.27\textwidth]{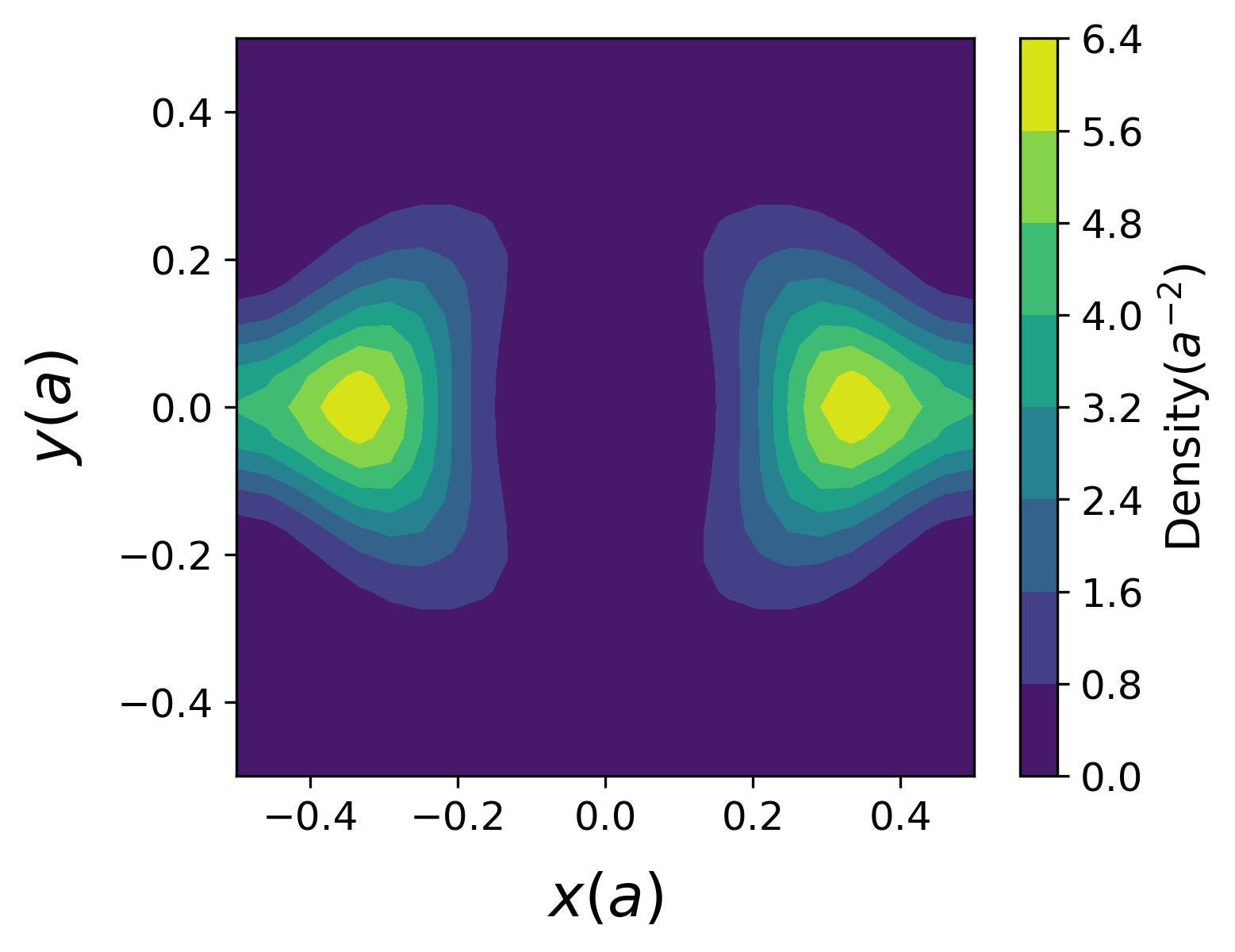} \\
        \includegraphics[width=0.27\textwidth]{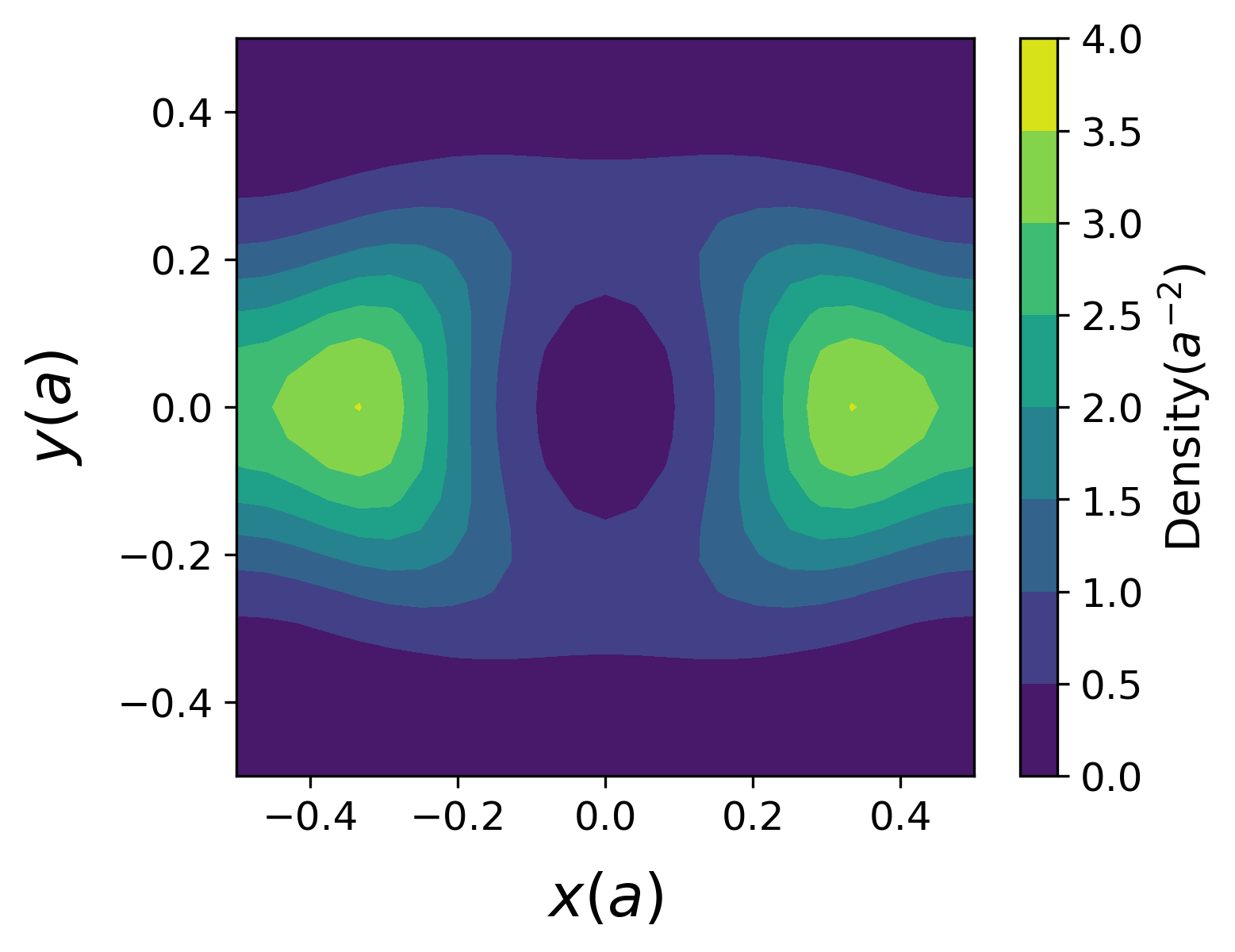} &
        \includegraphics[width=0.27\textwidth]{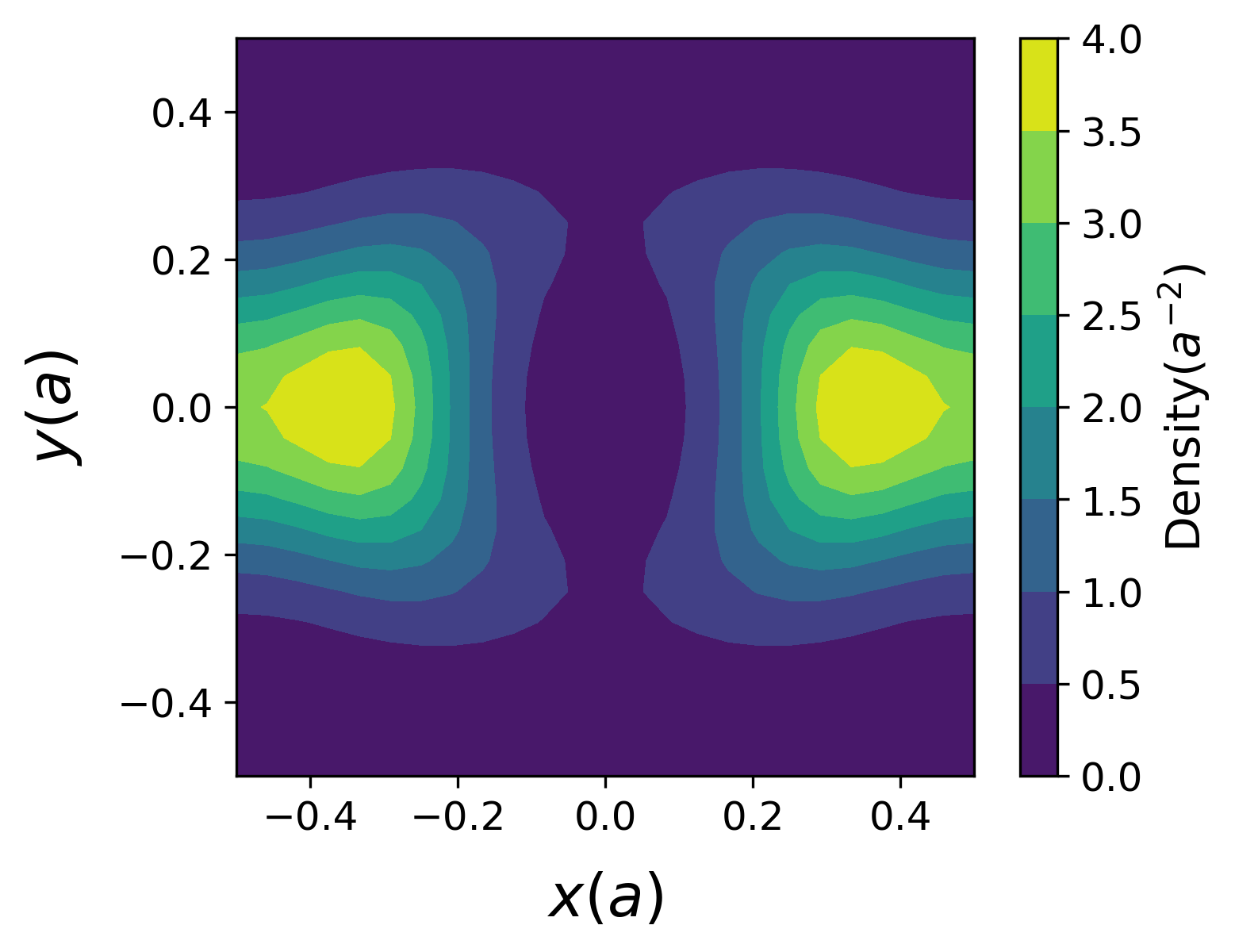} &
        \includegraphics[width=0.27\textwidth]{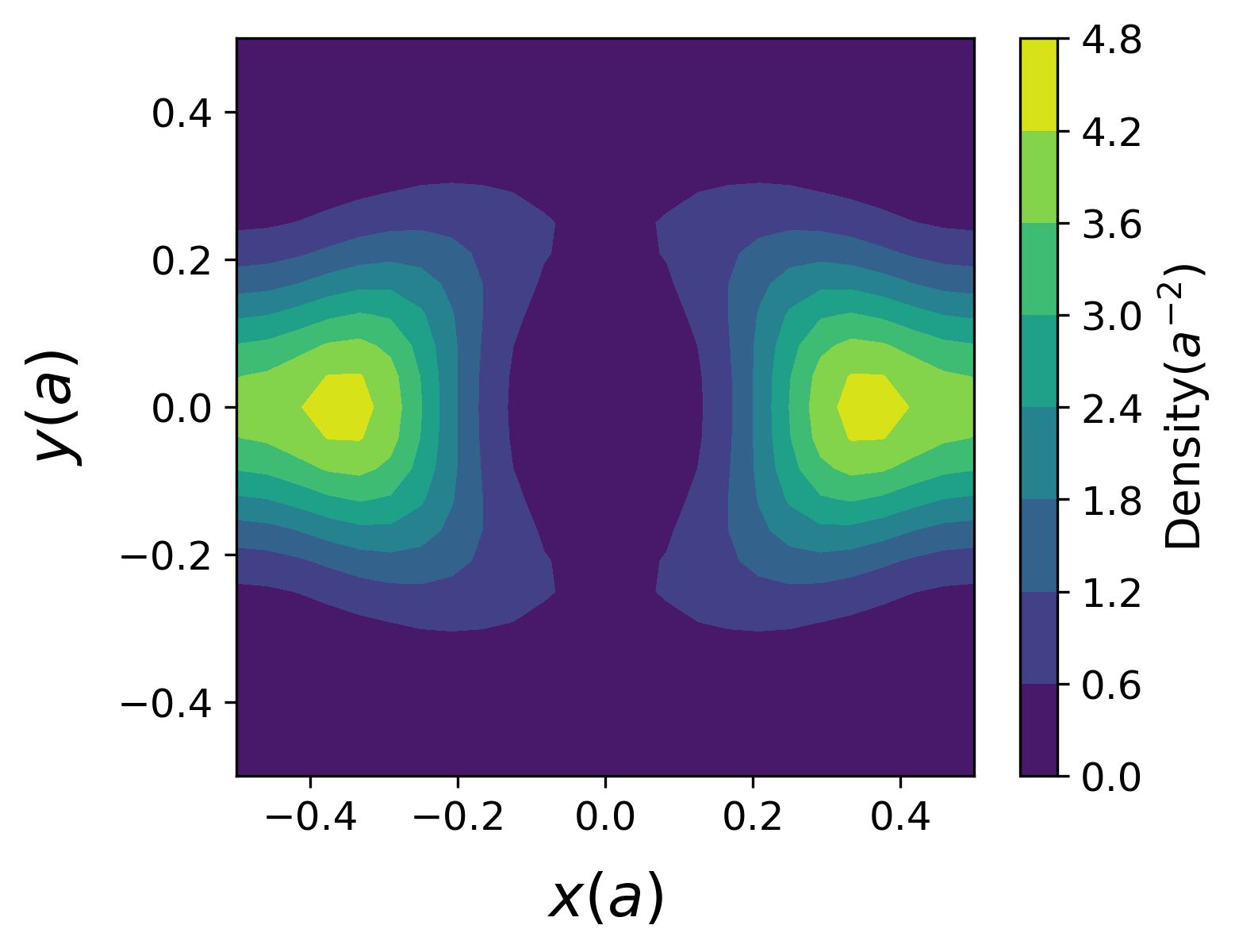} &
        \includegraphics[width=0.27\textwidth]{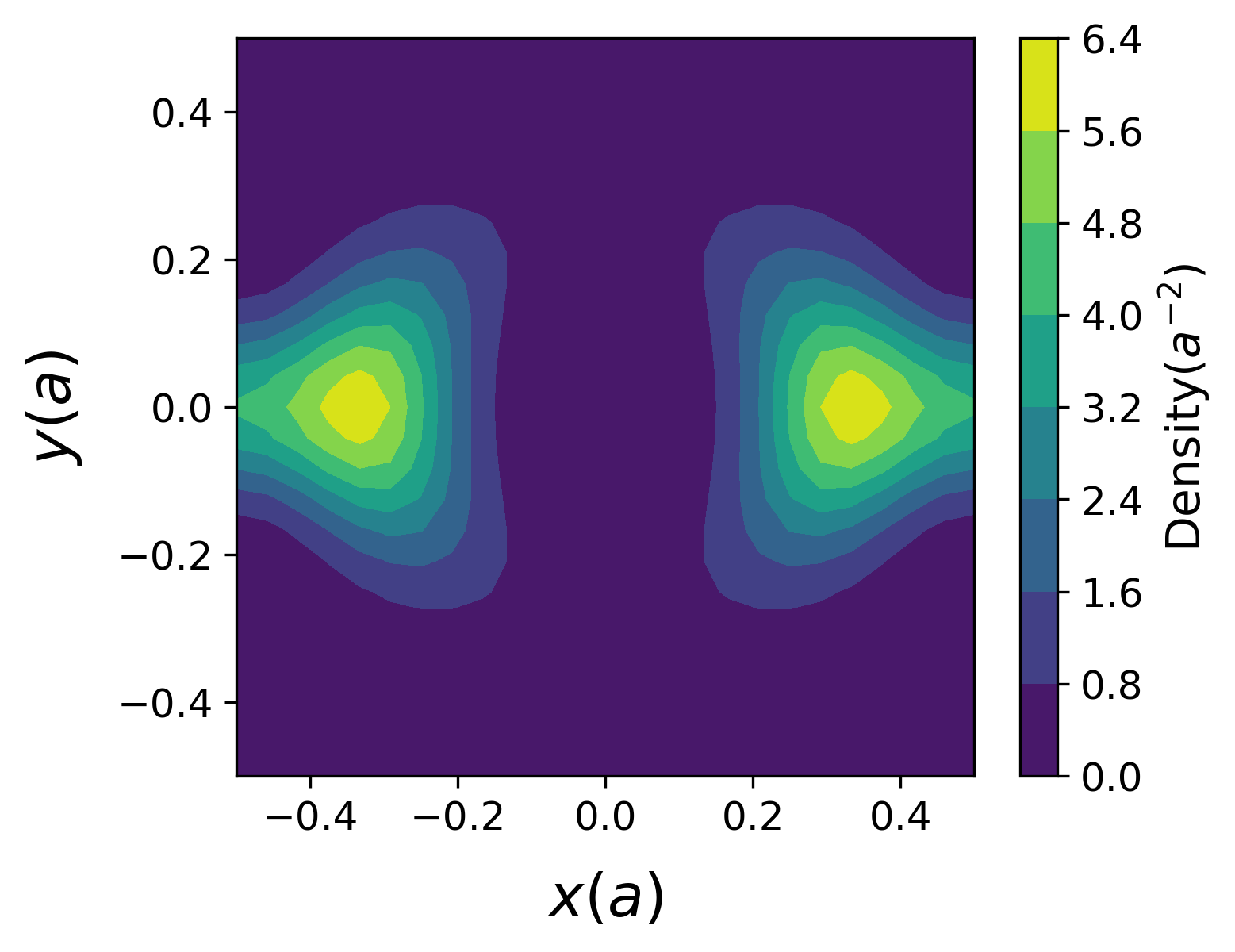} \\
    \end{tabular}
    
    \caption{Electron density distribution in a unit cell for Rashba coupling constants of $0$ (top row) and $40~\text{meV}\cdot\text{nm}$ (bottom row). Columns from left to right correspond to magnetic fields of $0$, $5$, $10$, and $23~\text{T}$, respectively.}
\label{fig:density}
\end{figure}

The electron density plots in Fig.~\ref{fig:density} show that the parameters used yield very similar density profiles, due to similar chemical potential values across different Rashba coupling constants.

For clarity, we present only the cases with Rashba coupling constants of $0$ and $40~\mathrm{meV{\cdot}nm}$. An apparent difference of electron density distributions can be observed for the magnetic field values corresponding to each column. The lobes (local maxima of the electron density), become more sharply defined from left to right, and the density increases in the regions of localization. Consistent with previous studies, the density remains primarily confined along the $x$-axis due to significant confinement in the $y$-direction. 

\begin{figure}[H]
    \centering
    \setlength{\tabcolsep}{1pt} 
    \renewcommand{\arraystretch}{0.5} 
    \begin{tabular}{cccc}
        \includegraphics[width=0.27\textwidth]{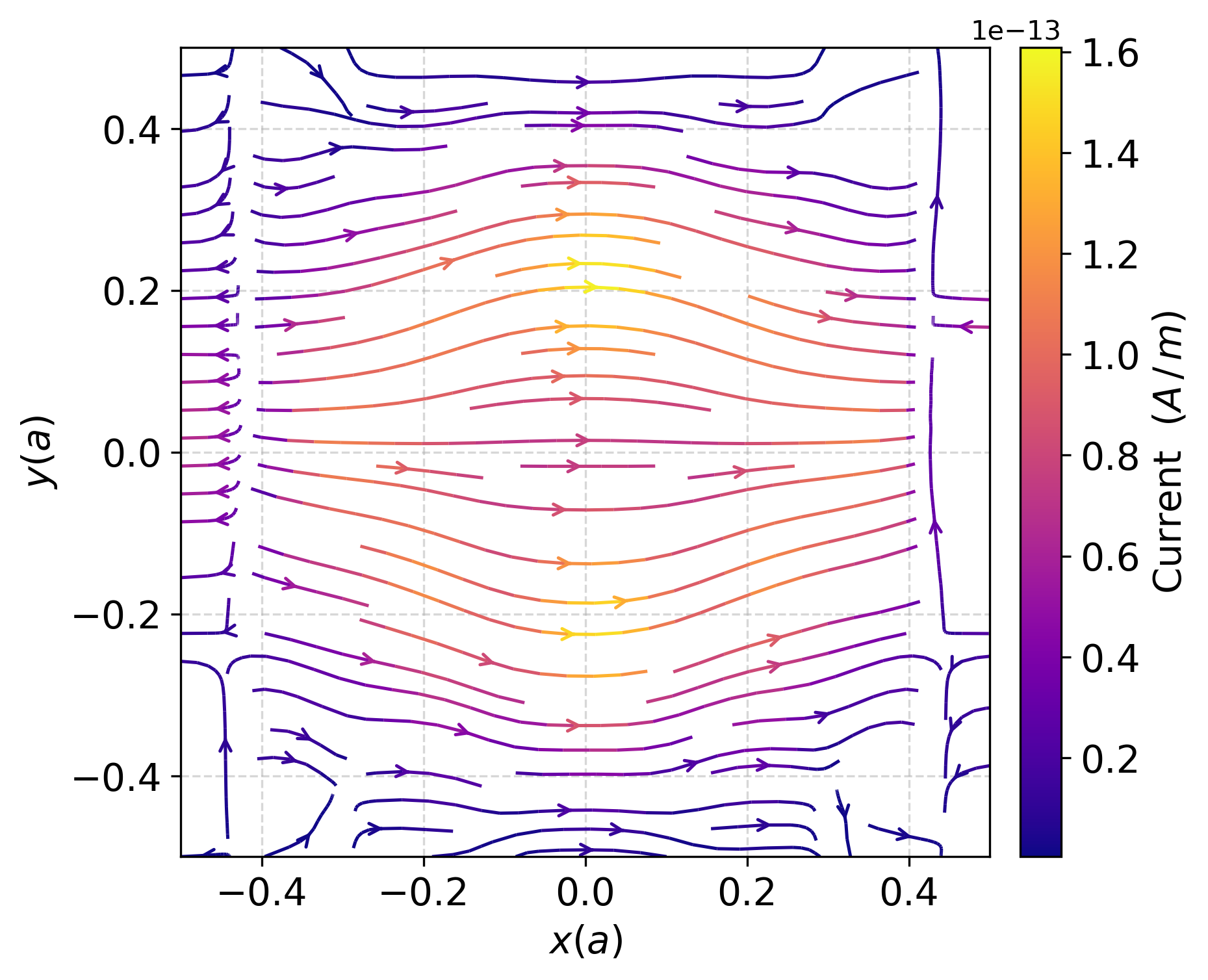} &
        \includegraphics[width=0.27\textwidth]{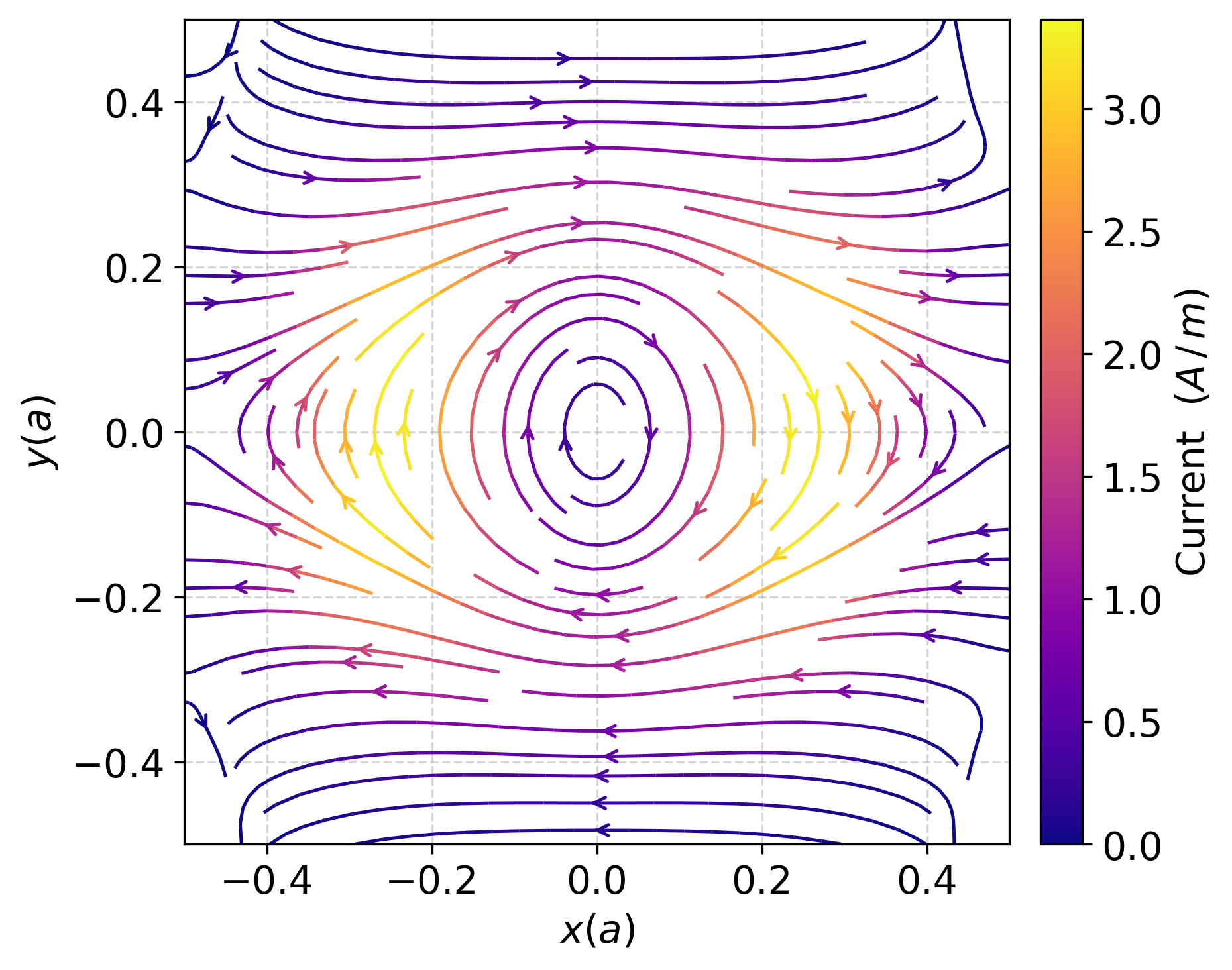} &
        \includegraphics[width=0.26\textwidth]{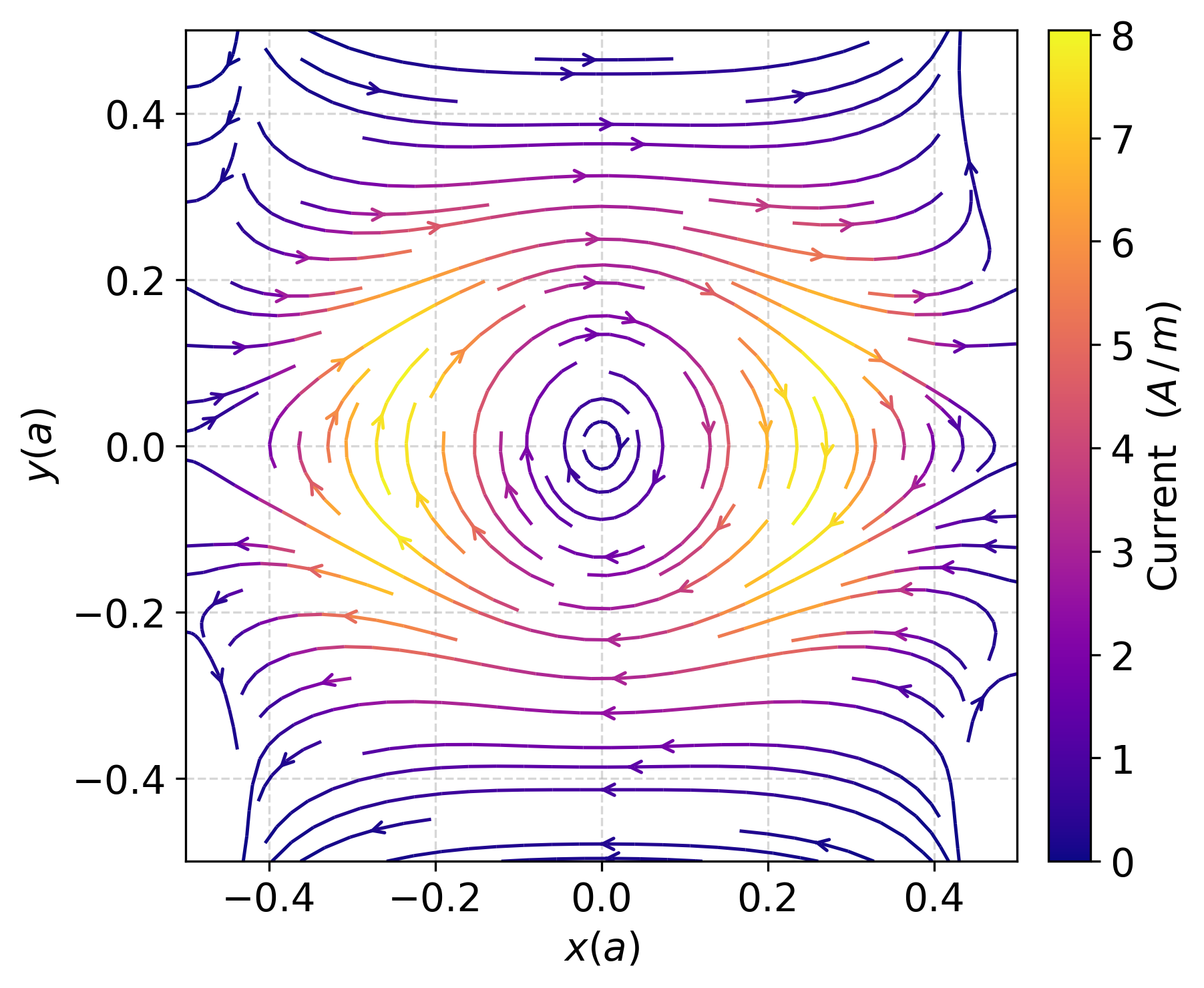} &
        \includegraphics[width=0.27\textwidth]{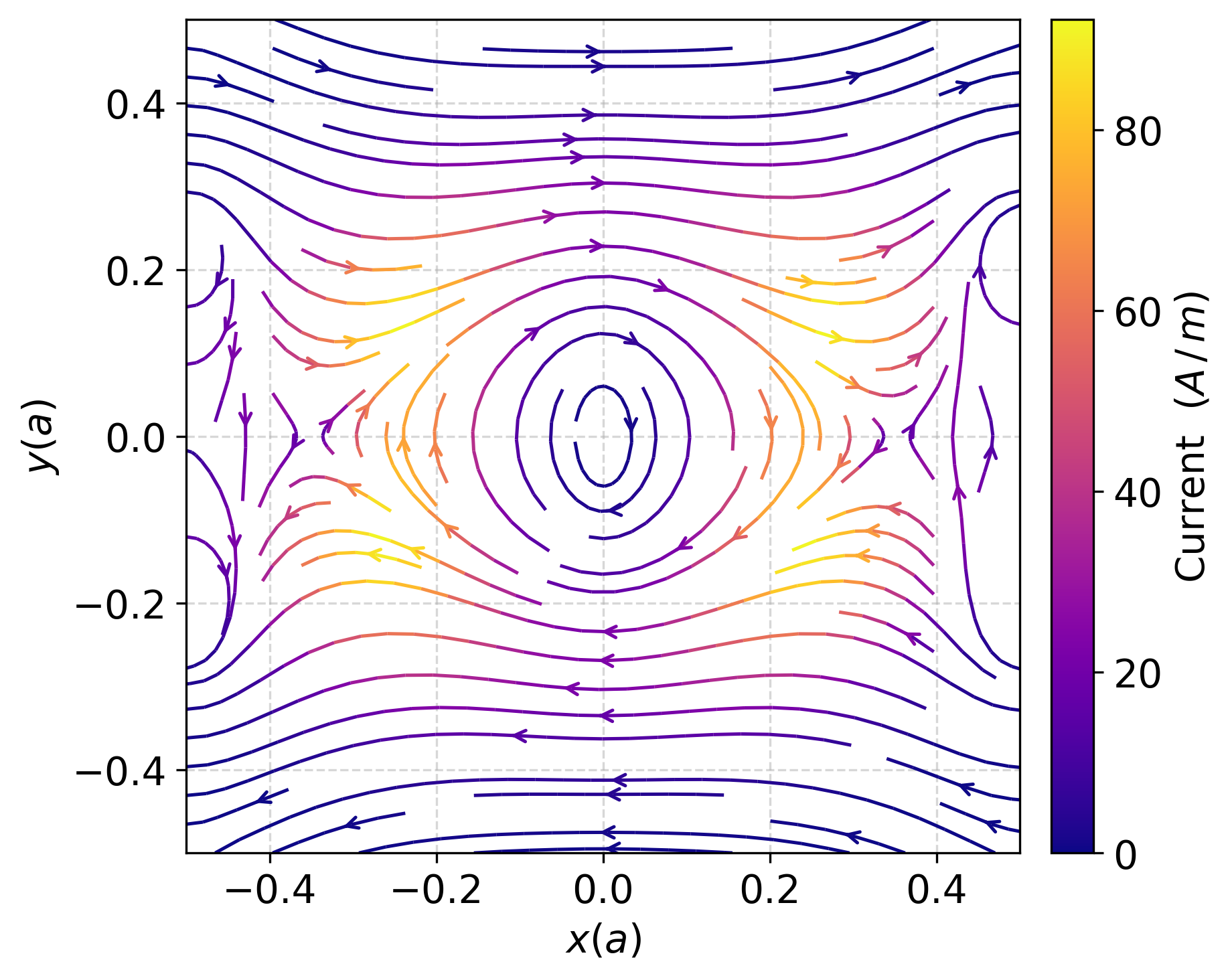} \\
        \includegraphics[width=0.27\textwidth]{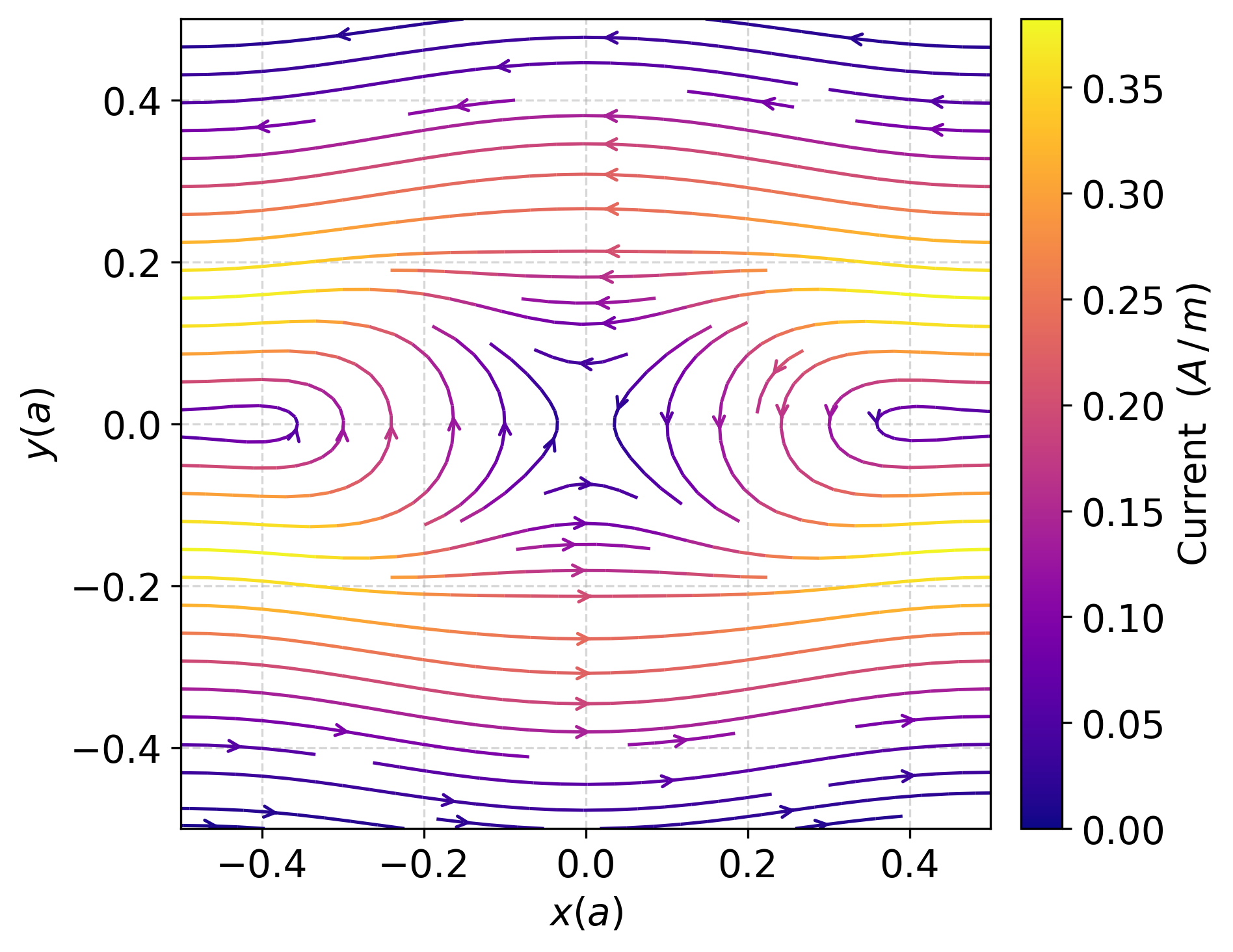} &
        \includegraphics[width=0.27\textwidth]{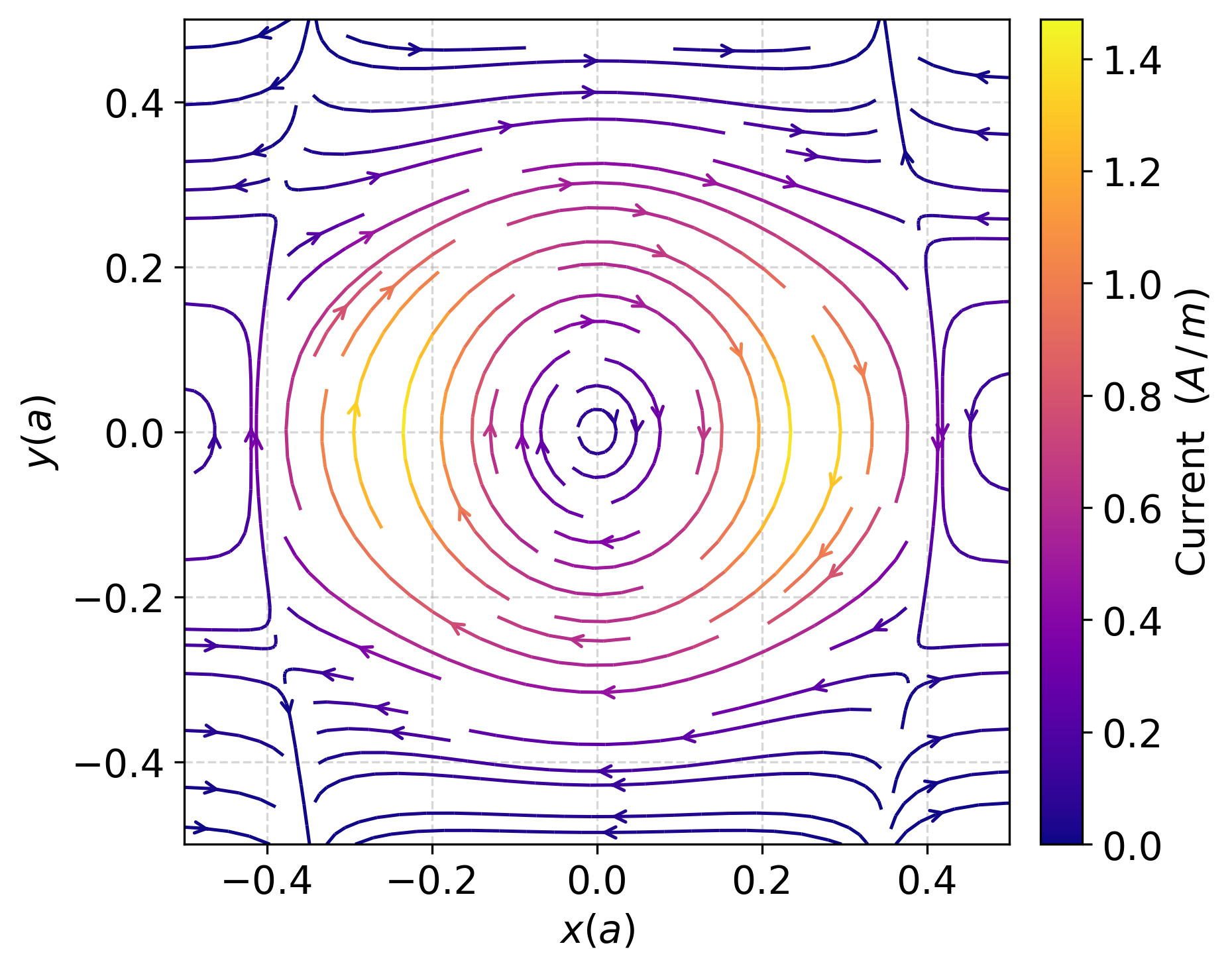} &
        \includegraphics[width=0.27\textwidth]{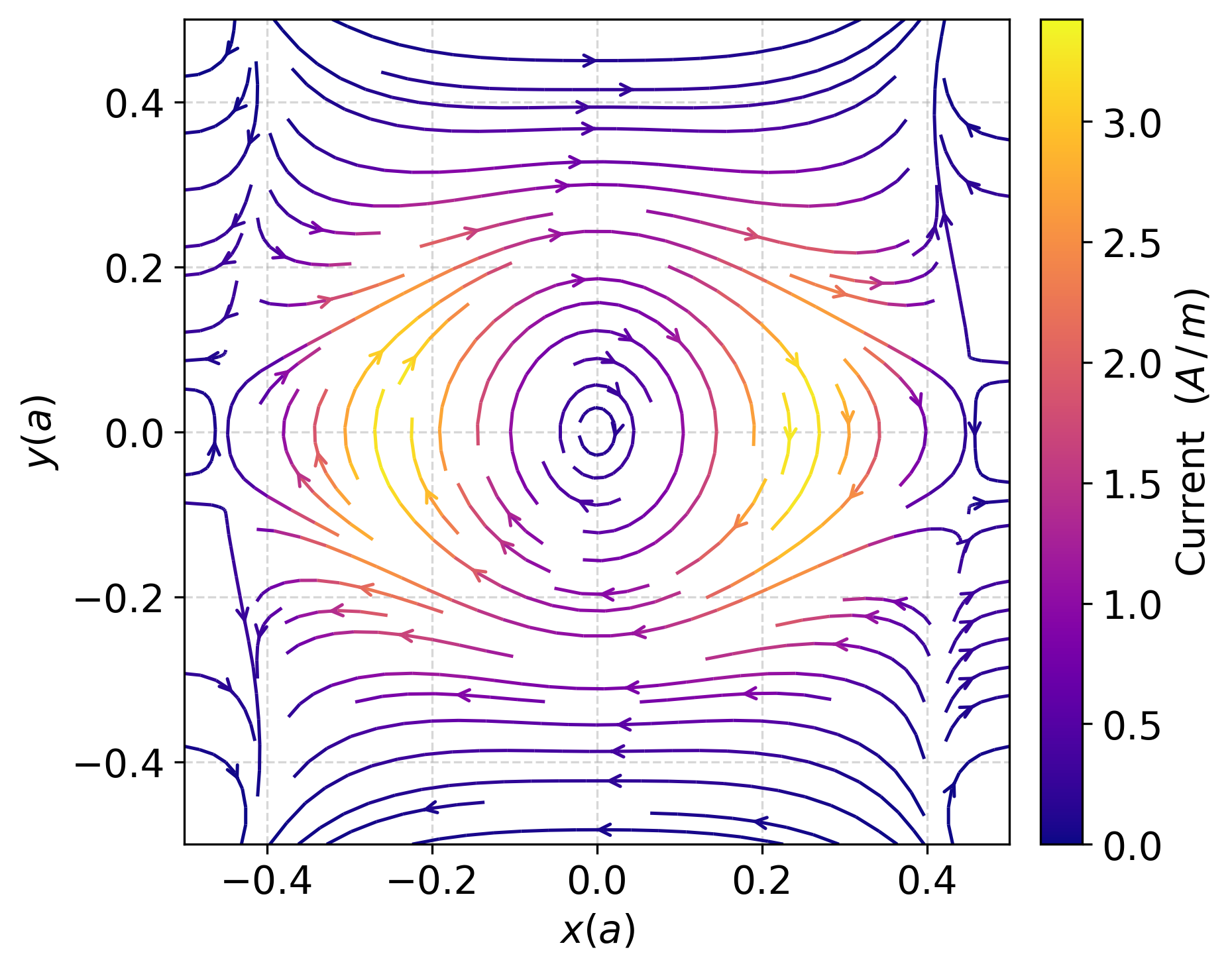} &
        \includegraphics[width=0.27\textwidth]{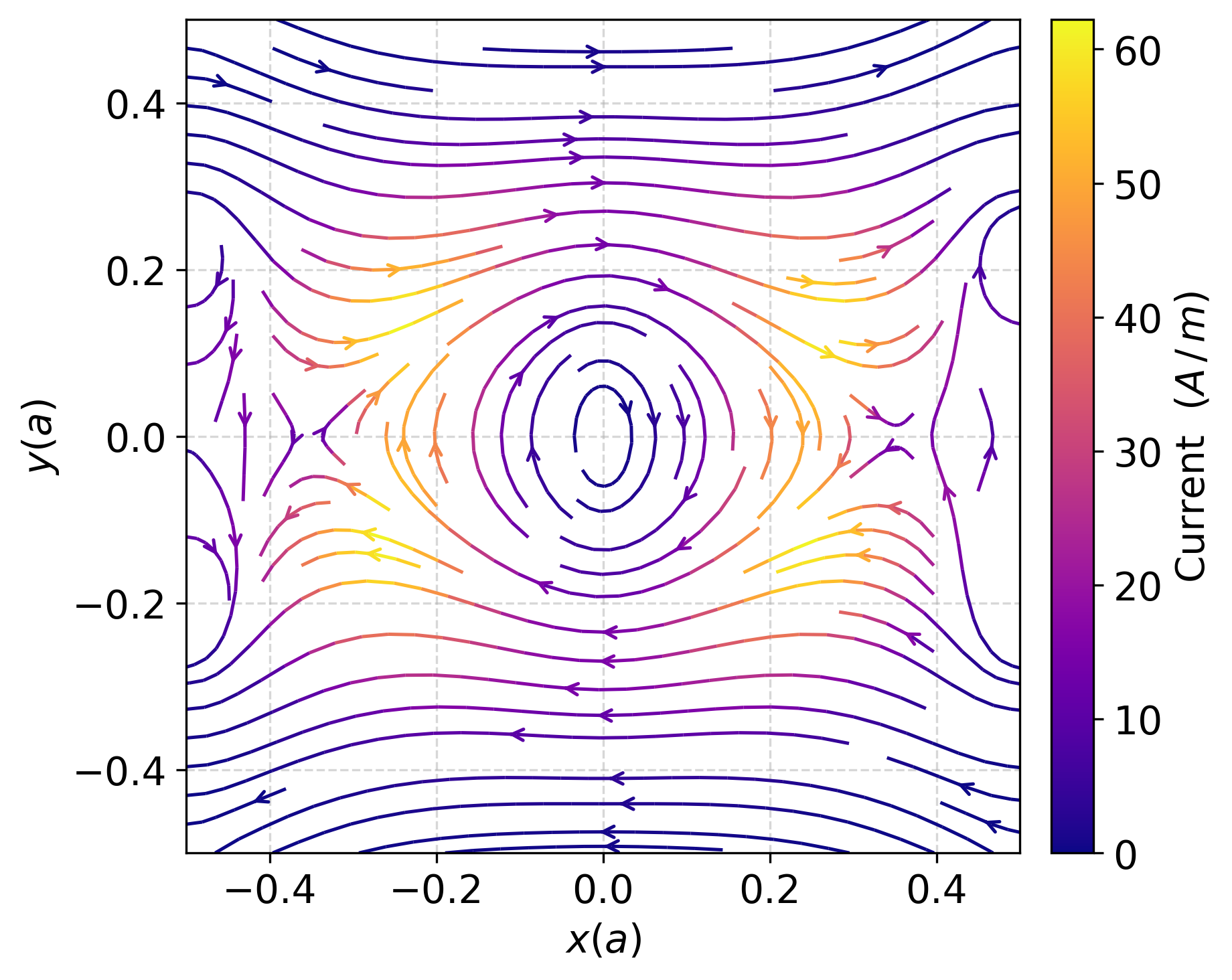} \\
        \includegraphics[width=0.27\textwidth]{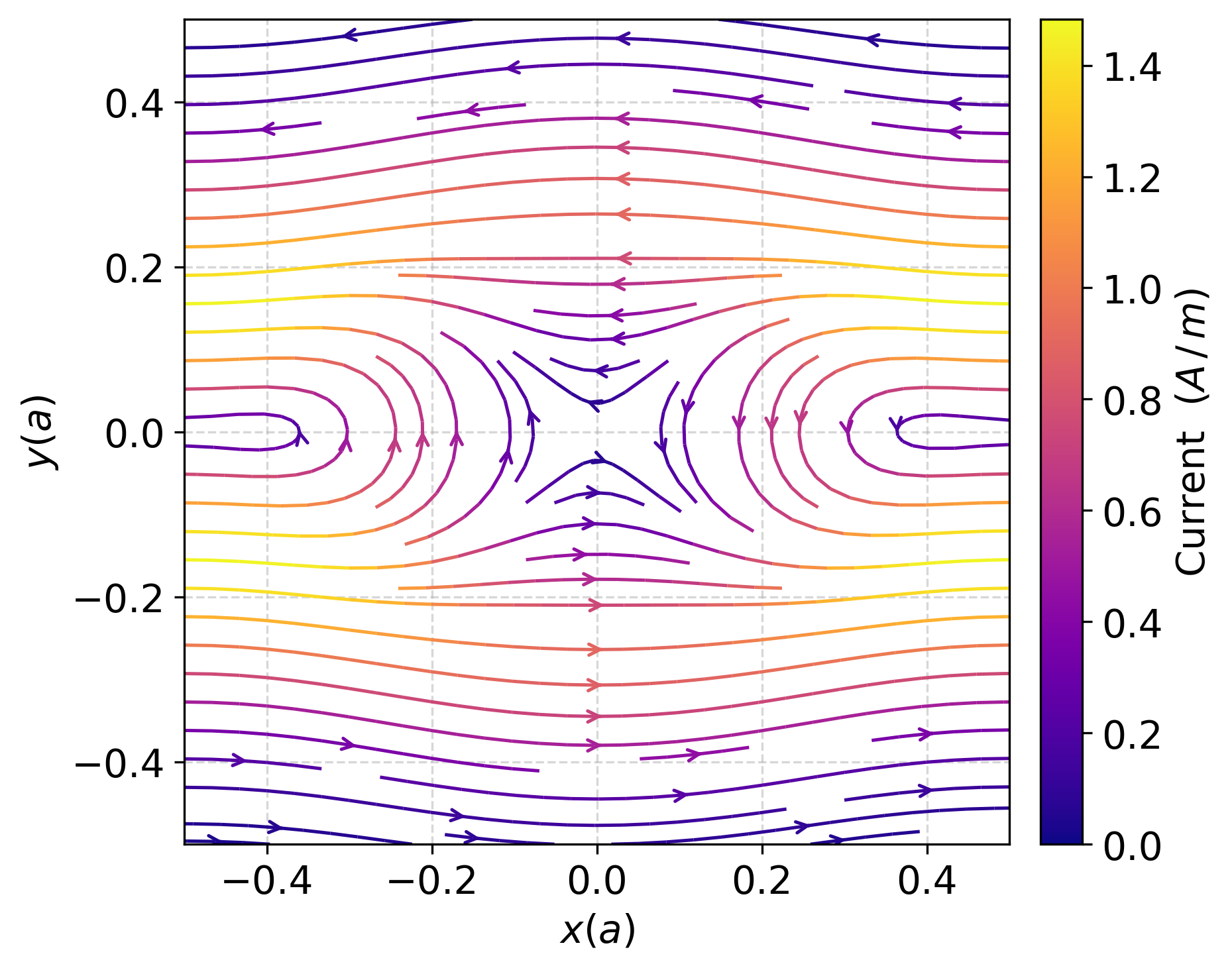} & 
        \includegraphics[width=0.27\textwidth]{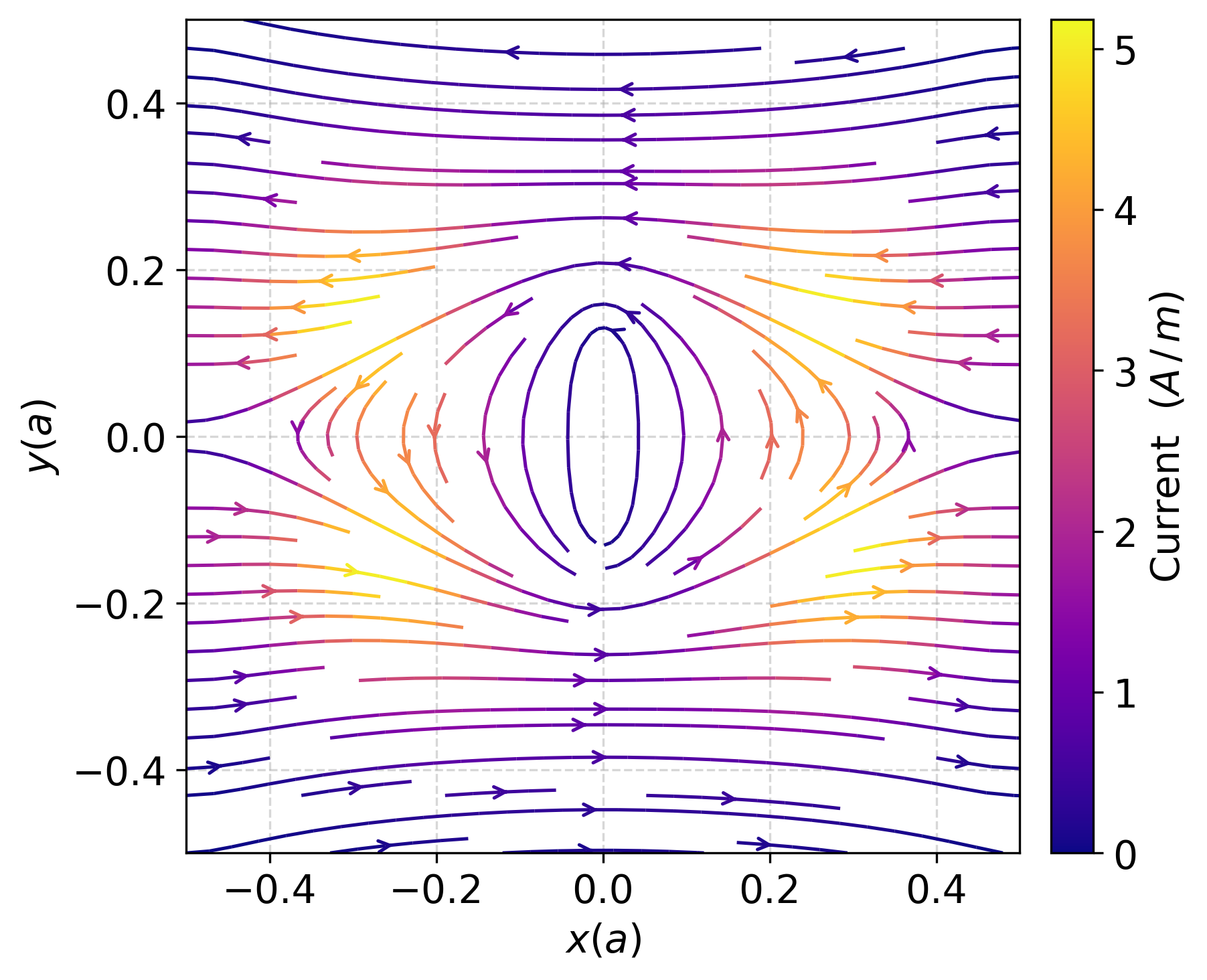} & 
        \includegraphics[width=0.27\textwidth]{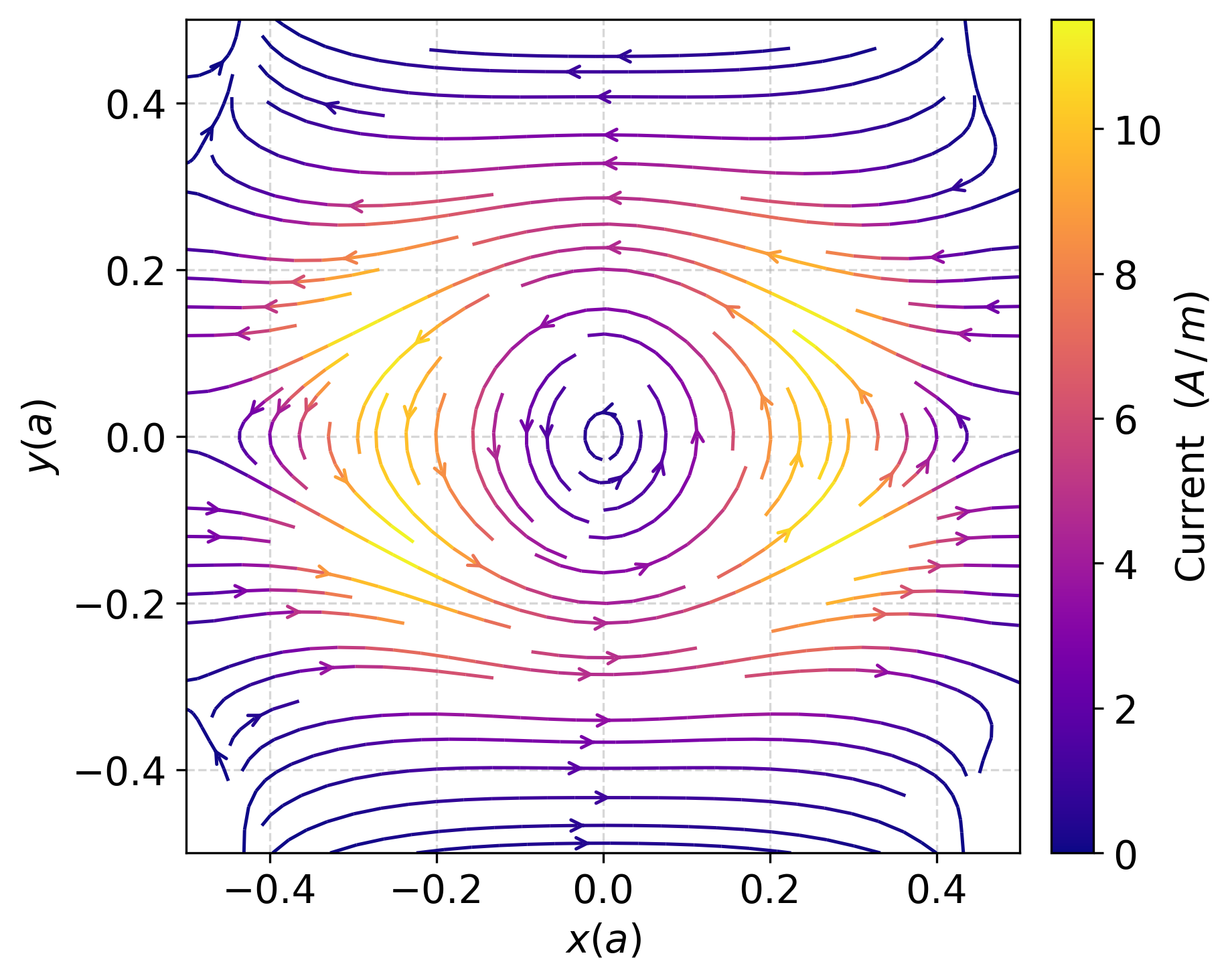} & 
        \includegraphics[width=0.27\textwidth]{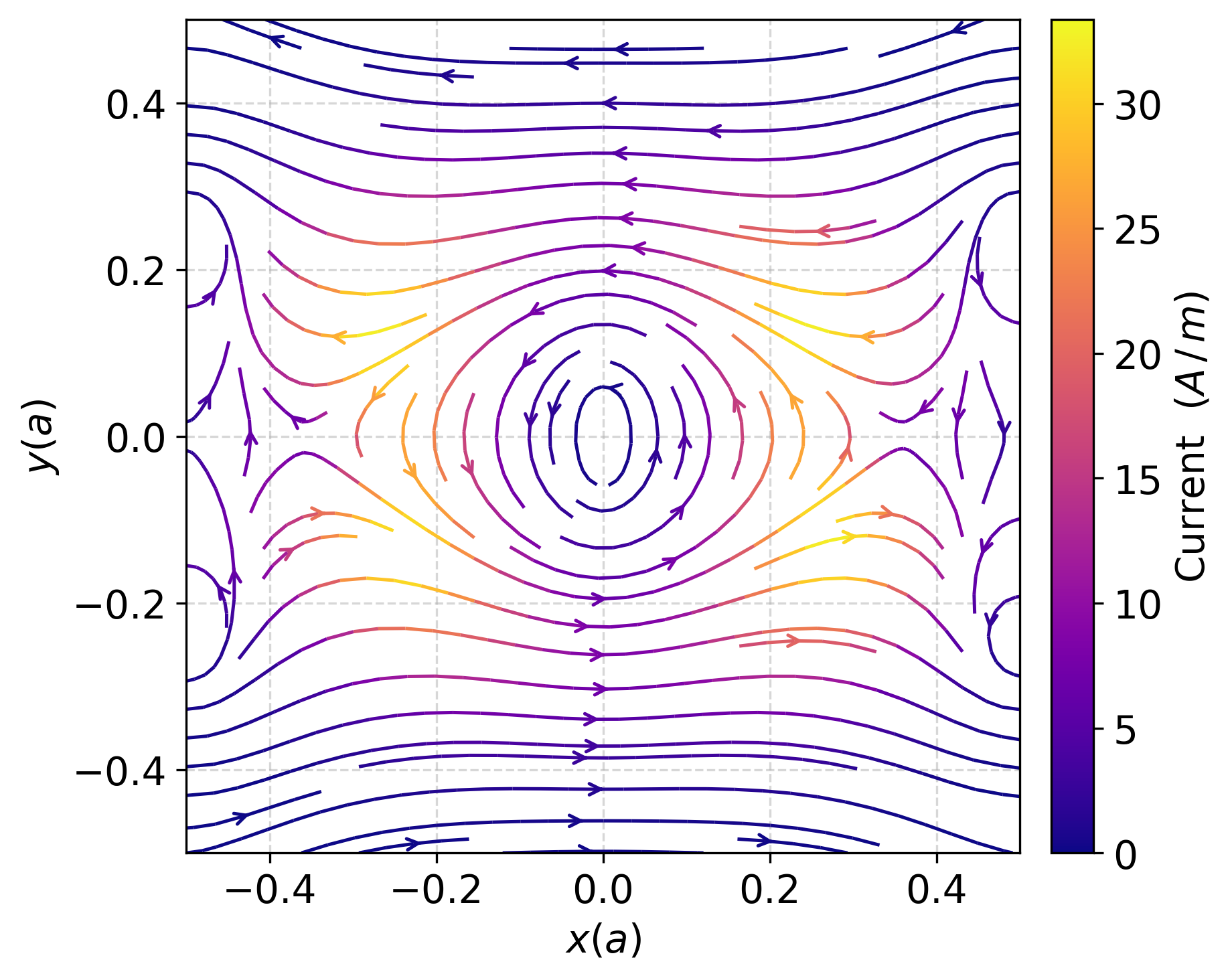} \\
    \end{tabular}
    \caption{Difference between spin-up and spin-down persistent currents in the system. Rows correspond to Rashba coupling constants of $0$, $20$, and $40~\text{meV}\cdot\text{nm}$ (top to bottom), and columns to magnetic fields of $0$, $5$, $10$, and $23~\text{T}$ (left to right).}
\label{fig:persistent_minus}
\end{figure}

\begin{figure}[H]
    \centering
    \setlength{\tabcolsep}{1pt} 
    \renewcommand{\arraystretch}{0.5} 
    \begin{tabular}{cccc}
        \includegraphics[width=0.27\textwidth]{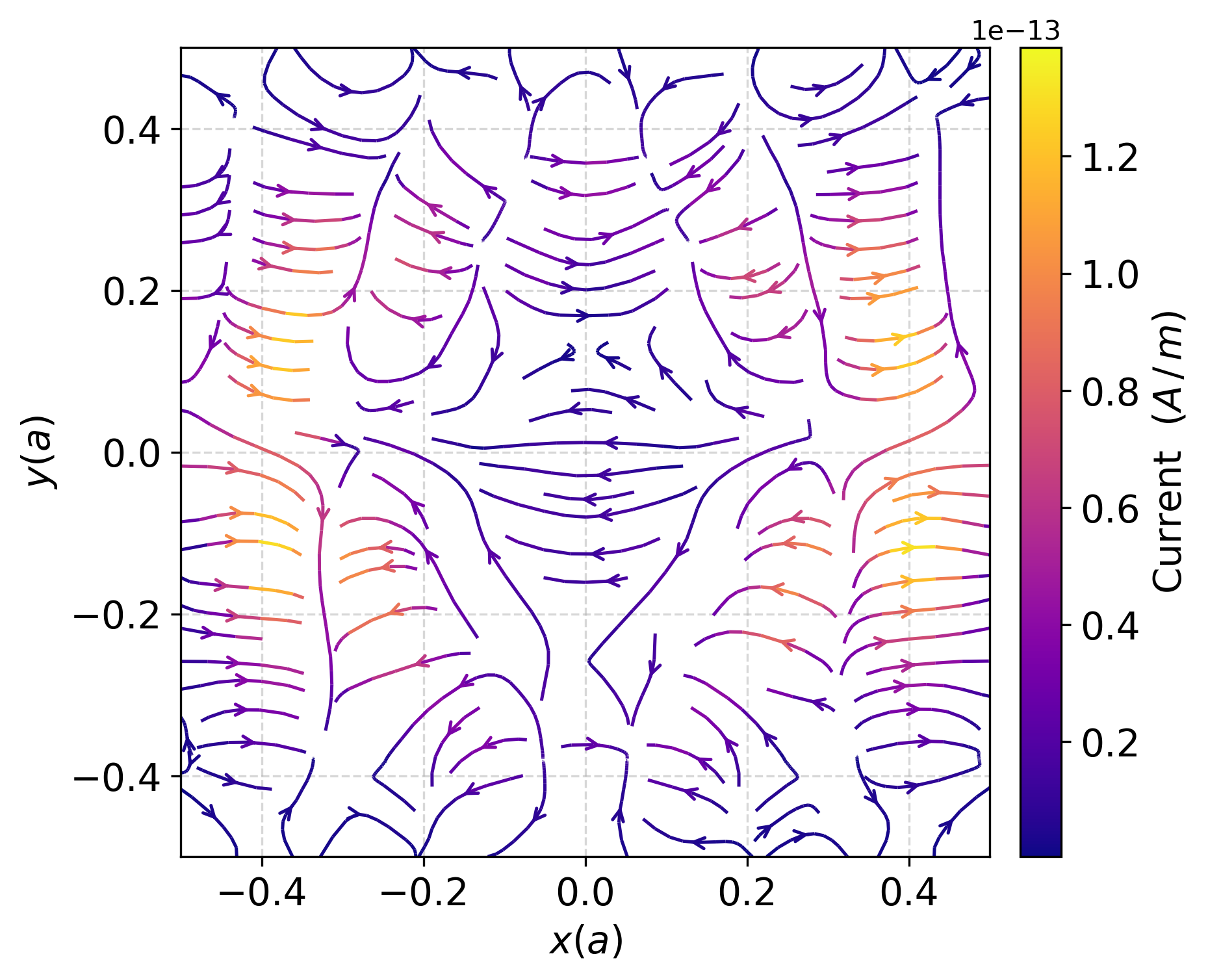} &
        \includegraphics[width=0.27\textwidth]{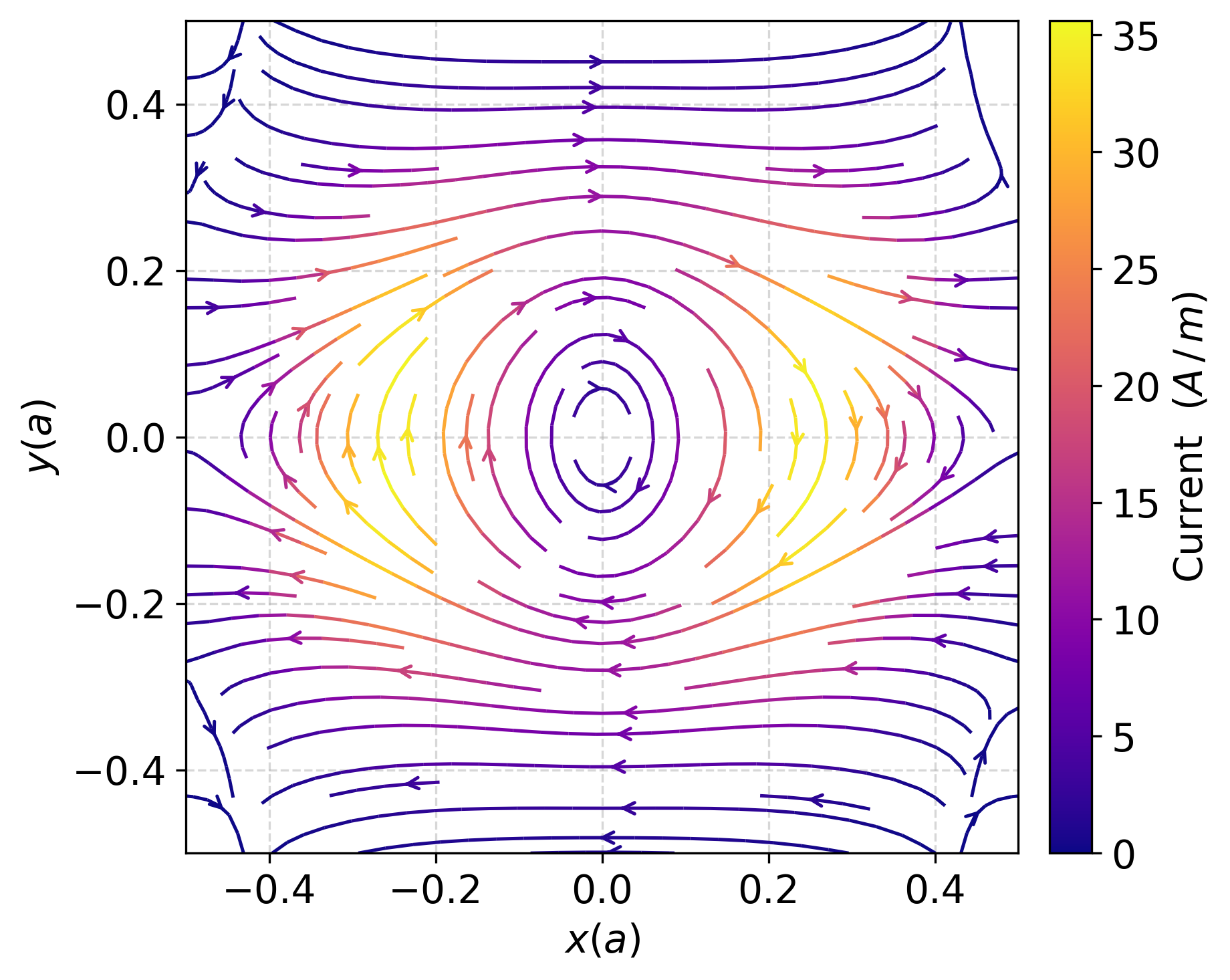} &
        \includegraphics[width=0.27\textwidth]{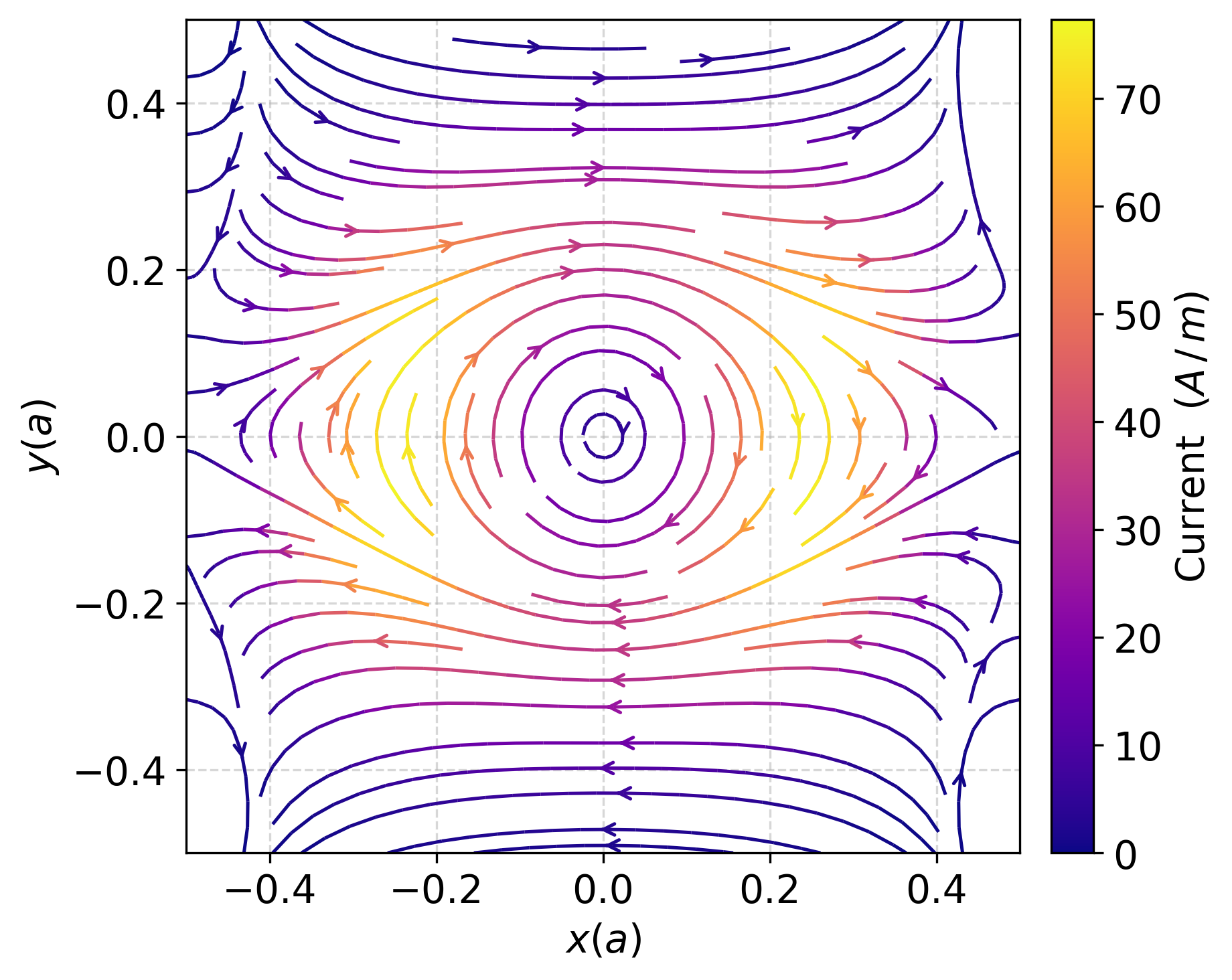} &
        \includegraphics[width=0.27\textwidth]{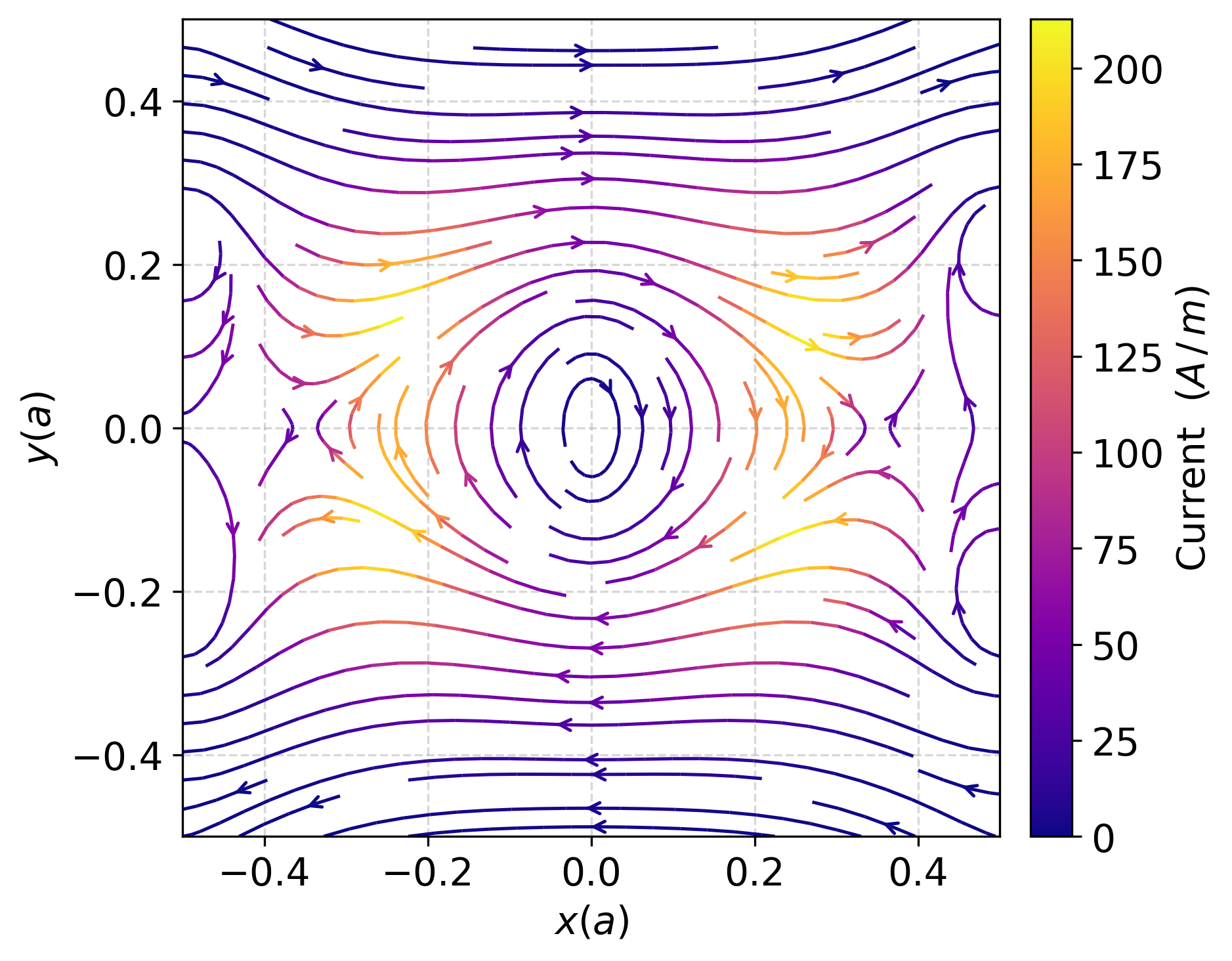} \\
        \includegraphics[width=0.27\textwidth]{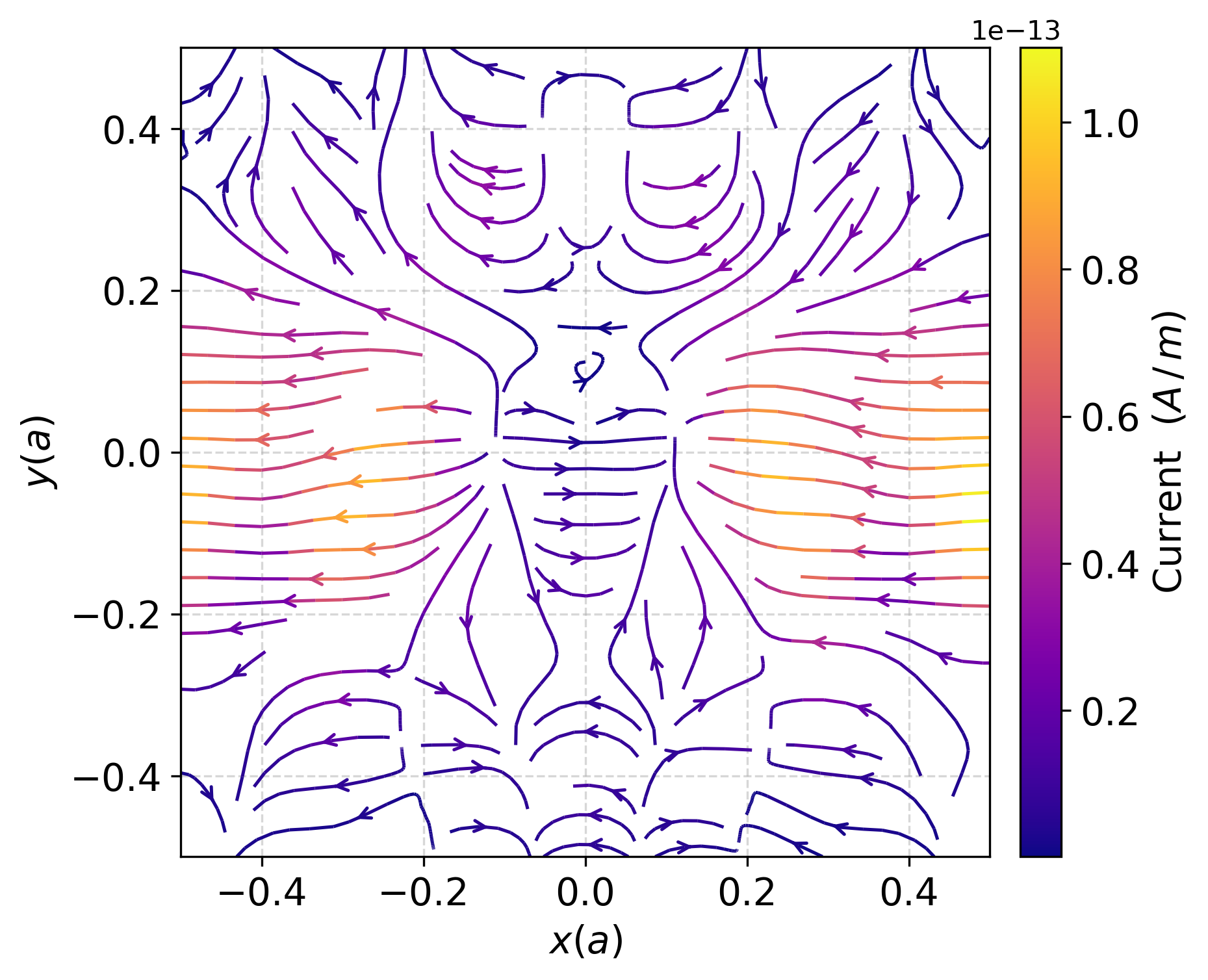} &
        \includegraphics[width=0.27\textwidth]{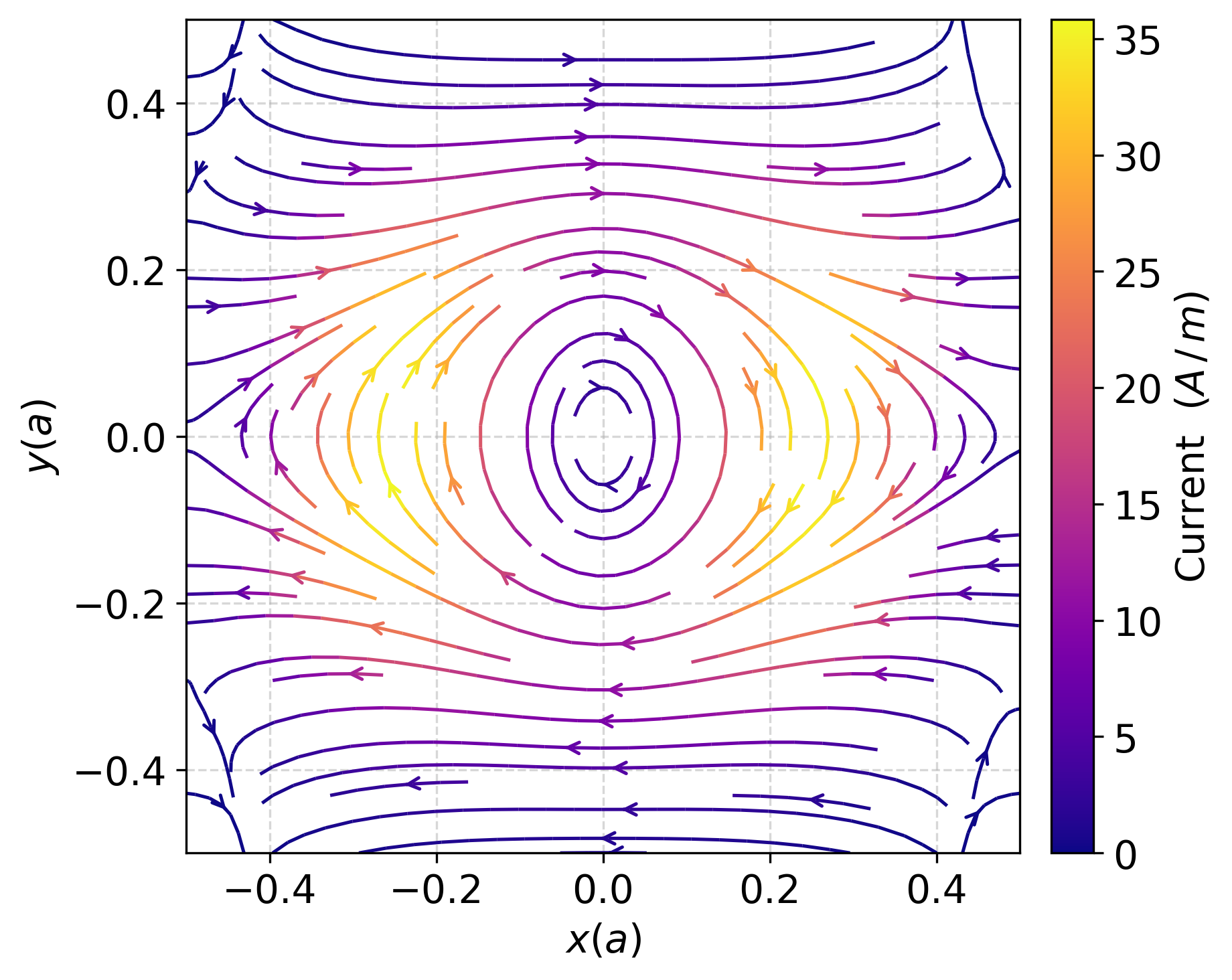} &
        \includegraphics[width=0.27\textwidth]{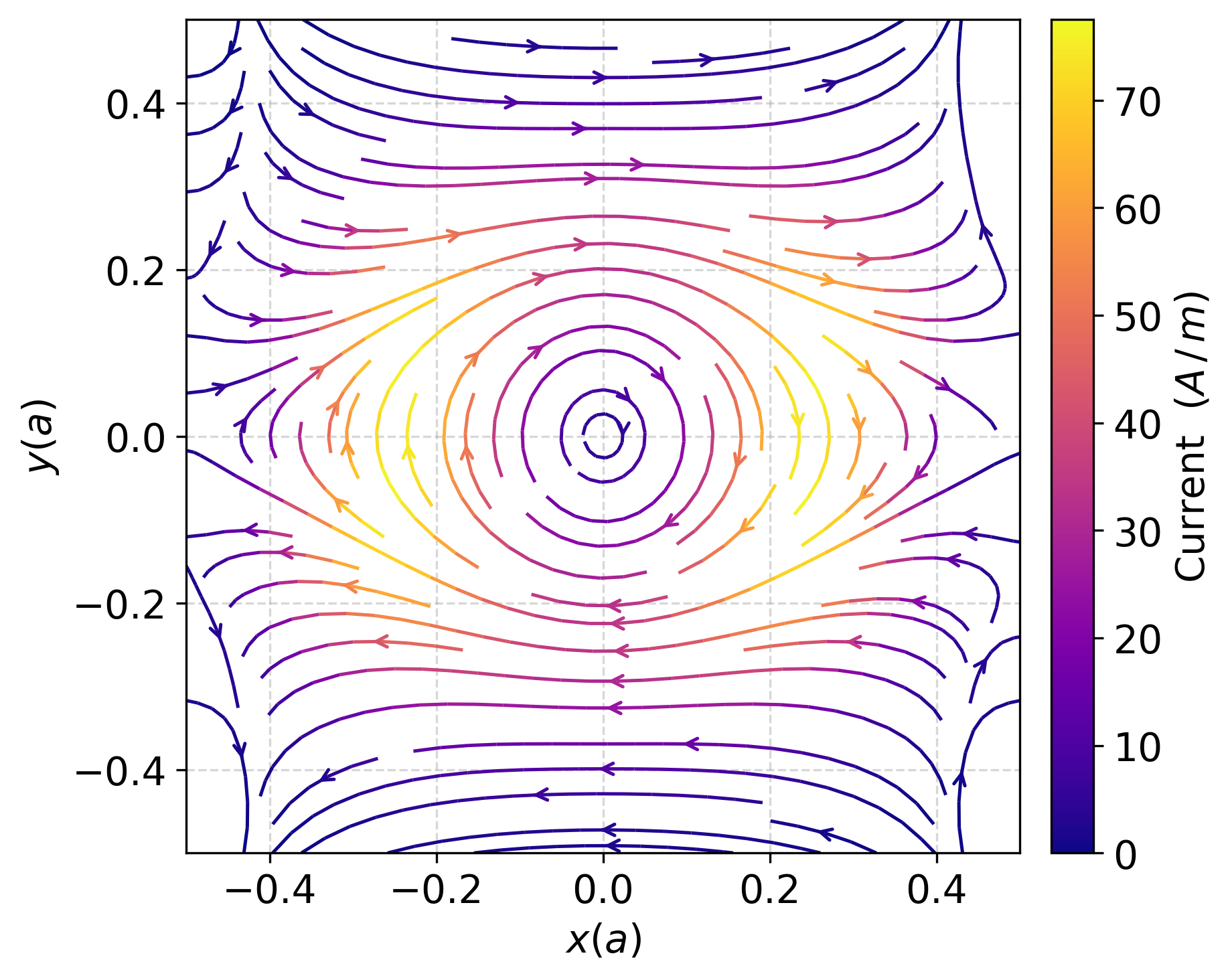} &
        \includegraphics[width=0.27\textwidth]{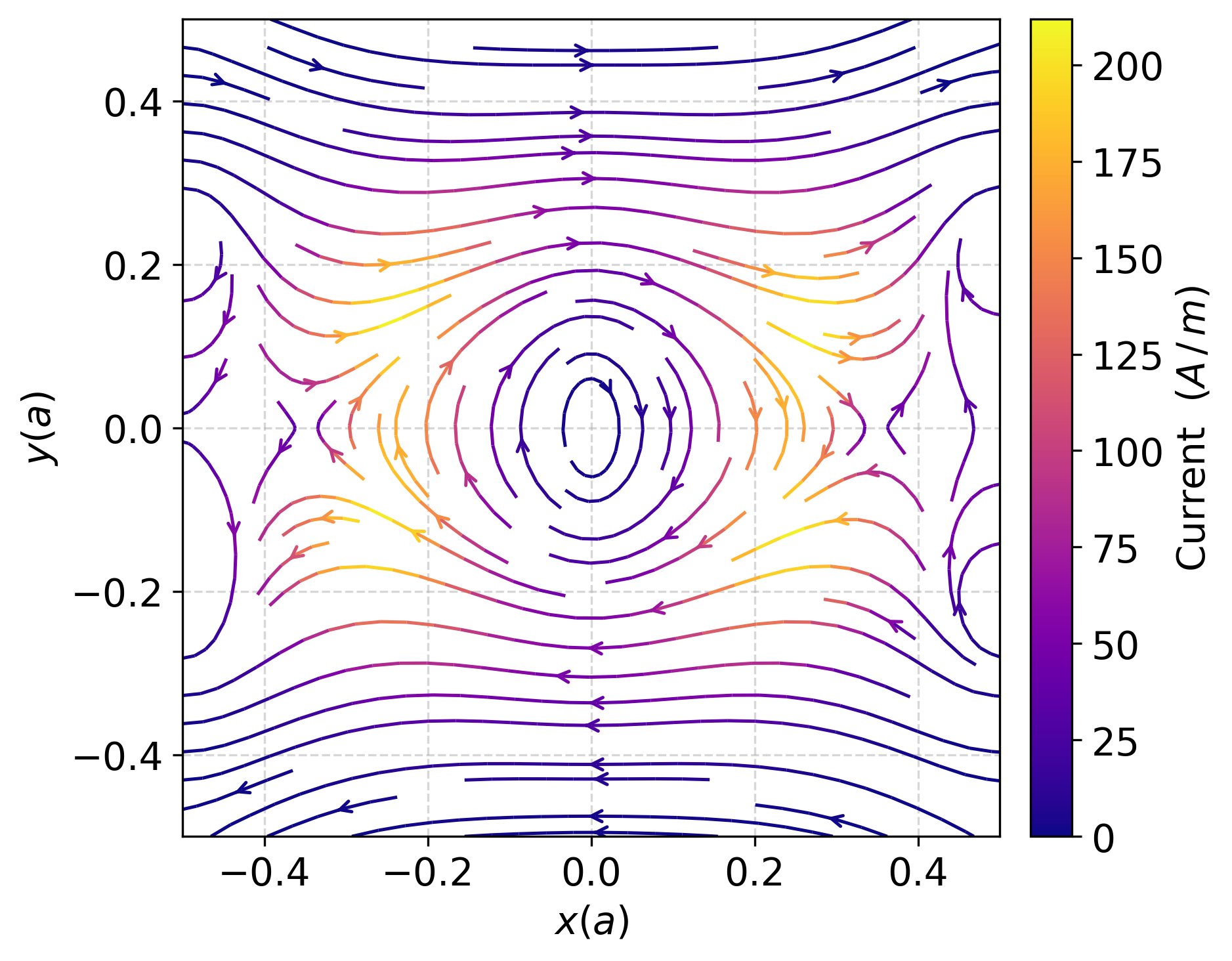} \\
    \end{tabular}
    \caption{Overall persistent currents (spin-up + spin-down) in the system. Rows correspond to Rashba coupling constants of $0$ and $40~\text{meV}\cdot\text{nm}$ (top to bottom), and columns to magnetic fields of $0$, $5$, $10$, and $23~\text{T}$ (left to right).}
    \label{fig:persistent_plus}
\end{figure}

Fig.~\ref{fig:persistent_minus} and Fig.~\ref{fig:persistent_plus} represent the difference ($e\cdot (j_{\uparrow} - j_{\downarrow})$) and the sum ($e\cdot (j_{\uparrow} +  j_{\downarrow})$) of the current densities for the spin-up and spin-down states. Four magnetic field values ($0$, $5$, $10$ and $23~\text{T}$) and three Rashba coupling constants ($0$, $20$, and $40~\text{meV} \cdot \text{nm}$) are considered here. In all cases, the total divergence of the current satisfies this $\int \nabla \cdot \mathbf{j} \, d^2\mathbf{r} = 0$ condition. Both the total (spin-up + spin-down) and spin-difference (spin-up - spin-down) persistent current densities increase locally with increasing magnetic field, indicating both confining and polarizing effects of the magnetic field. For the total current, variations in the Rashba coupling constant do not affect the local current magnitudes, and no local currents are observed for any of the selected values of the Rashba coupling constant in the absence of a magnetic field, as was expected. In contrast, the magnitude of the spin-difference current increases with the increase of Rashba parameter, in the absence of a magnetic field. We also emphasize that at $23\,\mathrm{T}$, the current magnitudes decrease along with increase of coupling constant. This is because of the dominance of the magnetic field effect over the Rashba coupling effect. 
Due to the interplay between the magnetic field and the coupling constant, the subfigures in the last row and second and third columns of Fig.~\ref{fig:persistent_minus}, show a distinctive localization of currents compared to those in the rows above, where the currents are mostly localized in circular patterns. In the former case, the currents appear to be localized in four corners and the circular pattern is significantly squeezed. This effect is more pronounced at \(5\,\mathrm{T}\) in the plot data, corresponding to a node in the miniband structure where the energy dispersion is suppressed. Interestingly, the magnitude of the current does not consistently decrease within the corresponding columns. As mentioned above regarding the total spin current, no difference is observed with changes in the Rashba coupling constant. Therefore,
\noindent
in Fig.~\ref{fig:persistent_plus} we include only two rows of data for $\alpha=0$ and $\alpha=40$\(\,\mathrm{meV} \cdot \mathrm{nm}\). The circular pattern is conserved across all plots. However, in the first column, no net current is observed, the appeared arrows correspond to currents as small as $10^{-13} A/m$, which lies in the range of the computational error,  indicating that the spin-up and spin-down currents cancel each other out in the absence of a magnetic field.

\begin{figure}[H]
    \centering
    \begin{minipage}[t]{0.50\textwidth}
        \centering
        \includegraphics[width=\linewidth]{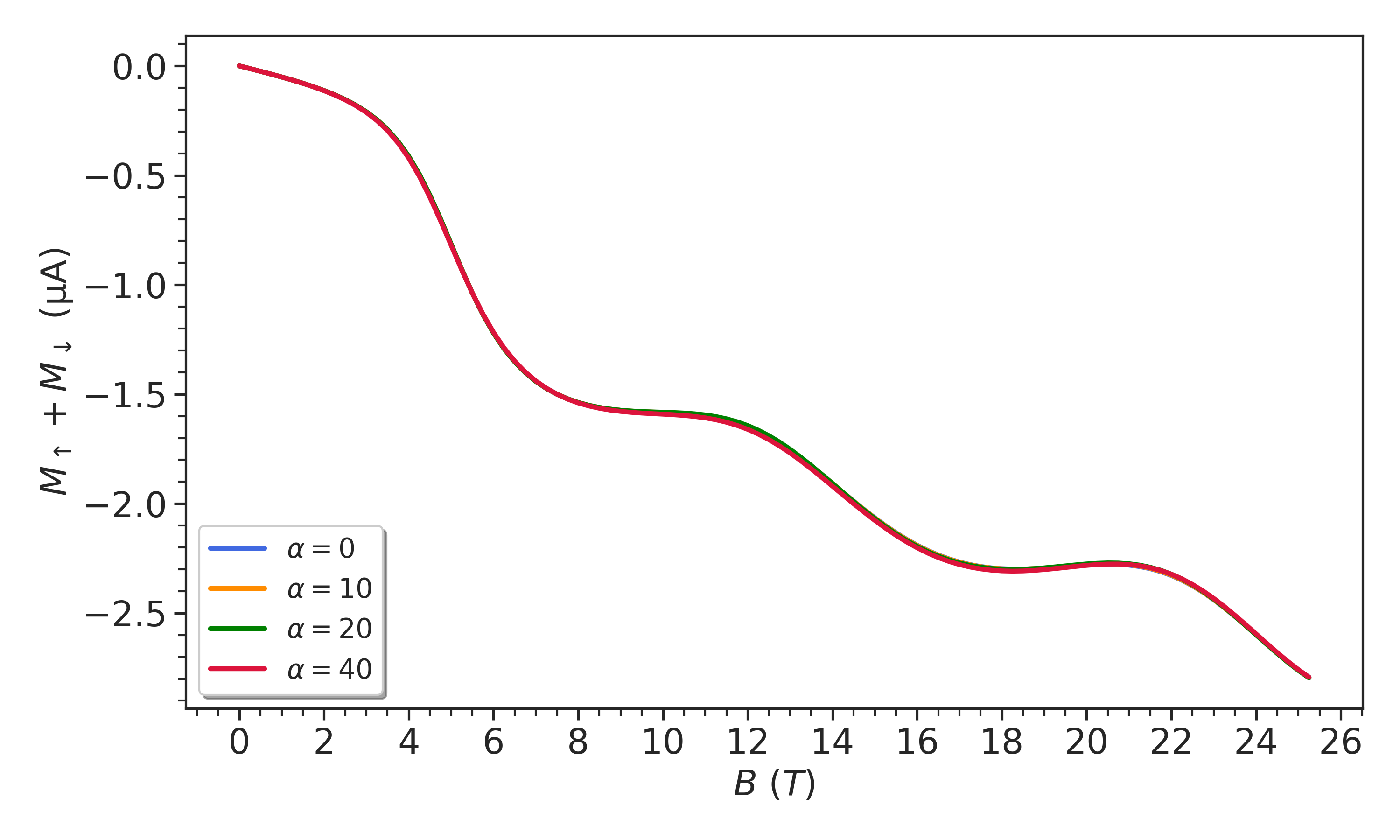}
    \end{minipage}%
    \hfill
    \begin{minipage}[t]{0.50\textwidth}
        \centering
        \includegraphics[width=\linewidth]{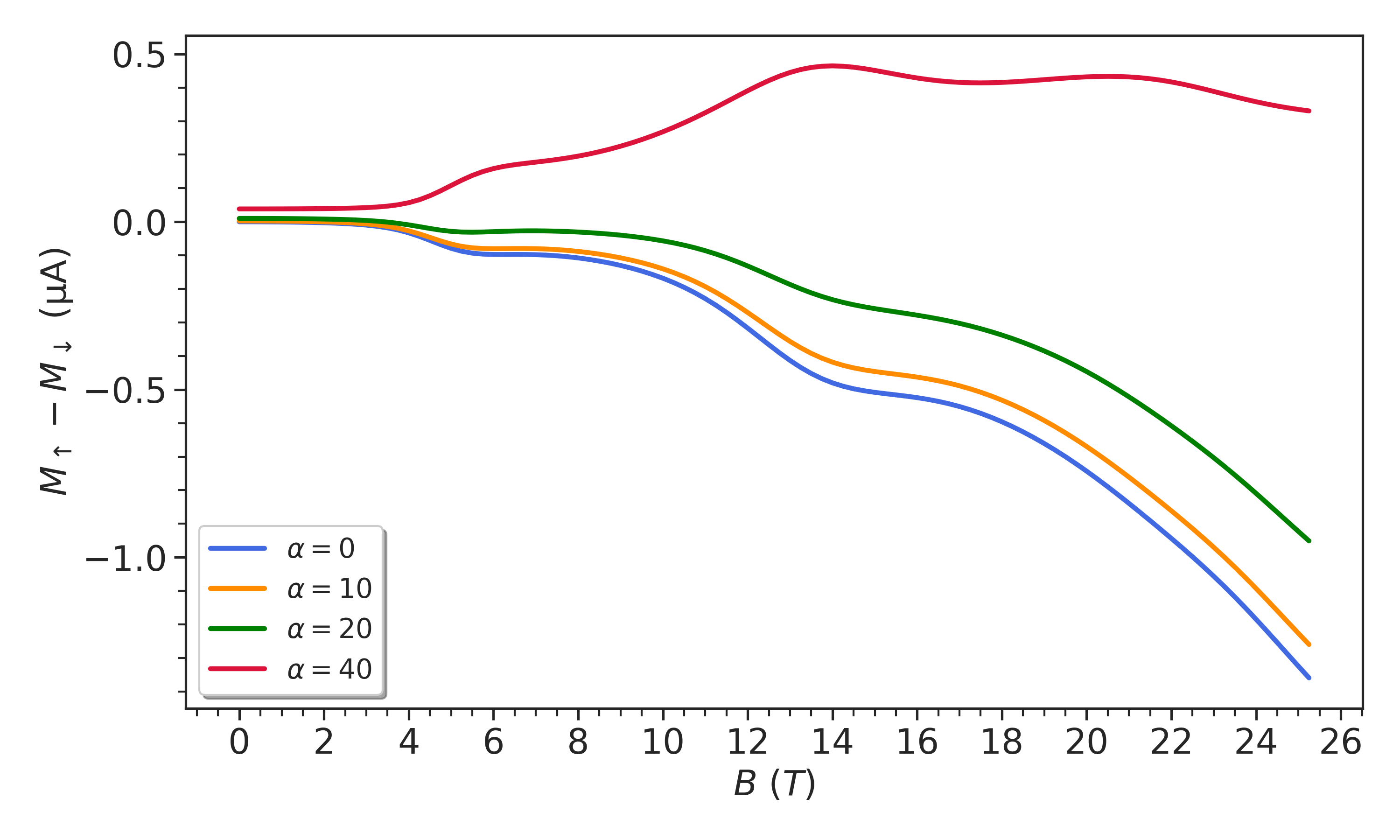}
    \end{minipage}
    \\[0.5cm]
    
    \caption{The left panel shows the spin-difference magnetization (spin-up - spin-down), while the right panel presents the spin-summed magnetization (spin-up + spin-down).}
    \label{fig:magnetization}
\end{figure}

The orbital magnetization of the system and the spin-difference magnetization, defined as the difference between the magnetizations of spin-up and spin-down states, both plotted as functions of the magnetic field, are shown in Fig.~\ref{fig:magnetization}. A diamagnetic response is observed as in our previous works \cite{harutyunyan2025magneto, mansoury2022signature}. It can be noticed that the Rashba coupling constant has no significant effect on the total magnetization (the left panel of Fig.~\ref{fig:magnetization}). The slight oscillations in total magnetization are related to the miniband nodes. It is evident that both \( M_{\uparrow} \) and \( M_{\downarrow} \) have negative values individually. The spin resolved magnetization has a monotonic negative slope; when the Rashba SOI is not very large (as is seen from the right panel of Fig.~\ref{fig:magnetization}) that is, increasing the magnetic field enhances the dominance of spin-down states. As the Rashba coupling constant $\alpha$ increases, the curve shifts upward, and even for $\alpha = 40\, \mathrm{meV \cdot nm}$, the spin-difference magnetization becomes positive, demonstrating Rashba-induced inversion. The sign flip of the magnetization is attributed to the competition between the Rashba and Zeeman effects, connected to the feature we observed in the  spin polarization of the persistent current, where at certain values of the coupling constant and magnetic field, we observe the change in the direction of the current circulation (compare the second and third subfigures of the second row with the second and third subfigures of the third row Fig.~\ref{fig:persistent_minus}).

\begin{figure}[H]
    \centering
    \begin{minipage}[t]{0.5\textwidth}
        \centering
        \includegraphics[width=\linewidth]{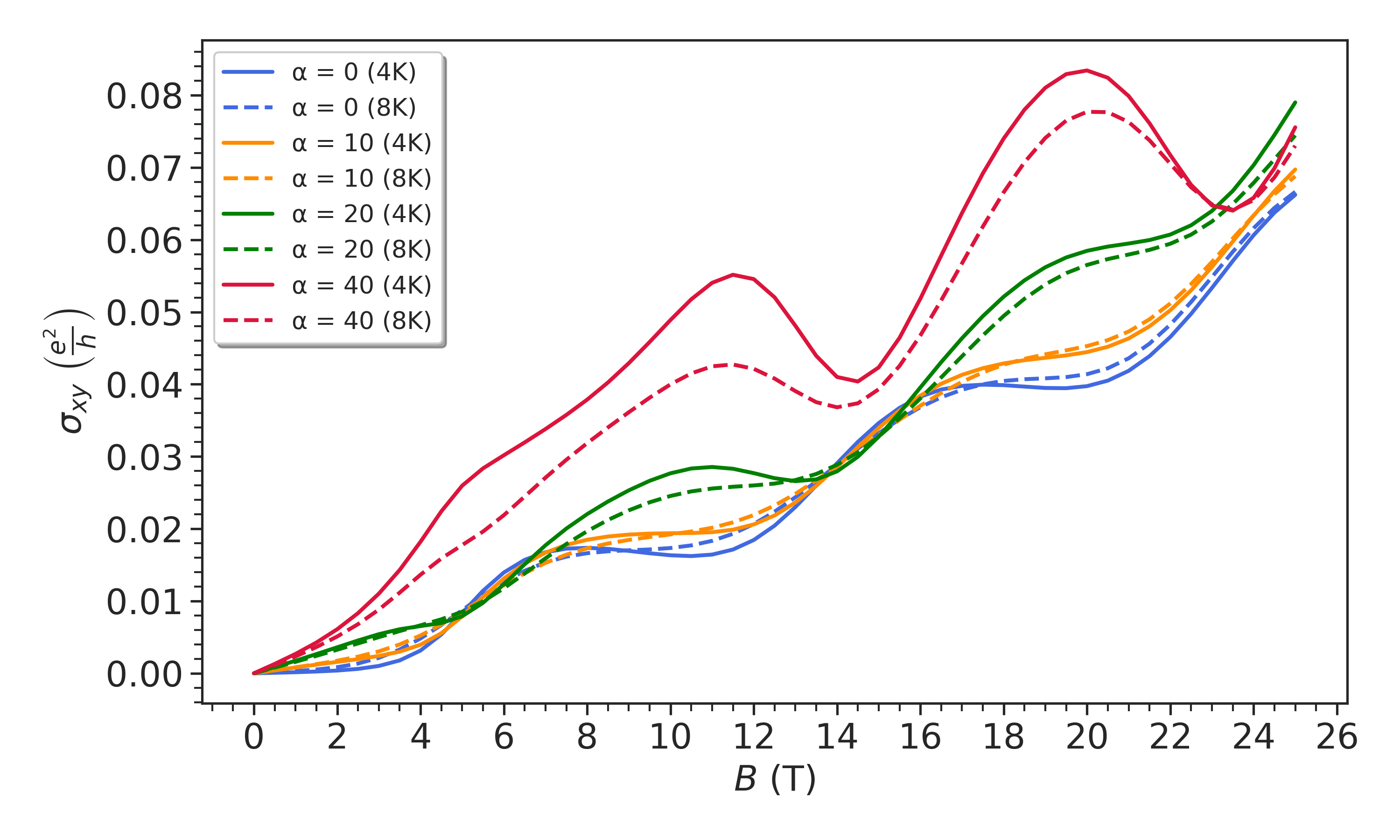}
    \end{minipage}%
    \hfill
    \begin{minipage}[t]{0.5\textwidth}
        \centering
        \includegraphics[width=\linewidth]{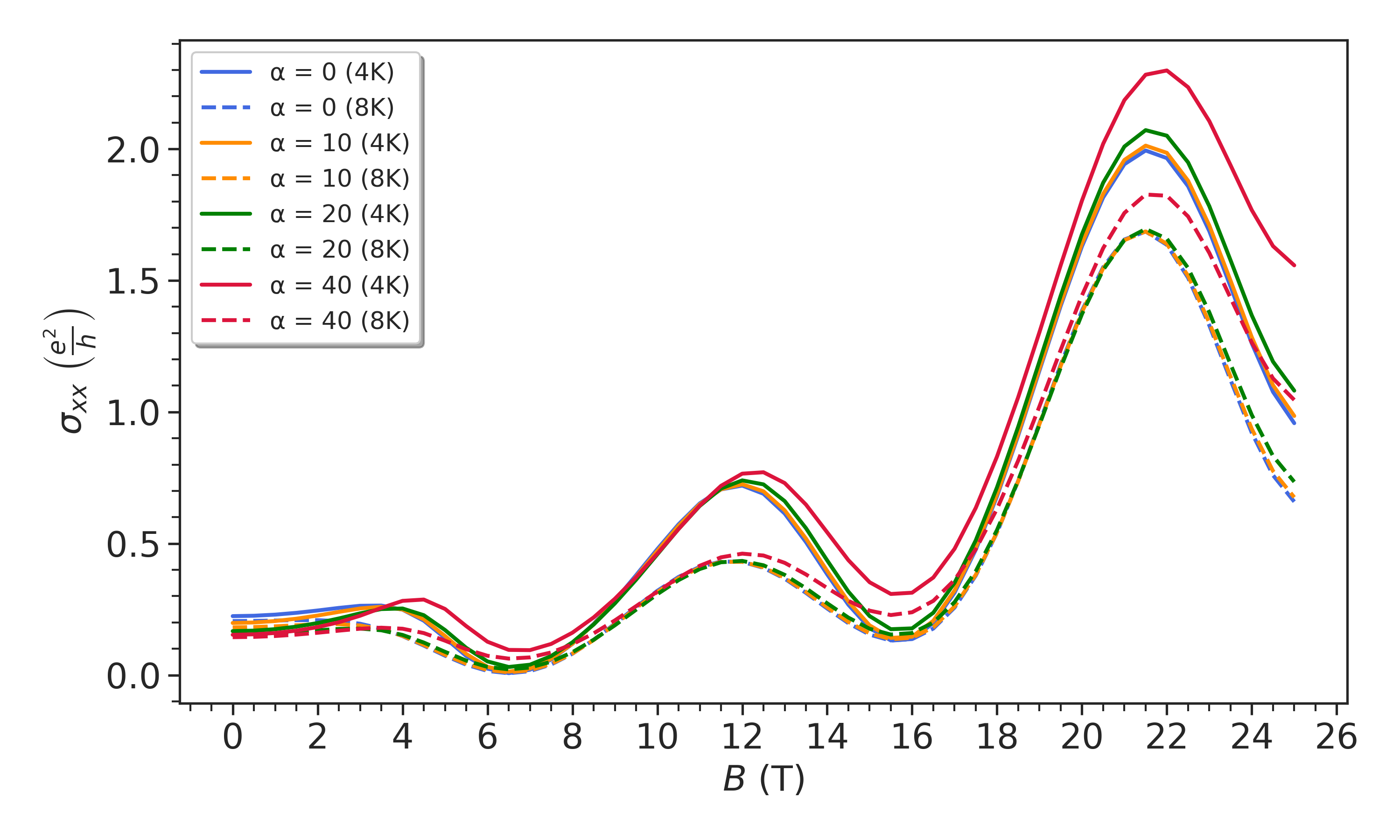}
    \end{minipage}
    \caption{Transverse ($\sigma_{xy}$, left) and longitudinal ($\sigma_{xx}$, right) conductance as functions of magnetic field, for temperatures of 4\,K (solid lines) and 8\,K (dashed lines).}
    \label{fig:conductivity}
\end{figure}

In Fig.~\ref{fig:conductivity}, the conductivity calculated using the Kubo-Greenwood formula is presented. Note that, the Hall conductivity regime is not captured. In other words, our calculations include all possible values of the quasi-momentum \( k_x \). This, in turn, results in a magnetic-field-dependent chemical potential, in contrast to the constant chemical potential in the quantum Hall regime. The ladder-like structure of \( \sigma_{xy} \), as well as the Shubnikov de Haas-
\noindent
like oscillatory behavior of \( \sigma_{xx} \) depicted in Fig.~\ref{fig:conductivity}, are more closely associated with the nodes of the minibands than with the quantum Hall effect. The behavior of $\sigma_{xx}$ emphasizes the significant increase in conductivity at magnetic field values corresponding to these nodes.

\noindent
At these points, the chemical potential coincides with a highly degenerated energy level (see Fig.~\ref{fig:bandstructure} which includes all the miniband states, leading to a significant increase in the intraminiband transitions, and hence in \( \sigma_{xx} \). Note, that the central region of the miniband (where the density of states varies significantly with the increase of magnetic field) primarily contributes to $\sigma_{xx}$, while the miniband edges (where the density of states varies slightly with the magnetic field) mostly contribute to $\sigma_{xy}$. As a result, the change in the magnetic field leads to a clear oscillation in \( \sigma_{xx} \), while the oscillation of \( \sigma_{xy} \), for not very large values of the Rashba coupling parameter, mostly exhibits a ladder-like behavior.

\begin{figure}[h]
    \centering
    \includegraphics[width=0.5\textwidth]{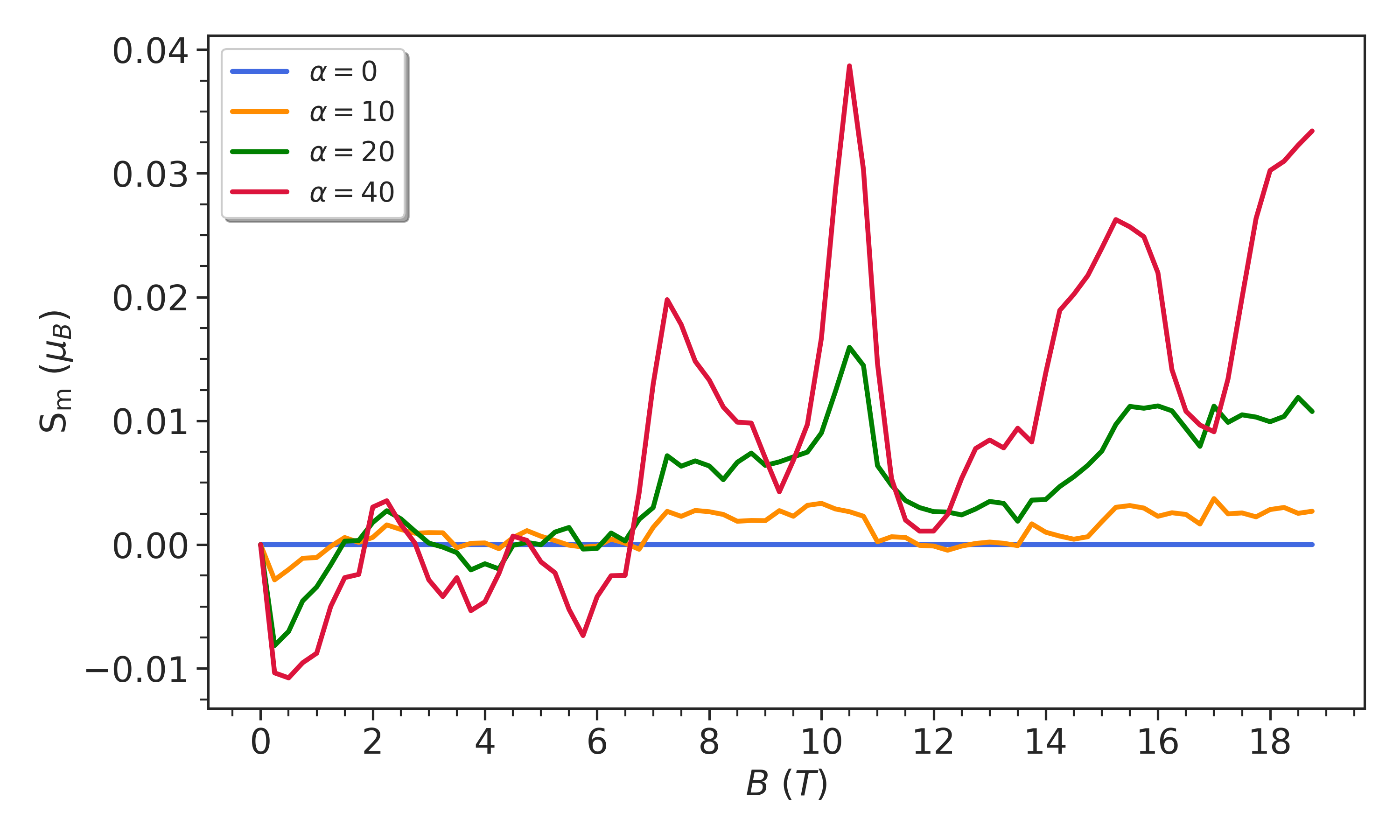}
    \caption{Spin polarization as a function of magnetic field.}
    \label{fig:spin_magnetization}
\end{figure}

The spin polarization is calculated using Eq.~\ref{eq:spin}. The results, which are represented in Fig.~\ref{fig:spin_magnetization}, show that without Rashba interaction, there is no spin polarization. However, for considerable Rashba SOI, we observe pronounced shifts, for certain ranges of magnetic field induction. A similar pattern of the spin polarization is described in \cite{lyapilin2007oscillations} for a two dimensional Rashba system. Thus, our system exhibits expected behavior under the Rashba effect: as the Rashba coupling constant increases, there is a stronger net preference for one spin orientation, for stronger external magnetic fields.
\section*{\normalsize Summary}
In conclusion, we have investigated the magneto-transport properties of a two-dimensional non-interacting electron gas in a 1D chain of planar quantum rings, under the influence of the Rashba SOI and a transverse magnetic field. An external potential employed is designed to capture the key features of the structure, namely periodicity along the $x$-direction, confinement in the $y$-direction, and the topology associated with quantum rings. Within this framework, we reproduce highly degenerate energy levels at specific magnetic field values obtained in our previous work. This degeneracy, which arises due to translational symmetry and magnetic phase interference, is now observed in the presence of the Rashba effect.
The competition between the Zeeman and Rashba splittings gives rise to nontrivial behavior of considered properties of the system.

Despite changes in Rashba coupling, the electron density distribution remains almost constant. However, it is significantly modified by variations in the magnetic field. The increase in the magnetic field give rise to an enhanced electron density around $x$-axis.
The signature of miniband nodes observed in previous work is preserved in the total magnetization patterns, which displays oscillatory behavior. A shift from diamagnetic to paramagnetic behavior in the spin-difference orbital magnetization is observed at high Rashba coupling. In particular, at sufficiently large values of the Rashba coupling constant, the spin-polarized currents undergo an inversion. 

The components of the magnetoconductance tensor, $\sigma_{xy}$ and $\sigma_{xx}$, display ladder-like and oscillatory patterns, respectively, reminiscent of the Hall effect and Shubnikov–de Haas oscillations under varying magnetic field, yet they relate to the periodic collapse of minibands, resulting in significant oscillations in the density of states.

Finally, we present the spin polarization as a function of the Rashba coupling constant and the magnetic field. The results reveal rich oscillatory behavior; as the Rashba coupling constant increases, one spin orientation becomes dominant, especially at larger magnetic field values.
The obtained results provide physical insights into experimentally relevant electronic, transport, and spin characteristics of modulated quasi-2D semiconductor structures, which is an important contribution in the field of advanced 2D-based materials for magneto-transport and spintronics applications.

\section*{\normalsize Acknowledgements}
The author would like to thank Dr. Vram Mughnetsyan and Prof. Vidar Gudmundsson for their valuable discussions and insightful suggestions throughout the course of this work. This research was supported by Higher Education and Science Committee of RA
under Grants No. 24AA-1C001 and .


\end{document}